\documentclass[prd,preprint,tightenlines,11pt,amsmath,amssymb]{revtex4}

\usepackage{graphicx}       
\usepackage{dcolumn}        
\usepackage{multirow}
\usepackage{bm}             
\usepackage{psfrag}
\usepackage{color}

\newcommand{\beq}{\begin{equation}}
\newcommand{\eeq}{\end{equation}}
\newcommand{\mass}{\mu}

\newcommand{\p}{\rho}

\newcommand{\dr}{\delta r}
\newcommand{\dth}{\delta \theta}
\newcommand{\dph}{\delta \phi}
\newcommand{\Prr}{P_{rr}}

\newcommand{\MM}{\hat{\mathcal{M}}}
\newcommand{\Mtil}{\tilde{\mathcal{M}}}

\newcommand{\nn}{\nonumber}

\newcommand{\sinq}{\sin \theta}
\newcommand{\sinqsq}{\sin^2 \theta}
\newcommand{\cosq}{\cos \theta}

\newcommand{\half}{\tfrac{1}{2}}
\newcommand{\drs}{\partial_{r*}}
\newcommand{\dt}{\partial_t}
\newcommand{\dq}{\partial_\theta}

\newcommand{\tmax}{t_\text{max}}

\mathchardef\mhyphen="2D

\providecommand{\e}[1]{\ensuremath{\times 10^{-#1}}}

\newcommand{\ellK}{\text{elK}}
\newcommand{\ellE}{\text{elE}}

\newcommand{\alp}{\alpha}
\newcommand{\bet}{\beta}
\newcommand{\gam}{\gamma}

\newcommand{\eps}{\epsilon}

\newcommand{\Delt}{\triangle t}
\newcommand{\Delr}{\triangle r_\ast}
\newcommand{\Delq}{\triangle \theta}

\providecommand{\e}[1]{\ensuremath{\times 10^{-#1}}}
\newcommand{\msp}{\phantom{-}}

\newcommand{\hh}{{\overline{{h}}}}

\newcommand{\Y}{Y^*_{lm}}

\newcommand{\Chi}{\hat{\chi}}

\newcommand{\ZZ}{\mathcal{Z}}

\newcommand{\Hplus}{\mathcal{H}^+}

\newcommand{\ghat}{\hat{\gamma}}

\newcommand{\Dop}{\hat{\mathcal{D}}}

\begin{document}

\preprint{}

 \title{Self-force via $m$-mode regularization and 2+1D evolution: \\ 
 III. Gravitational field on Schwarzschild spacetime}

\author{Sam R. Dolan}
 \email{s.dolan@shef.ac.uk}
 \affiliation{%
Consortium for Fundamental Physics, School of Mathematics and Statistics, 
University of Sheffield, Sheffield S3 7RH, United Kingdom.
  \\
}%
\author{Leor Barack}
 \email{l.barack@soton.ac.uk}
 \affiliation{%
 School of Mathematics, University of Southampton, Southampton SO17 1BJ, United Kingdom. \\
 }%

\date{\today} 

\begin{abstract}
This is the third in a series of papers aimed at developing a practical time-domain method for self-force calculations in Kerr spacetime. The key elements of the method are (i) removal of a singular part of the perturbation field with a suitable analytic ``puncture'', (ii) decomposition of the perturbation equations in azimuthal ($m$-)modes, taking advantage of the axial symmetry of the Kerr background, (iii) numerical evolution of the individual $m$-modes in 2+1-dimensions with a finite difference scheme, and (iv) reconstruction of the local self-force from the mode sum. Here we report a first implementation of the method to compute the {\em gravitational} self-force. We work in the Lorenz gauge, solving directly for the metric perturbation in 2+1-dimensions. The modes $m=0,1$ contain nonradiative pieces, whose time-domain evolution is hampered by certain gauge instabilities. We study this problem in detail and propose ways around it. In the current work we use the Schwarzschild geometry as a platform for development; in a forthcoming paper---the fourth in the series---we apply our method to the gravitational self-force in Kerr geometry. 
\end{abstract}

\maketitle

%
%
%
%

\section{Introduction\label{sec:intro}}
The study of self-forces in curved spacetimes is enjoying a surge of activity, motivated in part by the tantalising prospect of detecting gravitational-wave radiation generated by compact bodies in orbit around black holes \cite{LIGO, eLISA}. The roots of the self-force program can be traced back to Dirac's treatment of radiation reaction in classical electromagnetism \cite{Dirac}, and its extension to curved spacetimes by DeWitt and Brehme \cite{DeWitt:Brehme}. Contemporary interest was ignited in 1997 with the first derivations of an expression for the gravitational self-force (GSF), now known as the MiSaTaQuWa formula \cite{Mino:Sasaki:Tanaka:1997, Quinn:Wald:1997}. In subsequent years, there has been careful work on the foundations of the theory \cite{Barack:Ori:2001, Detweiler:Whiting:2003, Gralla:Wald:2008, Harte:2008, Gralla:Harte:Wald:2009, Harte:2009, Harte:2010, Pound:2010, Pound:2010b}, and on various practical schemes for numerical computations \cite{Diaz-Rivera:2004, Barack:Lousto:2005, Haas:Poisson:2006, Haas:2011, Barack:Golbourn:Sago:2007, Barack:Sago:2007, CDOW1, CDOW2, Canizares:Sopuerta:2009,  Canizares:Sopuerta:Jaramillo:2010, Thornburg:2009, Thornburg:2010, Warburton:Barack:2010, Hopper:Evans:2010, Keidl:Shah:Friedman:2010, Shah:Keidl:Friedman:2011, Warburton:Akcay:2011, Vega:Wardell:Diener:2011, Wardell:Vega:Thornburg:Diener:2011, Akcay:2011, Shah:2012gu} (see reviews \cite{Barack:2009, Poisson:Pound:Vega:2011}).

The self-force program is broadly motivated by the two-body problem in General Relativity, and, more specifically, by the Extreme Mass Ratio Inspiral (EMRI) system in astrophysics \cite{Barack:Cutler:2004, Thornburg:gwnotes, eLISA}, in which a compact object of mass $\mass$ (e.g., a stellar-mass black hole or neutron star) is in a strong-field bound orbit around a massive black hole of mass $M$, such that the mass ratio $\eta \equiv \mass / M$ is very small. This scenario is difficult to handle in Numerical Relativity (but see, e.g.,~\cite{Lousto:Nakano:Zlochower:Campanelli:2010a, Lousto:Nakano:Zlochower:Campanelli:2010b, Lousto:Zlochower:2010, Nakano:2011} for recent progress). On the other hand, it is tailor-made for black hole perturbation theory \cite{Regge:Wheeler:1957, Zerilli:1970, Vishveshwara:1970, Teukolsky:1972, Chandrasekhar:1983}, which is based around expansions in powers of the mass ratio $\eta$. The compact mass generates a metric perturbation (MP) $h_{\alp \bet}$ at order $\mathcal{O}(\eta)$, which in turn leads to a back-reaction on the motion of the compact mass. This may be interpreted as a self-force $F_{\text{self}}^{\alp} [\propto \mathcal{O}(\eta^2)]$, which acts to deflect the motion away from a geodesic of the background black hole spacetime. Two key issues arise naturally. The first is that of \emph{regularization}: how may a meaningful GSF be obtained from a metric perturbation which diverges at the location of the compact mass? The second is that of \emph{gauge}: which predictions of the theory are independent of the choice of gauge of the MP? Both issues have been addressed in some depth by the community (see Refs.~\cite{Mino:Sasaki:Tanaka:1997, Barack:Ori:2000, Barack:Mino:2002,Detweiler:Whiting:2003, Detweiler:Poisson:2004, Pound:2010, Pound:2012nt, Gralla:2012db,Pound:2012dk} and \cite{Barack:Ori:2001, Detweiler:2008, Sago:Barack:Detweiler:2008, Barack:Sago:2009, Barack:Sago:2011, Gralla:2011ke,Gralla:2011zr}), and the self-force program has undoubtedly now come of age.

A significant milestone was passed in 2008, with the first comparison of conservative GSF effects with Post-Newtonian (PN) theory \cite{Detweiler:2008}, and the comparison between two independent GSF implementations, based on distinct gauges \cite{Sago:Barack:Detweiler:2008}. Comparison of strong-field GSF effects with Post-Newtonian (PN) theory \cite{Favata:2010, Blanchet:Detweiler:LeTiec:Whiting:2010a, Blanchet:Detweiler:LeTiec:Whiting:2010b, Barack:Damour:Sago:2010} and with Numerical Relativity (NR) \cite{LeTiec:Mroue:Barack:2011} have also been made. Some recent highlights in the self-force program include (for instance) using the GSF for calibrating the Effective One-Body formalism \cite{Damour:2009, Barack:Damour:Sago:2010, Barausse:Buonanno:LeTiec:2011,Akcay:2012ea} and other syntheses \cite{LeTiec:Blanchet:Whiting:2011, LeTiec:2011, Damour:Nagar:Pollney:Reisswig:2011}; the first long-term evolutions of self-forced orbits \cite{Warburton:Akcay:2011, Diener:Vega:Wardell:Detweiler:2011}; formulations of the second-order-in-mass-ratio problem~\cite{Rosenthal:2006, Detweiler:2011, Galley, Pound:2012nt,Gralla:2012db,Pound:2012dk}; and the study of the implications of self-force for the cosmic censorship hypothesis \cite{Barausse:Cardoso:Khanna:2010, Barausse:Cardoso:Khanna:2011, Isoyama:Sago:Tanaka:2011, Zimmerman:Vega:Poisson:Haas:2012}. Nevertheless, and despite rapid progress, all calculations of the GSF in the literature are rooted to the non-rotating (Schwarzschild) black hole background. A practical formulation for computing the GSF on the rotating black hole spacetime, represented by Kerr's solution, is now a high priority for the community. 

This paper is the third in a series aimed at developing a practical time-domain scheme for Kerr GSF calculations. Our approach is based on $m$-mode regularization in the time domain, a method first proposed in Refs.~\cite{Barack:Golbourn:2007, Barack:Golbourn:Sago:2007}. In Papers I \cite{Dolan:Barack:2011} and II \cite{DBW:2011} of this series, we established a proof-of-principle by applying the method to a simpler toy model, the problem of the scalar-field self-force on Schwarzschild and Kerr spacetimes. In the present work we apply the method to compute the GSF itself, and we take the opportunity to validate our results against well-known Schwarzschild results. Details of a GSF calculation on Kerr will follow in Paper IV.

Our method combines several steps: (i) formulation of the linearized Einstein equations as a Z4 system with constraint damping, using a (generalized) Lorenz gauge, (ii) removal of a certain singular part of the MP with a suitable puncture; (iii) the decomposition of the system into azimuthal ($m$-)modes; (iv) numerical evolution of the resulting 2+1D equations [one set of (generally) 10 coupled equations for each $m$-mode] with a finite difference scheme, and finally (v) reconstruction of the GSF from convergent sums over $m$-modes. 

In line with the original philosophy, the aim is to develop a scheme in which the regularization of the MP is conducted in the Lorenz gauge. A key advantage is that, in this gauge, the singular part of the MP, which informs our choice of puncture, is (in some sense) ``isotropic'' in the vicinity of the compact mass, and consequently the regularization procedure is well-understood. On the Schwarzschild spacetime, the linearized Einstein equations for the Lorenz-gauge MP are fully separable into Fourier-harmonic modes (using tensor spherical harmonics) \cite{Akcay:2011}. As far as we know this is \emph{not} possible in Kerr spacetime, our ultimate goal. Hence a key motivation for working in the frequency domain has been lost: the Lorenz-gauge system cannot be reduced to a set of ordinary differential equations. Time-domain schemes offer distinct advantages (as explained in Papers I and II), as they are well-suited to the study of highly-eccentric or unbound orbits, and they allow the orbital evolution of self-forced orbits to proceed in a self-consistent way \cite{Diener:Vega:Wardell:Detweiler:2011} (unlike frequency-domain schemes, which must resort to use a ``quasi-equilibrium'' approach, based on osculating geodesics \cite{Pound:Poisson:2008, Warburton:Akcay:2011}). Furthermore, the linearized Einstein equations in Lorenz gauge provide a natural starting point for any time-domain scheme, because they form a Z4 system (Sec.~\ref{subsec:gauge-choice} and Ref.~\cite{Bona:2003, Bona:Palenzuela:2004}), which is hyperbolic in character. 

Notwithstanding, a frequency-domain approach to the problem is also being developed, with significant progress over past few years. In the approach of Shah {\it et al.~}\cite{Keidl:Shah:Friedman:2010, Shah:Keidl:Friedman:2011}, the MP in a (modified) radiation gauge is reconstructed from Hertz potentials obeying ordinary differential equations, which offers significant computational advantage. This approach has recently led to a first computation of a GSF effect in Kerr spacetime \cite{Shah:2012gu}. It is hoped that the same method could eventually allow computation of the GSF itself, for generic bound orbits.

The Kerr spacetime is axisymmetric, and hence it is natural to decompose the MP into azimuthal $m$-modes. This decomposition has two key advantages: firstly, it reduces the computational burden for the time domain scheme, which now proceeds in 2+1 dimensions (2+1D) rather than in 3+1D. Secondly, it allows us to consider the $m=0$ and $m=1$ modes separately from the rest of the system. These modes contain non-radiative degrees of freedom that require more careful consideration of initial data and conservation laws. In addition, in our Lorenz gauge formulation it turns out that the $m=0$ and $m=1$ modes are susceptible to gauge instabilities which disrupt the numerical evolutions. Controlling these gauge modes is a theme of this work, and it motivates the use of a generalized version of the Lorenz gauge. The cost of $m$-mode decomposition is two-fold: the $m$-mode versions of the puncture (and effective source) are found by evaluating integrals, which can be computationally costly, and the self-force must be reconstructed from a sum over modes. The latter does not seem to pose a practical problem, as the mode sum converges rather rapidly with currently available punctures (the convergence properties of the $m$-mode sum were carefully investigated in Ref.~\cite{Barack:Golbourn:Sago:2007} and in Papers I and II). 

The remainder of this paper is organized as follows. In Sec.~\ref{sec:fundamentals} we present a preview of the fundamentals of our approach, giving details of the linearized Einstein equations (\ref{subsec:linearized-eqns}), the Z4 scheme in (generalized) Lorenz gauge with constraint damping (\ref{subsec:gauge-choice}), and the $m$-mode regularization scheme (\ref{subsec:regularization}). This section also includes a discussion of relevant conservation laws (Sec.~\ref{subsec:conservation}), which, though presented elsewhere \cite{Abbott:Deser:1982}, may be somewhat unfamiliar to many in the self-force community. In Sec.~\ref{sec:implementation} we describe some features of our implementation for circular orbits on Schwarzschild spacetime. Here we give explicitly the $m$-mode field equations (\ref{subsec:m-mode-eqs}), the $m$-mode decomposition of the puncture and effective source (\ref{subsec:mmode-reg}), the physical boundary conditions (\ref{subsec:bc}), and the details of the numerical implementation (\ref{subsec:impdetails}). In Sec.~\ref{sec:results:1} we present a selection of initial numerical results (\ref{subsec:slices}), validate the results of our code for the modes $m \ge 2$ (\ref{subsec:validation}), and describe the manifestation of gauge mode instabilities in the $m=0$ and $m=1$ modes (\ref{subsec:m01-numerical}). In Sec.~\ref{sec:low-multipoles} we study the low multipoles analytically, and present closed-form solutions for certain Lorenz-gauge modes that we diagnose as implicated in our numerical instabilities. 
In Sec.~\ref{sec:stabilization} we describe two practical methods for mitigating these gauge-mode instabilities. The first involves the use of a particular generalized Lorenz gauge, which stabilizes the $m=0$ system, and which reduces to the Lorenz gauge at late time. The second is a frequency filter for the $m=1$ mode, which eliminates the undesirable gauge modes. In Sec.~\ref{sec:results:2} we present results for the total GSF, and compare with the literature. We conclude in Sec.~\ref{sec:outlook} with a discussion of the way forward towards Kerr calculations. The appendices contain some technical details of our calculations. 

Throughout this work we use geometrized units, with $G=c=1$.

\section{Fundamentals\label{sec:fundamentals}}

In this section we describe the elements of our approach. Subsection \ref{subsec:motion} is a  recap of the bare essentials of GSF formulation. In Sec.~\ref{subsec:linearized-eqns} we state the field equations, and in Sec.~\ref{subsec:gauge-choice} we describe the Z4 formulation with gauge constraint damping that underpins our time-domain approach. Subsection \ref{subsec:regularization} describes our puncture scheme and its $m$-mode implementation. Finally, in Sec.~\ref{subsec:conservation} we show that the linearized Einstein equations give rise to one quasilocal conservation law for each Killing vector that the (Ricci-flat) background admits. It seems that this construction, which can help us to understand the physical content of the non-radiative multipoles, has not been widely appreciated in the self-force literature so far. 

\subsection{Gravitational self-force formulation\label{subsec:motion}}

The aim of the GSF program is to provide an effective description of the general-relativistic motion of a ``small'' compact body, of mass $\mu$ and characteristic size $l_s\sim\mu$, on a dominant background spacetime with typical length scale $\mathcal{R}\gg\mu$. 
For an EMRI, expansion in the point-particle limit $l_s \rightarrow 0$ corresponds (approximately) to an expansion in powers of the small mass ratio $\eta$. At leading order, the small body behaves as a point particle moving on a geodesic of the background spacetime. At subleading order, the point particle acts as a source for a MP $h_{\alp \bet} \sim \mathcal{O}(\eta)$, which exerts its influence back on the body to generate a GSF. Some care is needed when handling expansions, because the MP $h_{\alp \bet}$ generated by a point-particle is singular, diverging like $\sim \mu / r$ in the limit $r\rightarrow0$ (where $r$ is a measure of the spatial distance to the body). Mino, Sasaki and Tanaka obtained the first expression for the GSF \cite{Mino:Sasaki:Tanaka:1997} by applying a method of matched asymptotic expansions, in which ``inner'' and ``outer'' expansions for the MP were matched in a buffer region defined by $\mu / M \ll r / \mathcal{R} \ll 1$. The key expression was also obtained using an axiomatic approach by Quinn and Wald \cite{Quinn:Wald:1997}. Further rigorous work has subsequently put the theory of GSF on a firm footing \cite{Gralla:Wald:2008, Pound:2010b} (see \cite{Poisson:Pound:Vega:2011} for an introduction to the literature). Below we state the key results.

The motion of the compact body may be described by a worldline $\gamma$ on the background spacetime. Let us parameterize this worldline by $x^{\alp}=z^{\alp}(\tau)$, with four-velocity $u^{\alp} \equiv dz^{\alp} / d\tau$, where $\tau$ is proper time along $\gamma$. For finite $\eta$, the motion is governed by the self-forced equation 
\beq
\mu a^{\alp} = F_{\text{self}}^{\alp} \sim \mathcal{O}(\eta^2) ,
\label{eom}
\eeq
where $a^{\alp} = u^{\alp}_{\ ;\beta}u^{\beta}$ is the self-acceleration. We use a semicolon to denote covariant differentiation with respect to the background spacetime, and indices are raised and lowered using the background metric $g_{\alp \bet}$. 
At leading order in $\eta$, the GSF $F_{\text{self}}^\alp$ is given by
\beq
F_{\text{self}}^\alp  =  - \frac{\mu}{2} \left( g^{\alp \bet} + u^{\alp} u^{\bet} \right) \left( 2 h^{R}_{\bet \gamma ; \delta} - h^{R}_{\gamma \delta ; \bet} \right) u^{\gamma} u^{\delta}  . \label{a-sf} \\
\eeq 
Here $h_{\alp \bet}^{R}$ is the Detweiler--Whiting ``R-field'' \cite{Detweiler:Whiting:2003}, which is a particular locally-defined smooth vacuum solution of the MP equations, obtained from the full (retarded) MP via a procedure described in Sec.~\ref{subsec:regularization} below. The retarded MP itself, $h_{\alp \bet}$, is a solution to the linearized Einstein equations (Sec.~\ref{subsec:linearized-eqns}) with a point-particle source, i.e.
\beq
T_{\alp \bet}(x^\nu) = \mu \int_{-\infty}^{\infty} (-g)^{-1/2} \delta^4 \left[ x^{\nu} - z^\nu(\tau) \right] u_{\alp} u_{\bet} d\tau ,  \label{point-particle}
\eeq 
where $g$ is the metric determinant for the background.

In this work we will not solve Eq.\ (\ref{eom}) for $z^{\alpha}(\tau)$. Rather, we will compute the GSF $F_{\text{self}}^\alp$ along a {\em fixed} geodesic of the background geometry, ignoring the back reaction from the GSF on the motion [i.e., we assume $z^{\nu}(\tau)$ in Eq.\ (\ref{point-particle}) represents a geodesic trajectory]. We leave for future work the important task of implementing a self-consistent evolution scheme \cite{Pound:Poisson:2008, Diener:Vega:Wardell:Detweiler:2011}, in which the GSF information is fed back at each time step in order to compute the evolving orbit.

\subsection{Linearized Einstein Equations\label{subsec:linearized-eqns}}

As described above, we split the spacetime metric into a background $g_{\alp \bet}$ at order $\eta^0$, and a  MP $h_{\alp \bet}$ at order $\eta^1$, and (in the first-order formulation) we neglect higher-order terms. Further key quantities, such as the affine connection ${\Gamma}^{\nu}_{\alp \bet}$ may be expanded in a similar way, i.e., $ \Gamma^{(0)\nu}_{\alp \bet} + \eta \Gamma^{(1)\nu}_{\alp \bet} + \mathcal{O}(\eta^2)$. We note in passing that, although 
 the affine connection is not a tensor, its first-order variation,
\beq
\eta \Gamma^{(1)\mu}_{\alp \bet} = \tfrac{1}{2} g^{\mu \nu} \left( h_{\nu \alp ; \bet} + h_{\nu \bet ; \alp} - h_{\alp \bet ; \nu} \right) ,
\eeq
behaves like a tensor with respect to the background spacetime \cite{Regge:Wheeler:1957, Zerilli:1970}. The Einstein tensor is expanded in a similar fashion, $G^{}_{\alp \bet}=G^{(0)}_{\alp \bet} + \eta G^{(1)}_{\alp \bet}$, with $G^{(0)}_{\alp \bet}$ for a vacuum background. At first order in the mass ratio the Einstein equations take the form
\beq
-2 G^{(1)}_{\alp \bet} = \mathcal{A}_{\alp \bet} + \mathcal{B}_{\alp \bet} = - 16 \pi T_{\alp \bet} , \label{AB-eq}
\eeq 
where
\begin{eqnarray}
\mathcal{A}_{\alp \bet} &\equiv& {{\hh_{\alp \bet ; \nu}}}^{\nu} + 2 {{{R^{(0)\gamma}}_\alp}{}^\delta}_\bet \hh_{\gamma \delta} , \label{A-eq} \\
\mathcal{B}_{\alp \bet} &\equiv& g_{\alp \bet} {Z^\nu}_{;\nu} - Z_{\alp ; \bet} - Z_{\bet ; \alp} , \\
Z_{\alp} &\equiv& {\hh_{\alp \bet}}^{; \bet} .  \label{Zdef}
\end{eqnarray}
Here ${{{R^{(0)\gamma}}_\alp}{}^\delta}_\bet$ is the Riemann tensor for the background spacetime, 
and $\hh_{\alp \bet}$ is the trace-reversed MP defined by
\beq
\hh_{\alp \bet} \equiv h_{\alp \bet} - \tfrac{1}{2} g_{\alp \bet} h, 
\eeq
with $h \equiv h^{\ \alp}_\alp$. 

A gauge transformation $x^\alp \rightarrow x^{\alp\prime} = x^{\alp} - \xi^{\alp}(x)$, where $\xi^ \alp \sim \mathcal{O}(\eta)$, generates a MP
\beq
h^{(\xi)}_{\alp \bet} = \xi_{\alp ; \bet} + \xi_{\bet ; \alp} .  \label{hgauge}  
\eeq
If the gauge vector is at least twice-differentiable, then this ``pure gauge'' MP is automatically a solution of the vacuum equations, i.e.~Eq.~(\ref{AB-eq}) with $T_{\alp \bet} = 0$. 


\subsection{Gauge choice, Z4 systems and constraint damping\label{subsec:gauge-choice}}



The standard prescription for GSF regularization (see Sec.~\ref{subsec:regularization} below) is formulated in the \emph{Lorenz gauge}, defined through
\beq
Z_{\alp} = 0  \label{Lor-gauge}.
\eeq
This condition leads to the simplification $\mathcal{B}_{\mu \nu} = 0$, so that the linearized Einstein's equations (\ref{AB-eq}) becomes {\em hyperbolic}. 
Within the Lorenz gauge there remains a residual gauge freedom: Consider a gauge vector $\xi^{\alp}$ that satisfies ${\xi_{\alp ; \nu}}^\nu = 0$. Then the corresponding (trace-reversed) MP $\hh_{\alp \bet}^{(\xi)} = \xi_{\alp ; \bet} + \xi_{\bet ; \alp} - g_{\alp \bet} {\xi^\nu}_{;\nu}$ clearly satisfies the Lorenz-gauge condition (\ref{Lor-gauge}), assuming, as we do here, that the background is Ricci-flat. 


In our work we shall also define a generalized version of the Lorenz gauge,
\beq
Z_{\alp}  = H_\alp( h_{\bet\gamma} , x) ,   \label{glg}
\eeq
where $H_{\alp}$ are four ``gauge-driver'' functions, to be specified. In order to preserve the hyperbolic character of Eq.~(\ref{AB-eq}), we permit $H_{\alp}$ to depend only on the MP, and \emph{not} on its derivatives. In addition, we insist that $H_{\alp}$ is regular everywhere, even on the worldline. The trivial choice $H_{\alp}=0$ corresponds to the Lorenz gauge. (The idea of using gauge-driver functions is not a new one \cite{Bona:2003}, and it is being applied in many formulations of numerical relativity.) 

A key feature of the linearized set (\ref{AB-eq}) with $Z_{\alpha}=0$ is that its initial-value formulation preserves the Lorenz-gauge condition: If a solution has $Z_{\alpha}=0$ on an initial Cauchy surface, then this condition will be satisfied at any time. However, since suitable initial data (i.e., Lorenz-gauge data compatible with a moving point-particle source) are not available, and since finite-differencing numerical error is inevitable, in practice any numerical implementation of Eq.~(\ref{AB-eq}) would lead to gauge-condition violations. A practical solution, outlined in Refs.~\cite{Gundlach:etal:2005, Barack:Lousto:2005}, employs ideas from the Z4 formulation of Numerical Relativity \cite{Bona:2003, Bona:Palenzuela:2004}. 
The Z4 system is obtained in its 4-dimensional covariant
form by replacing the vacuum Einstein equations $G_{\alp \bet} = 0$ by
$G_{\alp\bet} + \nabla_{\alp} \mathbb{Z}_{\bet} + \nabla_{\bet} \mathbb{Z}_{\alp} - g_{\alp \bet} \nabla^{\gamma} \mathbb{Z}_{\gamma} = 0$. 
Here $\mathbb{Z}_{\alp}$ is a supplementary four-vector of constraints. A solution of the Z4 equations is a solution of the Einstein equations if and only if $\mathbb{Z}_\alp = 0$.

In our linearized system, we take the constraint vector to be 
\beq
\mathbb{Z}_{\alp} = -\half\left(Z_{\alp} - H_{\alp} \right) ,
\eeq
leading to the new system 
\beq
\mathcal{A}_{\alp \bet} + \hat{\mathcal{B}}_{\alp \bet} = - 16 \pi T_{\alp \bet} , \label{AB-eq2}
\eeq 
where $\mathcal{A}_{\alp \bet}$ is as in Eq.~(\ref{AB-eq}), and 
\beq
\hat{\mathcal{B}}_{\alp \bet} \equiv g_{\alp \bet} {H^\nu}_{;\nu} - H_{\alp ; \bet} - H_{\bet ; \alp} . \label{B-eq-alt}
\eeq
Note that, after this replacement, the constraint vector $\mathbb{Z}_\alp$ obeys a wave equation $\Box \mathbb{Z}_\alp \equiv {\mathbb{Z}_{\alp ; \bet}}^{\bet} = 0$. [to see this, consider the divergence of Eq.~(\ref{AB-eq2}) and make use of Ricci-flatness and stress-energy conservation ${T_{\alp \bet}}^{;\bet} = 0$.] The set (\ref{AB-eq2}) is only equivalent to (\ref{AB-eq}) if $\mathbb{Z}_{\alp}=0$ (i.e., if $H_{\alpha}=Z_{\alpha}$). In order to drive $\mathbb{Z}_{\alp}$ towards zero, we supplement the set (\ref{AB-eq2}) with a ``constraint damping'' term $\mathcal{C}_{\alp \bet}$ \cite{Gundlach:etal:2005}. In principle, any choice of $\mathcal{C}_{\alp \bet}$ that vanishes as $\mathbb{Z}_{\alp} \rightarrow 0$ is justified. We will employ a constraint damping term of the form
\beq
\mathcal{C}_{\alp \bet} = -2 \kappa (n_{\alp} \mathbb{Z}_{\bet} + n_{\bet} \mathbb{Z}_{\alp}) , \label{C-eq}
\eeq
where $\kappa$ and $n_\alp$ are scalar and vector fields to be specified in Sec.~\ref{subsec:gcd}.

In summary, in the following sections we will implement the modified system
\beq
\mathcal{A}_{\alp \bet} + \hat{\mathcal{B}}_{\alp \bet} + \mathcal{C}_{\alp \bet} = -16 \pi T_{\alp \bet} , \label{ABC-eq}
\eeq
where the left-hand terms are given by Eq.~(\ref{A-eq}), (\ref{B-eq-alt}) and (\ref{C-eq}). 
The key requirement of any successful numerical implementation is that the constraint violation dissipates with time, $\mathbb{Z}_{\alp}\rightarrow 0$, so that we recover a valid solution of the linearized equations (\ref{AB-eq}).
In Secs.~\ref{sec:implementation} and \ref{sec:results:1}, we describe a Lorenz-gauge ($H_{\alp} = 0$) implementation that works well for the modes $m \ge 2$, but which suffers from linear-in-$t$ gauge mode instabilities in the modes $m=0$ and $m=1$. In Sec.~\ref{sec:stabilization} we describe a generalized Lorenz-gauge implementation that goes some way towards curing the instability.


\subsection{Regularization\label{subsec:regularization}}

As discussed in Sec.~\ref{subsec:motion}, a point-like source generates a MP which diverges along the particle's worldline, and a method of regularization is required in order to extract the correct regular field $h_{\alpha\beta}^R$ that enters the GSF construction formula (\ref{a-sf}). Detweiler and Whiting \cite{Detweiler:Whiting:2003} gave a prescription for constructing $h_{\alpha\beta}^R$ through a subtraction 
\beq
h_{\alpha\beta}^R=h_{\alp \bet} - h_{\alp \bet}^{S},
\eeq
where $h_{\alp \bet}$ is the retarded solution of Eq.~(\ref{AB-eq}) with the point-particle source (\ref{point-particle}), and $h_{\alp \bet}^{S}$ (the $S$ field, for ``singular/symmetric'') is locally defined in terms of a Green function $G^{S\ \ \alp \bet}_{\alp'\bet'}(x,x')$, which is (i) symmetric in its indices and arguments, (ii) a solution of the inhomogeneous equation (\ref{AB-eq}), and (iii) zero when $x$ and $x'$ are connected by a timelike geodesic. The Green function has a local definition in terms of a Hadamard parametrix. The terms in this parametrix can be expanded in powers of the coordinate separation of the points. This leads to a local expansion for the S field \cite{Detweiler:Whiting:2003, Barack:Golbourn:2007, Vega:Detweiler:2008, Wardell:thesis, DBW:2011, Wardell:Vega:Thornburg:Diener:2011}. 

\subsubsection{The puncture scheme}
Following \cite{Barack:Golbourn:Sago:2007} and Papers I and II, let us introduce the $\mathcal{P}$ field $\hh_{\mu \nu}^{\mathcal{P}}$ (the ``puncture''), which has the same local expansion as the S field, $\hh_{\mu \nu}^{S}$, up to a certain order, and which has a smooth global continuation. 
Corresponding to this puncture field is an $\mathcal{R}$ (``residual'') field, defined through
\beq 
\hh_{\alp \bet}^{\mathcal{R}} = \hh_{\alp \bet} - \hh_{\alp \bet}^\mathcal{P} .  \label{h-res}
\eeq
 If the puncture field agrees with the S field in the local vicinity of the worldline up to a suitably high order (see Paper I), then (i) the residual $\mathcal{R}$ and Detweiler-Whiting R fields agree on the worldline, and, (ii) the GSF can be obtained from the gradient of the $\mathcal{R}$ field using
\beq
F_{\text{self}}^{\alp} =  \left. \mu k^{\alp \bet \gamma \delta} \hh^{R}_{\bet \gamma ; \delta} \right|_{x=z(\tau)} =  \lim_{x \rightarrow z(\tau)} \mu k^{\alp \bet \gamma \delta} \hh^{\mathcal{R}}_{\bet \gamma ; \delta} ,
\eeq
where
\beq
k^{\alp \bet \gamma \delta}(x) =  \tfrac12 g^{\alp \delta} \hat u^{\bet} \hat u^{\gamma}  - g^{\alp \bet} \hat u^{\gamma} \hat u^{\delta} -  \tfrac12 \hat u^{\alp} \hat u^{\bet} \hat u^{\gamma} \hat u^{\delta}  + \tfrac14 \hat u^{\alp} g^{\bet \gamma} \hat u^{\delta} + \tfrac14 g^{\alp \delta} g^{\bet \gamma}. \label{k-def}
\eeq 
Here $\hat u^{\alpha}=\hat u^{\alpha}(x)$ is any smooth extension of the four-velocity $u^{\alpha}$ off the particle's worldline.
The so-called ``red shift'' variable $\tilde{H}$ \cite{Detweiler:2005, Detweiler:2008, Sago:Barack:Detweiler:2008, Shah:Keidl:Friedman:2011} may also be found from the residual field in a straightforward way:
\beq
\tilde{H} \equiv \left. \tfrac{1}{2} u^{\alp} u^{\bet}  h^{R}_{\alp \bet}  \right|_{x=z(\tau)} =  \lim_{x \rightarrow z(\tau)}  \tfrac{1}{2} \hat u^{\alp} \hat u^{\bet} h^{\mathcal{R}}_{\alp \bet}  .  \label{Htil-def}
\eeq

\subsubsection{$m$-mode decomposition\label{subsec:mmode}}

Following the approach of \cite{Barack:Golbourn:2007, Barack:Golbourn:Sago:2007} and Papers I and II, we take advantage of the axisymmetry of the Kerr spacetime (i.e., the existence of the azimuthal Killing vector $X^\alpha_{(\phi)} \equiv \partial x^{\alpha}/\partial \phi$), to decompose key quantities into azimuthal $m$-modes. That is, we let
\beq
\mathcal{X}(t,r,\theta,\phi) = \sum_{m=-\infty}^{\infty} \mathcal{X}^{(m)}(t,r,\theta) e^{i m \phi} ,  \label{m-mode}
\eeq
where $\mathcal{X}$ is any relevant physical quantity---for example, a component of the retarded MP $h_{\alp \bet}$---and $(t,r,\theta,\phi)$ are Boyer-Lindquist (BL) coordinates. If $\mathcal{X}$ is real, then it follows that $\mathcal{X}^{(m)}$ satisfies the complex-conjugation symmetry,
\beq
\mathcal{X}^{(-m)} = \mathcal{X}^{(m)\ast} .
\eeq
To obtain the $m$-modes, we may apply the inverse transformation,
\beq
\mathcal{X}^{(m)}(t,r,\theta) = \frac{1}{2\pi} \int_{-\pi}^\pi \mathcal{X}(t,r,\theta,\phi) e^{- i m \phi} d \phi .
\label{m-inv}
\eeq
Here some care is needed if $\mathcal{X}^{(m)}$ is not a smooth function; in particular, when applying the transform to the puncture field, we should first ensure that it is smooth everywhere off the particle's worldline.

Key quantities, such as the value of the R-field $\hh_{\alp \bet}^{R}$ at the particle, or the GSF $F^{\alpha}_{\rm self}$ exerted on the particle, may then be recovered from a sum over modes:
\beq \label{hm}
\hh_{\alp \bet}^{R}[z(\tau)] = 
\sum_{m=0}^{\infty} \hat \hh^{\mathcal{R}(m)}_{\alp \bet},
\quad\text{where}\quad
\hat \hh^{\mathcal{R}(m)}_{\alp \bet}\equiv
\epsilon_m\, \text{Re} \lim_{x \rightarrow z(\tau)}  \hh^{\mathcal{R}(m)}_{\alp\bet} e^{i m \phi}  ,
\eeq

\beq \label{Fm}
F_{\text{self}}^{\alp}[z(\tau)] = 
\sum_{m=0}^{\infty} \hat F_{\text{self}}^{\alp(m)},
\quad\text{where}\quad
\hat F_{\text{self}}^{\alp(m)}\equiv
\mu \, \epsilon_m\, \text{Re} \lim_{x \rightarrow z(\tau)}  k^{\alp \bet \gamma \delta}  \nabla_{\delta} \left( \hh^{\mathcal{R}(m)}_{\bet \gamma} e^{i m \phi}  \right).
\eeq
Here $\epsilon_{m} = 2$ for $m \neq 0$ and $\epsilon_{m} = 1$ for $m = 0$; we have folded over the $m<0$ contributions onto the $m>0$ ones, with an overhat indicating the combined contribution from the two $\pm m$-modes for given $m$ (which forms a real quantity). Some care is needed to show that reversing the order of the sum and limit is justified; this analysis is given in Ref.~\cite{Barack:Golbourn:Sago:2007}.

\subsection{Conservation laws\label{subsec:conservation}}

In this section we will construct quasilocal conserved quantities of the linearized system (\ref{AB-eq}) corresponding to symmetries of the background metric. These will play an important role in our discussion of low multipoles in Sec.~\ref{sec:low-multipoles}. 

Let us suppose that the background spacetime admits a Killing vector $X^{\alpha}$ (for the following discussion we only assume that the background is vacuum; we will specialize to Kerr later). Then, following Abbott and Deser \cite{Abbott:Deser:1982}, we introduce the antisymmetric two-form
\beq
F_{\alp \bet} \equiv - \frac{1}{8 \pi} \left( X^\lambda \hh_{\lambda [\alp ; \bet]} + {X^\lambda}_{;[\alp} \hh_{\bet] \lambda} + X_{[\alp} Z_{\bet]}  \right),  \label{Fdef}
\eeq
where $Z_{\alp}$ is defined in Eq.~(\ref{Zdef}) and square brackets denote antisymmetrization: $Y_{[\alp ; \bet]} \equiv \tfrac12 (Y_{\alp;\bet} - Y_{\bet;\alp})$. It is reasonably straightforward to verify that the divergence of $F_{\alp \bet}$ is proportional to the left-hand side of Eq.~(\ref{AB-eq}) contracted with the Killing vector, so that
\beq
{F_{\alp \bet}}^{;\bet} = T_{\alp \bet} X^\bet \equiv j_{\alp}.  \label{jdef}
\eeq
The divergence of the current $j_{\alp}$ is zero, ${j_{\alp}}^{;\alp} = 0$, due to the conservation of stress-energy, ${T_{\alp \bet}}^{;\bet} = 0$, in conjunction with the Killing property $X_{\alp ; \bet} + X_{\bet ; \alp} = 0$. 
This means that the total ``charge''
\beq \label{Qdef}
Q(X)\equiv \int_{\Sigma} j^{\alp} d\Sigma_{\alp}
\eeq
contained in a spacelike hypersurface $\Sigma$ extending to infinity is conserved ($\Sigma$-independent), assuming $j^{\alpha}=0$ at spatial infinity (see, e.g., Chap.~3 of Ref.~\cite{Poisson-toolkit}; we follow here the notation of \cite{Poisson-toolkit}, with $d\Sigma_{\alp}$ representing the appropriate vector area element on $\Sigma$). If the background spacetime is stationary, i.e., admits a time-translation Killing vector $X^{\alpha}_{(t)}$, than $Q(X^{\alpha}_{(t)})$ gives a Komar-like definition of the {\it mass-energy} content of the MP. Similarly, if the background spacetime is axially symmetric, with a rotational Killing vector $X^{\alpha}_{(\phi)}$, than $Q(X^{\alpha}_{(\phi)})$ gives a Komar-like definition of the {\it angular-momentum} content of the MP. [These definitions are Komar-like in that they bear on the time-translation and rotation symmetries of spacetime. However, here the relevant symmetries are these of a {\it background} spacetime; the full spacetime (background+perturbation) need not have any symmetry for our definitions to hold.]

\begin{figure} 
 \begin{center}
  \includegraphics[width=8cm]{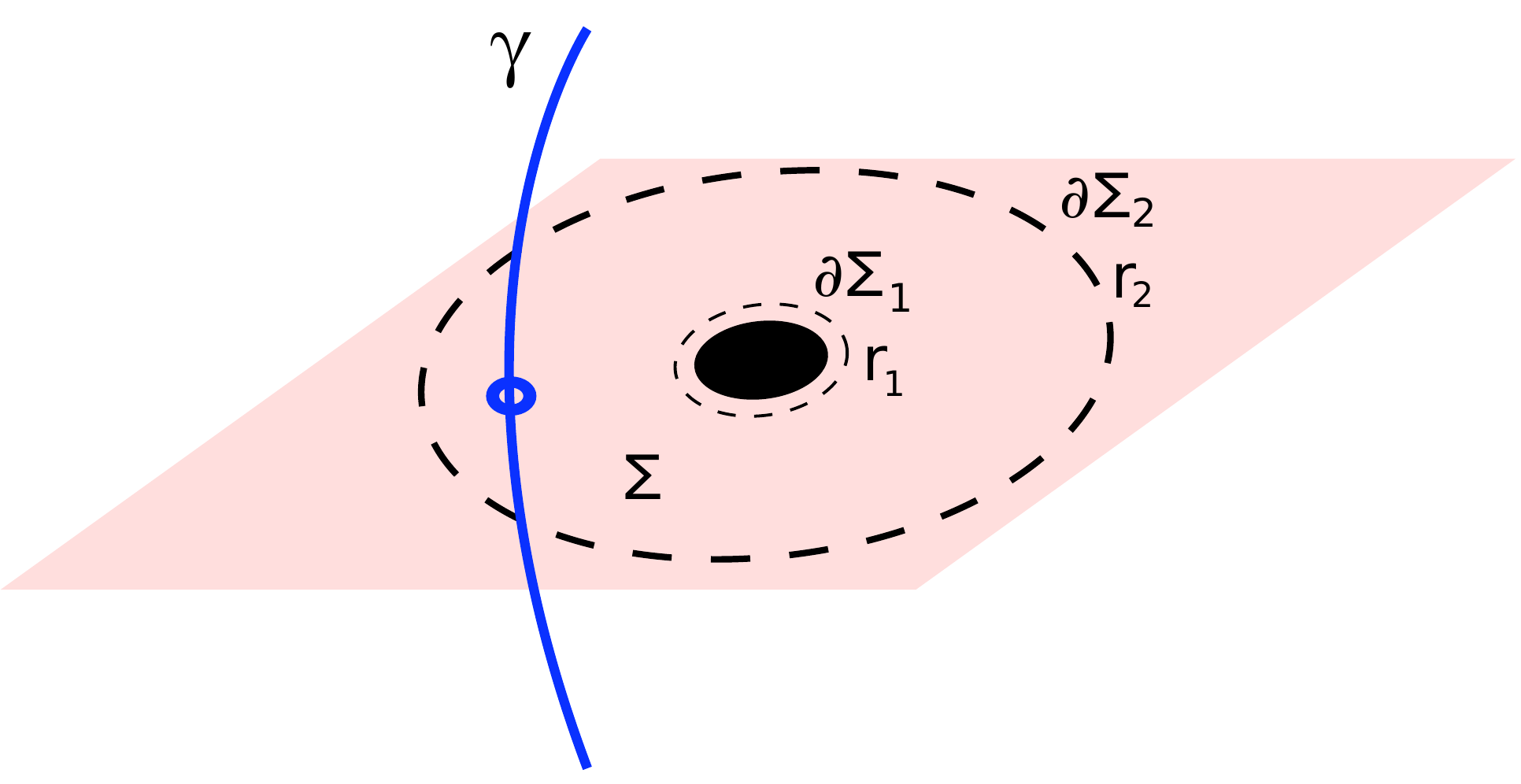}
 \end{center}
 \caption{Diagram illustrating the construction discussed in the text. $\Sigma$ is a closed spacelike 3-surface, defined by $t = {\rm const}$ and $r_1 < r < r_2$, outside a Kerr black hole.  The hypersurface is penetrated by the particle worldline $\gamma$. Stokes' theorem is used to relate the particle's energy and angular momentum to surface integrals of the MP over the boundaries $\partial \Sigma_1$ and $\partial \Sigma_2$ at $r=r_1$ and $r=r_2$, respectively. 
 }
 \label{fig:hypersurface}
\end{figure}

Let us now specialize to a Kerr background and a point particle with stress-energy given by Eq.~(\ref{point-particle}). 
Consider a closed spacelike 3-volume $\Sigma$, defined by $t=t_{\Sigma}(={\rm const})$ and $r_h<r_1 < r < r_2$, for some $r_1$ and $r_2$ and with $r=r_h$ being the BL radius of the event horizon (see Fig.~\ref{fig:hypersurface}). We find that the ``charge'' contained in $\Sigma$
is
\beq \label{Q}
Q(X) = \begin{cases} 
   \mu X^{\alp} u_{\alp}, & r_1 < r_p(t_{\Sigma}) < r_2, \\
    0, & \text{otherwise} , 
    \end{cases}  
\eeq
where $r_{p}(t_{\Sigma})$ it the particle's radius at time $t_{\Sigma}$.
With the Killing vectors $X_{(t)}^{\alp}\equiv \partial x^{\alp}/ \partial t$ and $X_{(\phi)}^{\alp}\equiv \partial x^{\alp}/\partial \phi$ of the Kerr geometry, the quantity $\mu X^{\alp} u_{\alp}$ on the right-hand side is, respectively, (minus) the energy $\mu\mathcal{E}\equiv -\mu u_t$ and the angular momentum $\mu \mathcal{L}\equiv \mu u_{\phi}$, which are conserved along the particle's geodesic worldline.

Using Stokes' theorem we may express $Q(X)$ more usefully in terms of quantities on the 2-dimensional boundaries of $\Sigma$. Let $\partial \Sigma_{1}$ and $\partial \Sigma_{2}$ denote the inner and outer boundaries of $\Sigma$, at $r=r_1$ and $r=r_2$, respectively. Then from Eqs.\ (\ref{jdef}) and (\ref{Qdef}) it follows (cf., e.g., Sec.~3.3.3 of Ref.~\cite{Poisson-toolkit}) that 
\beq \label{calFdiff}
Q(X)={\cal F}(X,\partial\Sigma_2)-{\cal F}(X,\partial\Sigma_1),
\eeq
where
\beq
{\cal F}(X,\partial\Sigma)\equiv \frac12  \int_{\partial\Sigma} F^{\alp \bet} d\Sigma_{\alp \bet} ,  \label{stokes}
\eeq
in which $d\Sigma_{\alp \bet}$ is an appropriate 2-surface element on $\partial\Sigma$ \cite{Poisson-toolkit}. 
We note that the surface integral ${\cal F}(X,\partial\Sigma)$ (unlike $F_{\alpha\beta}$ itself) is {\it gauge invariant}, in the sense that it vanishes for any (smooth) pure-gauge MP of the form (\ref{hgauge}). This can be verified by noting that, for $h_{\alpha\beta}=h_{\alpha\beta}^{(\xi)}$, $F_{\alp \bet}$ can be written as the divergence,
$F^{(\xi)}_{\alp \bet} = \eta_{\alp \bet \gam} ^{\ \ \ ; \gam}$, of a 3-form 
\beq
\eta_{\alp \bet \gam} = -\frac{3}{8\pi} \left( X_{[\alp} \xi_{\bet ; \gam]} + X_{[\alp ; \bet} \xi_{\gam]} \right) \label{eta-def}.
\eeq
The integral of $F^{(\xi)}_{\alp \bet}$ over a 2-surface of constant $t,r$ then works out to be proportional to $\int(A_{\theta,\phi}-A_{\phi,\theta})d\theta d\phi$, with $A_\delta \equiv \eta^{\alp \bet \gam} \epsilon_{\alp \bet \gam \delta}$. This surface integral vanishes for any $A_\delta$ (assuming $\xi^{\alpha}$, and hence $A_\delta$, are smooth over the 2-surface), leading to ${\cal F}^{(\xi)}(X,\partial\Sigma)=0$.


Eqs.\ (\ref{Q}) and (\ref{calFdiff}) relate the conserved quantities $\cal E$ and $\cal L$ to the {\em difference} between two surface integrals $\cal F$ enclosing the relevant volume. In fact, a stronger result can be established, in the form of a statement about $\cal F$ itself, for any 2-surface $\partial\Sigma_r$ of constant $t=t_{\Sigma}$ and constant $r$:
\beq 
{\cal F}(X,\partial\Sigma_r) = \begin{cases} 
    \mu X^{\alp} u_{\alp}, & r> r_p(t_\Sigma), \\
    0, & r< r_p(t_{\Sigma}) . \end{cases}  \label{calF}
\eeq
This result is valid in any gauge for either $X=X^{\alpha}_{(t)}$ or $X=X^{\alpha}_{(\phi)}$. It follows from the following argument. Consider the surface integral ${\cal F}(X,\partial\Sigma_\infty)$ for a 2-sphere $\partial\Sigma_\infty$ at $r\to\infty$. Using the $m$-mode decomposition described above we may formally split the physical MP into a stationary, axially-symmetric piece ($m=0$) and a non-axially-symmetric piece ($m>0$), with the latter averaging to zero upon integration over $\partial\Sigma_\infty$ and hence not contributing to $\cal F$. The remaining, $m=0$ contribution is made up of two pieces, one coming from the particle's monopole-like mass perturbation, and the other from the particle's dipole-like angular-momentum perturbation; higher multipoles decay fast enough at $r\to\infty$ for their contribution to ${\cal F}(X,\partial\Sigma_\infty)$ to vanish. It is straightforward to construct explicit expressions for the mass and angular-momentum perturbations in a particular gauge: Let $g^{\rm (BL)}_{\alpha\beta}(x^{\alpha};M,J)$ denote the Kerr metric in BL coordinates, with $M$ and $J \equiv a M$ being the black hole's mass and spin. Then the linear variations
\beq
h^{(\delta M)}_{\alp \bet} \equiv  \mu {\cal E}\left. \frac{ \partial g^{\rm (BL)}_{\alp \bet} }{\partial M} \right|_{J} \quad \quad \text{and} \quad \quad 
h^{(\delta J)}_{\alp \bet} \equiv  \mu {\cal L}\left. \frac{ \partial g^{\rm (BL)}_{\alp \bet} }{\partial J} \right|_{M}
\eeq
(with fixed BL coordinates) describe, respectively, the mass and angular-momentum perturbations associated with a particle of orbital energy $\mu{\cal E}$ and angular momentum $\mu{\cal L}$. An explicit calculation gives $-{\cal F}(X^{\alpha}_{(t)},\partial\Sigma_\infty) = \mu{\cal E}$ and $-{\cal F}(X^{\alpha}_{(\phi)},\partial\Sigma_\infty) = 0$ for $h^{(\delta M)}_{\alp \bet}$, and ${\cal F}(X^{\alpha}_{(t)},\partial\Sigma_\infty) = 0$ and ${\cal F}(X^{\alpha}_{(\phi)},\partial\Sigma_\infty) = \mu{\cal L}$ for $h^{(\delta J)}_{\alp \bet}$. Therefore, for the complete MP we find ${\cal F}(X,\partial\Sigma_{\infty})=\mu X^{\alp} u_{\alp}$, which should hold irrespective of the gauge chosen for the MP, by virtue of the gauge invariance of ${\cal F}$. The result (\ref{calF}) then follows directly from the ``jump'' condition (\ref{calFdiff}) with (\ref{Q}).



The relation in Eq.\ (\ref{calF}) is useful in that it provides us with a simple criterion by which to determine if a given numerically-constructed MP has the correct mass and angular-momentum contents. In Sec.\ \ref{sec:low-multipoles} we will show how this criterion can be used constructively to obtain the correct (physical) nonradiative pieces of the MP. 


\section{Implementation: Circular orbits on Schwarzschild spacetime\label{sec:implementation}}

In this section we describe the details of the first implementation of the $m$-mode regularization method for computing the GSF along geodesics of a black hole spacetime, focusing on the specific case of circular orbits on Schwarzschild spacetime.

\subsection{Setup\label{subsec:setup}: Circular geodesics on Schwarzschild spacetime}

In the region exterior to the black hole, the Schwarzschild geometry is described, using Schwarzschild coordinates $\{t,r,\theta,\phi\}$, by the line element
\beq
ds^2 = g_{\mu \nu} dx^\mu dx^\nu = -f(r) dt^2 + f^{-1}(r) dr^2 + r^2 (d \theta^2 + \sin^2 \theta d\phi^2 ),
\eeq
where 
\beq
f(r) = 1 - r_h / r,
\eeq
$M$ is the mass of the black hole, and $r=r_h = 2M$ is the radius of the event horizon. To describe the region beyond the future event horizon $\Hplus$, it is common to introduce the advanced time coordinate $v = t + r_\ast(r)$, where the ``tortoise'' radial coordinate $r_\ast$ is defined via
\beq
r_\ast = r + r_h \ln( r / r_h- 1 ), \label{rstar-def}
\eeq
with ${dr_\ast}/dr = f^{-1}(r)$.
This leads to the (``advanced'') Eddington--Finkelstein  line element,
\beq
ds^2 = -f(r) dv^2 + 2 dv dr + r^2 (d \theta^2 + \sin^2 \theta d\phi^2) .  \label{metric-aef}
\eeq

We consider a pointlike particle of mass $\mu$ in a circular orbit around the black hole. Neglecting GSF effects, the particle follows a geodesic $z^{\alp}(\tau)$ with a four-velocity $u^\alp = d z^\alp / d \tau$ with respect to proper time $\tau$. Without loss of generality, we take the orbit to be in the equatorial plane, so that
(in Schwarzschild coordinates)
\beq
z^\alp(\tau) = \left[ t(\tau), r_0, \pi/2, \phi(\tau) = \Omega t(\tau) \right] .
\eeq
The orbital frequency is
\beq
\Omega \equiv d\phi / dt = \sqrt{M / r_0^3} ,
\eeq
and the four-velocity is 
\beq
u^\alp = \left(\mathcal{E} / f_0 \right) \left[1, 0, 0, \Omega \right] .  \label{uvec}
\eeq
Such a geodesic has (conserved) specific energy and angular momentum given by
\begin{eqnarray}
\mathcal{E} = -u_t &=& f_0 \left( 1 - 3M / r_0 \right)^{-1/2} ,  \label{specific-energy} \\
\mathcal{L} = u_\phi &=& (M r_0)^{1/2} \left( 1 - 3M / r_0 \right)^{-1/2} , \label{specific-angmom}
\end{eqnarray}
where $f_0 \equiv 1-2M/r_0$.

\subsection{2+1D field equations in vacuum\label{subsec:m-mode-eqs}}

We next describe our method for solving the linearized Einstein equations in 2+1D, subject to the Lorenz-gauge constraint $Z_{\mu} = 0$ (i.e.~$H_{\mu} = 0$). Focusing for now on the vacuum case ($T_{\alpha\beta}=0$) and specializing to the Lorenz gauge (${\cal B}_{\alpha\beta}=0$), we have the linear system of coupled PDEs, given by 
\beq \label{A=0}
{\cal A}_{\alpha\beta}=0
\eeq
[recall Eqs.\ (\ref{AB-eq}) with (\ref{A-eq})].

Consider the $m$-mode decomposition of this system. We write, formally,
\beq
\hh_{\alp\bet} =  \sum_{m = -\infty}^{\infty}  \hh_{\alp\bet}^{(m)}  e^{i m \phi} ,  \quad \quad \hh_{\alp\bet}^{(m)} =  \frac{\mu}{r} \, \ghat_{\alp\bet}(r,\theta)  u^{(m)}_{\alp\bet}(t,r,\theta)  \label{h-ansatz}
\eeq
(no summation over $\alpha\beta$),
with $\ghat_{\alp\bet}(r,\theta) \equiv \sqrt{\hat{g}_{\alp\bet} \hat{g}_{\alp\bet}}$ (again, no summation) and $\hat{g}_{\alp\bet} = \text{diag} \left[ 1, f^{-2}, r^2, r^2 \sin^2 \theta \right]$. Here we have chosen the $\ghat_{\alp\bet}$ prefactor to (i) scale the components of the MP in a similar way to the components of $g_{\alp\bet}$, and (ii) avoid $f^{1/2}$ terms in the equations that follow. Inserting  (\ref{h-ansatz}) into (\ref{A=0}) leads to a set of ten coupled equations
\beq
f\,\Box_{\text{sc}} u^{(m)}_{\alp \bet} + \MM^{(m)}_{\alp \bet} = 0 ,  \label{no-gcd}
\eeq
where $\MM^{(m)}_{\alp \bet}  = \MM^{(m)}_{\alp \bet}\left( \partial_t u_{\mu \nu}, \partial_{r_\ast} u_{\mu \nu}, \partial_\theta u_{\mu \nu}, u_{\mu \nu} \right) $ are given in Eqs.~(\ref{MM00})--(\ref{MM33}) of Appendix \ref{appendix:fieldeq}, and 
\beq
f \,\Box_{\text{sc}} \equiv -\frac{\partial^2}{\partial t^2} + \frac{\partial^2}{\partial r_*^2} + \frac{f}{r^2} \left[ \frac{\partial^2}{\partial \theta^2} + \cot \theta \frac{\partial}{\partial \theta} - \left( \frac{2M}{r} + \frac{m^2}{\sin^2 \theta} \right) \right] 
\eeq
is a scalar-like wave operator.
This set is complemented by an $m$-decomposed version of the four Lorenz-gauge equations $Z_\alp = 0$, 
which we give explicitly in Eq.~(\ref{gauge-t})--(\ref{gauge-p}) of Appendix \ref{appendix:fieldeq}. 

\subsubsection{Gauge constraint damping\label{subsec:gcd}}


As noted above (and in Refs.~\cite{Gundlach:etal:2005,Barack:Lousto:2005}), numerical gauge-constraint violations may be driven towards zero by adding suitable damping terms to the set (\ref{A=0}). In principle, we can add linear combinations of $Z_\alp$ in any convenient way. In practice, we have found the following covariant formulation to be a good choice:
\beq
\mathcal{A}_{\alp \bet} + \mathcal{C}_{\alp \bet} = 0, \quad \text{where} \quad \mathcal{C}_{\alp \bet} = \kappa \left( n_{\alp} Z_{\bet} + n_\bet Z_\alp \right) ,  \label{AC-eq}
\eeq
and $\kappa = df/dr = 2M/r^2$ with $n_{\alp} = [1, f^{-1}, 0, 0]$. Note that $n_{\alp}$ is a null vector (i.e., $n_\alp n^{\alp} = 0$), and our approach is similar, but slightly different, to that taken in Refs.~\cite{Barack:Lousto:2005,Barack:Sago:2007,Barack:Sago:2009}. Note also that, with this choice of constraint damping, all first-order $t$ and $r_*$ derivatives appear in Eq.\ (\ref{AC-eq}) only in the combination $\partial_v = \half(\dt + \drs)$. With this choice, the 2+1D vacuum equations become
\beq
f\,\Box_{\text{sc}} u^{(m)}_{\alp \bet} + \Mtil^{(m)}_{\alp \bet} = 0 ,  \label{sam-gcd}
\eeq
with $\Mtil^{(m)}_{\alp \bet}  = \Mtil^{(m)}_{\alp \bet}\left( \partial_v u_{\mu \nu}, \partial_\theta u_{\mu \nu}, u_{\mu \nu} \right) $ given explicitly in Eqs.~(\ref{Mtil00})--(\ref{Mtil33}) of Appendix \ref{appendix:fieldeq}. 


Alternative choices of constraint damping are possible, such as the version used by Barack and Lousto \cite{Barack:Lousto:2005} (and later Barack and Sago \cite{Barack:Sago:2007}) in the 1+1D setting. In the 2+1D setting, we find that implementing the Barack and Lousto choice leads to numerical evolutions which are susceptible to exponentially-growing instabilities at late times in modes $m \ge 2$, and at early times in modes $m=0$ and $m=1$.

\subsection{Sourced 2+1D field equations and the puncture scheme\label{subsec:mmode-reg}}

The stress-energy associated with a pointlike particle was given in Eq.~(\ref{point-particle}), with $(-g)^{1/2} = r^2 \sin \theta$. After taking the $m$-mode decomposition (see Sec.~\ref{subsec:mmode}) we reach the sourced $m$-mode equations 
\beq \label{EE}
f\,\Box_{\text{sc}} u^{(m)}_{\alp \bet} + \Mtil^{(m)}_{\alp \bet} = S^{(m)}_{\alp \bet},
\eeq
where
\beq
S^{(m)}_{\alp \bet} = -\frac{8 f_0}{\ghat_{\alp \bet} u^{t} r_0} e^{-i m \Omega t} \delta(r - r_0) \delta(\theta - \pi / 2)  u_{\alp} u_{\bet}   
\eeq
(no summation over $\alpha\beta$ implied).
Naively, we might attempt to solve this equation immediately on a 2+1D grid. However, the $\delta$-functions are difficult to implement, and, worse, the components of the retarded MP, $u_{\alp \bet}^{(m)}$, will in general diverge logarithmically as $(r,\theta) \rightarrow (r_0, \pi/2)$ \cite{Barack:Golbourn:2007}. 

\subsubsection{Second-order puncture}
In this work, we implement a puncture scheme first introduced in Ref.~\cite{Barack:Golbourn:Sago:2007}. There, it was shown that, under the $m$-mode decomposition, it is sufficient to use a puncture of second order (under the classification scheme of Paper I). This is in contrast to 3+1D schemes, in which the third-order puncture is the minimum requirement to obtain a residual field which is differentiable. A heuristic explanation for this is given in Sec.~II.G.2 of Paper I.

The Detweiler-Whiting S field may be defined locally for field points $x$ in the vicinity of a fixed point $z$ on the worldline. To promote the S field to a puncture function which is defined in the vicinity of the entire worldline, the next step is to make $z$ a function of the field point $x$. The simplest way to do this is to choose $z$ to be the point on the worldline with the same time $t$ coordinate as $x$. We may then define the coordinate differences,
\beq
\delta x^\alp \equiv x^{\alp} - z^{\alp}(t) = \left[ 0, \dr, \dth, \dph \right],
\eeq
where $\dr \equiv r - r_0$, $\dth \equiv \theta - \pi/2$ and $\dph \equiv \phi - \Omega t$. 
Now, following \cite{Barack:Golbourn:Sago:2007}, we introduce the puncture function 
\beq
\hh_{\alp \bet}^{\mathcal{P}}(x) = \frac{4 \mu}{\epsilon_{[2]}} \left[ \bar{u}_\alp \bar{u}_{\bet} + \left( \bar{\Gamma}^\lambda_{\alp \gamma} \bar{u}_{\bet} + \bar{\Gamma}^{\lambda}_{\bet \gamma} \bar{u}_{\alp} \right) \bar{u}_\lambda \delta x^{\gamma}  \right] ,  \label{punc-2nd}
\eeq
where $\bar{u}_\alp$ and $\bar{\Gamma}_{\alp \gamma}^\lambda$ are the tangent velocity and the affine connection evaluated at $z^{\alpha}(x)$, and $\epsilon_{[2]}$ is defined as follows:
\beq
\epsilon^2_{[2]} = \left. \left( g_{\alp \bet} + u_{\alp} u_\bet \right) \right|_{z} \delta x^{\alp} \delta x^\bet + 
\left. \left(u_\lambda u_\gam \Gamma^\lambda_{\alp \bet} + g_{\alp \bet , \gamma} / 2 \right) \right|_z \delta x^\alp \delta x^\bet \delta x^\gam .
\eeq
The above function $\hh_{\alp \bet}^{\mathcal{P}}(x)$ approximates the Detweiler--Whiting S field through $\mathcal{O}(\delta x^0)$.

There is one remaining problem: the puncture function defined in Eq.~(\ref{punc-2nd}) suffers a discontinuity at $\dph=\pm\pi$. To correct this, we replace the ``even'' and ``odd'' terms in $\dph$ as follows:
\begin{eqnarray}
\dph^2 &\rightarrow& 2 \left( 1 - \cos \dph \right)\ [= \dph^2 + \mathcal{O}(\dph^4)] , \\
\dph &\rightarrow& \sin \dph\ [= \dph + \mathcal{O}(\dph^3)] .
\end{eqnarray}
This keeps the local expansion of $\hh_{\alp \bet}^{\mathcal{P}}(x)$ near the particle intact through $\mathcal{O}(\delta x^0)$, so our puncture remains equal to the S field through this order. 
With the above replacement the puncture takes the form
\beq \label{punc}
\hh_{\alp \bet}^{\mathcal{P}} = \frac{4\mu \chi_{\alp \bet}}{\eps_{\mathcal{P}}},
\eeq
in which 
\beq
\eps_{\cal P} = \left[s_{200}\delta r^2+s_{020} \delta\theta^2+s_{300}\delta r^3 +s_{120}\delta r \delta\theta^2+ 2(s_{002}+s_{102}\delta r) (1 - \cos \dph) \right]^{1/2},
\eeq
and the nonvanishing components of $\chi_{\alpha\beta}$ are
\begin{equation}
\chi_{\alpha\beta} = 
\left\{
\begin{array}{ll}
u_{\alpha}u_{\beta} + D_{\alpha\beta} \dr  & {\rm for\ } \alpha\beta=tt,t\phi,\phi t,\phi\phi, \\
D_{\alpha\beta} \sin \dph		 & {\rm for\ } \alpha\beta=tr,rt,r\phi,\phi r .
\end{array}\right.
\end{equation}
The coefficients $s_{ijk}$, which depend on $r_0$ only, were given explicitly (in the more general Kerr case) in Appendix A of Paper II.
The coefficients $D_{\alpha\beta}$ read 
\begin{eqnarray}
D_{tt} &=& \frac{2{\cal E}^2 M}{r_0(r_0-2M)} ,   \label{Dtt} \\
D_{tr} &=& D_{rt}= -\frac{{\cal E}{\cal L}}{r_0} , \\
D_{t\phi} &=& D_{\phi t}=-\frac{{\cal E}{\cal L}(r_0-M)}{r_0(r_0-2M)}, \\
D_{r\phi} &=& D_{\phi r}=\frac{{\cal L}^2}{r_0} , \\
D_{\phi\phi} &=& \frac{2{\cal L}^2}{r_0} .  \label{Dpp}
\end{eqnarray}


\subsubsection{$m$-mode decomposition of the puncture function}
The next step is to compute the $m$-mode representation of the components of the puncture function. Using Eq.~(\ref{m-inv}) with $\phi = \dph + \Omega t$ we write
\begin{eqnarray}
\hh^{\mathcal{P}(m)}_{\alp \bet} =\frac{e^{-i m \Omega t}}{2\pi} \int_{-\pi}^{\pi} \hh^{\mathcal{P}}_{\alp \bet}(\dr,\dth,\dph) e^{-i m \dph} d(\dph) .
\end{eqnarray}
It turns out, as in the scalar-field case explored in Papers I and II, that all the relevant integrals over $\dph$, in the puncture and source (Sec.~\ref{subsec:mmode-source} below) can be represented in terms of complete elliptic integrals of the first and second kinds, as we detail in Appendix \ref{appendix:elliptic}. The $m$-mode representation of the puncture itself is straightforward:
we find
\beq
\hh^{\mathcal{P}(m)}_{\alp \bet} = \frac{4 \mu}{2 \pi}\, \Chi^{(m)}_{\alp \bet}\, e^{- i m \Omega t}  , \label{hmpunc}
\eeq
where the nonvanishing components of $\Chi^{(m)}_{\alp \bet}$ are
\begin{equation} \label{hatchi}
\Chi_{\alpha\beta}^{(m)} = 
\left\{
\begin{array}{ll}
(u_{\alpha}u_{\beta} + D_{\alpha\beta} \dr)I_0^m  & {\rm for\ } \alpha\beta=tt,t\phi,\phi t,\phi\phi, \\
D_{\alpha\beta} J_0^m		 & {\rm for\ } \alpha\beta=tr,rt,r\phi,\phi r ,
\end{array}\right.
\end{equation}
with $I_0^m$ and $J_0^m$ each being a combination of complete elliptic integrals given explicitly  in Eqs.~(\ref{integralI0}) to (\ref{integralJ0}) of Appendix \ref{appendix:elliptic}. Note that $I_0^m$ is purely real and logarithmically divergent at the particle, whereas $J_0^m$ is purely imaginary and it is continuous and differentiable at the particle (although its second derivatives diverge logarithmically there).

\subsubsection{$m$-mode decomposition of the effective source\label{subsec:mmode-source}}

The residual field $\hh_{\alp \bet}^{\mathcal{R}}$, defined in Eq.~(\ref{h-res}), obeys a wave equation with an ``effective source'',
\beq
\Dop \hh_{\alp \bet}^{\mathcal{R}} = \mathcal{T}^{\text{eff}}_{\alp \bet},
\eeq
where here we use $\Dop$ as a general shorthand notation for our choice of operator on the left-hand side of Eq.~(\ref{ABC-eq}), which depends on the choice of gauge and constraint damping.
The effective source is found from the action of the operator on the puncture function, i.e.,
\beq
\mathcal{T}^\text{eff}_{\alp \bet} = -16 \pi T_{\alp \bet} - \Dop \hh_{\alp \bet}^{\mathcal{P}} .
\eeq
Here, recall that $\hh_{\alp \bet}^{\mathcal{P}}$ is expressed as a function of $\{\dr, \dth, \dph\}$, and we may express the partial derivatives in $\Dop$ in terms of these variables, e.g., $\partial/\partial \phi = \partial / \partial (\dph)$ and $\partial/\partial t = - \Omega^{-1} \partial / \partial (\dph)$.
Let us now choose the left-hand side operator as in  Eq.~(\ref{AC-eq}). The $m$-mode decomposed equations governing the residual field are then
\beq
f\,\Box_{\text{sc}} u^{\mathcal{R}(m)}_{\alp \bet} + \Mtil^{\mathcal{R}(m)}_{\alp \bet} = S^{\text{eff}(m)}_{\alp \bet},
\eeq
where $\Mtil^{\mathcal{R}(m)}_{\alp \bet}$ is obtained by replacing $u_{\alp \bet}^{(m)} \rightarrow u_{\alp \bet}^{\mathcal{R}(m)}$ in Eqs.~(\ref{Mtil00})--(\ref{Mtil33}), with 
$u_{\alp \bet}^{\mathcal{R}(m)}$ defined from $\bar h_{\alp \bet}^{\mathcal{R}(m)}$ as in Eq.~(\ref{h-ansatz}). The effective source for the $m$-mode system is 
\beq
S^{\text{eff}(m)}_{\alp \bet} = \frac{f r}{\ghat_{\alp \bet}} \mathcal{T}_{\alp \bet}^{\text{eff}(m)}   \label{Seffm}
\eeq
(no sum implied), where $\mathcal{T}_{\alp \bet}^{\text{eff}(m)}$ are the $m$-modes of $\mathcal{T}_{\alp \bet}^{\text{eff}}$.
With our second-order puncture, the effective source diverges logarithmically on the particle's worldline (see Papers I \& II). 

For circular orbits, the $m$-mode effective source may be expressed in terms of complete elliptic integrals, as in the scalar case. To this end it is convenient to write $S^{\text{eff}(m)}_{\alp \bet}$ as a sum of two contributions, in the form 
\beq
S^{\text{eff}(m)}_{\alp \bet} = \frac{ 4 \mu f r}{2 \pi \ghat_{\alp \bet}} e^{-i m \Omega t} \left( Z^{{\rm sc}(m)}_{\alp \beta} + \Delta Z^{(m)}_{\alp \beta}    \right) ,  \label{SeffZeff}
\eeq
where $Z^{{\rm sc}(m)}_{\alp \beta}$ arises from the ``scalar'' part of the wave operator  (i.e, the term in $\hat {\cal D}$ involving $\Box_{\rm sc}$), and $\Delta Z^{1(m)}_{\alp \beta}$ represents all remaining terms. The first contribution can be expressed concisely in terms of quantities that were already computed in Paper I: we find 
\begin{equation}\label{Zsc}
Z^{{\rm sc}(m)}_{\alp \beta} = 
\left\{
\begin{array}{ll}
(u_{\alpha}u_{\beta} + D_{\alpha\beta} \dr)\sum_{k=1}^5 S_k I_k^m  & {\rm for\ } \alpha\beta=tt,t\phi,\phi t,\phi\phi, \\
D_{\alpha\beta} \sum_{k=1}^5 S_k J_k^m		 & {\rm for\ } \alpha\beta=tr,rt,r\phi,\phi r ,
\end{array}\right.
\end{equation}
where $I^m_{k}$ and $J^m_k$ are complete elliptic integrals given explicitly in Appendix \ref{appendix:elliptic} below, and the coefficients $S_k$ (which depend on $r$, $\theta$ and $r_0$ but not on $m$) were given explicitly in Appendix B of Paper I [Eqs.\ (B8)--(B12)]. The remaining part of the $m$-mode effective source may be computed with the assistance of a symbolic algebra package. We obtain the form 
\begin{eqnarray}\label{DeltaZ}
\Delta Z^{(m)}_{\alp \beta} =  \sum_{k=0,1,2,6,7} c^k_{\alp \beta} I_k^{m}  + \sum_{k=0,1,2} d^k_{\alp \beta} J_k^{m} ,
\end{eqnarray}
where the two extra elliptic integrals $I^m_{6,7}$ are also given in Appendix \ref{appendix:elliptic} [Eqs.\ (B11) and (B12)], and the various components of the ($m$-independent) coefficients $c^k_{\alp \beta}$ and $d^k_{\alp \beta}$ are given in Appendix \ref{appendix:source}.

\subsection{Boundary conditions\label{subsec:bc}}


We require that the advanced Eddington--Finkelstein (aEF) components of the MP are regular across the event horizon, i.e.\ for $R=2M$ and finite $v$, where $\{ v = t + r_\ast, R=r, \theta, \phi  \}$ are aEF coordinates. The aEF components are related to the Schwarzschild components via
\begin{eqnarray}
 \hh^{\text{(aEF)}}_{vv} &=& \hh_{tt} \label{bc-h00} \\
 \hh^{\text{(aEF)}}_{vR} &=& \hh_{tr} - f^{-1} \hh_{tt} \label{bc-h01} \\
 \hh^{\text{(aEF)}}_{RR} &=& \hh_{rr} - 2 f^{-1} \hh_{tr} + f^{-2} \hh_{tt} , \label{bc-h11} \\
 \hh^{\text{(aEF)}}_{R\theta} &=& \hh_{r\theta} - f^{-1} \hh_{t\theta} , \label{bc-h12}  \\
 \hh^{\text{(aEF)}}_{R\phi} &=& \hh_{r\phi} - f^{-1} \hh_{t\phi} , \label{bc-h13}  
\end{eqnarray}
and $\hh^{\text{(aEF)}}_{\alp \bet} = \hh_{\alp \bet}$ for all other components.
We hence require that the above algebraic combinations of Schwarzschild components remain regular (in particular, finite) as $r_\ast \to -\infty$. The same combinations are also required to be regular at the level of the $m$-mode MP $\hh_{\alp \bet}^{(m)}(t,r,\theta)$.


We also require boundary conditions at the poles, $\theta=0,\pi$. These are determined from the requirement that the MP is regular (in particular, continuously differentiable) when expressed in coordinates which are regular at the poles (e.g., locally Cartesian coordinates based at each pole). This, in turn, translates to a condition on the local behavior of the $m$-modes $\hh_{\alp \bet}^{(m)}(t,r,\theta)$ for $\theta\to 0^{+},\pi^{-}$ (with any fixed $t,r$). The form of this condition depends on the MP component in question. For the North pole ($\theta=0$) we find, for 
$m \ge 0$, 
\beq
\hh_{\alp \bet}^{(m)}(r,t,\theta ) \sim \begin{cases} 
\theta^m , & \alp \bet \in \{ t t, t r, r r  \} , \\
\theta^{|m-1|} , & \alp \bet \in \{ t \theta, t \phi, r \theta, r \phi  \} , \\
\theta^{|m-2|} , & \alp \bet \in \{ \theta\theta, \theta\phi, \phi\phi  \}.
\end{cases}  \label{bc-poles}
\eeq
The conditions at the South pole ($\theta=\pi$) can then be inferred from the reflection symmetry
\beq
\hh_{\alp \bet}^{(m)}(t,r, \pi-\theta ) = \begin{cases} 
- \hh_{\alp \bet} (t,r,\theta ), & \alp \bet \in \{ t \theta, r \theta, \theta \phi  \} , \\ 
+ \hh_{\alp \bet} (t,r,\theta ), \quad & \text{otherwise}.
\end{cases} \label{bc-equator}
\eeq

\subsection{Numerical implementation details\label{subsec:impdetails}}

\subsubsection{Numerical domain and the worldtube scheme \label{subsec:worldtube}}
The puncture function (\ref{hmpunc}) was constructed by truncating the series expansion of the Detweiler--Whiting S field at second order, and the effective source was formed by acting on the puncture with the relevant wave operators. Unfortunately the  global behaviour of the source (\ref{SeffZeff}) is not suitable for a numerical scheme. It generally diverges in the large-$r$ limit, and also at the poles (where $\sin \theta = 0$). Vega and collaborators \cite{Vega:Detweiler:2008, Vega:Diener:Tichy:Detweiler:2009, Vega:Wardell:Diener:2011, 
Wardell:Vega:Thornburg:Diener:2011, 
Diener:Vega:Wardell:Detweiler:2011} avoid this problem by multiplying the puncture by a windowing function, to smoothly moderate the behaviour of the effective source far from the worldline. By contrast, and following Ref.~\cite{Barack:Golbourn:2007}, Paper I (see Sec.~IIH) and Paper II (see Sec.~IVA), we apply here a ``worldtube'' scheme.

In the 2+1D domain, the worldtube $\mathcal{T}$ is defined to be a region of finite extent in $r$ and $\theta$ (or order $\sim M$), surrounding the worldline. Outside the worldtube, we evolve the vacuum wave equation for the physical (i.e. retarded) MP. Inside the worldtube, we evolve the sourced equation for the residual MP. Across the boundary of the tube, $\partial \mathcal{T}$, we convert between the two using the puncture field. Hence, our evolution equations consist of
\beq
\begin{cases}
f\,\Box_{\text{sc}} u^{(m)}_{\alp \bet} + \Mtil^{(m)}_{\alp \bet}  = 0, & \text{outside}\; \mathcal{T}, \\
f\,\Box_{\text{sc}} u^{\mathcal{R}(m)}_{\alp \bet} + \Mtil^{\mathcal{R}(m)}_{\alp \bet}  = S^{\text{eff}(m)}_{\alp \bet} , & \text{inside}\; \mathcal{T}, \\
u^{\mathcal{R}(m)}_{\alp \bet} = u^{(m)}_{\alp \bet}  - r (\mu \ghat_{\alp \bet})^{-1} \hh^{\mathcal{P}(m)}_{\alp \bet} , & \text{across}\; \partial \mathcal{T}.  
\end{cases} \label{u-eq}
\eeq
These equations are solved on a fixed uniform grid based on $t,r_\ast,\theta$ coordinates; see Fig.~1 in Paper II. 
Let us denote the corresponding grid spacings by $\Delt$, $\Delr$, $\Delq$. In the 2+1D domain, the particle's trajectory traces a straight line at $\theta=\pi/2$ and $r_*=r_{\ast0}[\equiv r_\ast(r_0)]$. We lay the grid so that the particle's trajectory on the grid crosses through a row of grid points (of fixed $r_*,\theta$). Then we define a worldtube ${\cal T}$ of fixed coordinate widths $\{\Gamma_{r_\ast}, \Gamma_\theta \}$ as follows: Consider a grid point with coordinates $(t,r_\ast, \theta)$. If $|r_\ast -  r_{\ast0}| \le \Gamma_{r_\ast} / 2$ and $|\theta - \pi/2| \le \Gamma_{\theta} / 2$ then the point is said to lie within the worldtube; otherwise it lies outside. 


In the finite difference scheme, the derivatives of the MP at a given grid point (say, point ``o'') are computed based on the values at several neighboring grid points. If point o lies close enough to $\partial{\cal T}$ that some of these points are ``in'' and others are ``out'' of the tube, we make the following adjustment. If the point o is ``out'', then we first demote all relevant ``in'' points to ``out'' points using
$u^{(m)}_{\alp \bet} = u^{\mathcal{R}(m)}_{\alp \bet} + r (\mu \ghat_{\alp \bet})^{-1} \hh^{\mathcal{P}(m)}_{\alp \bet}$, before applying the finite-difference formula. Conversely, if the point o is ``in'', we promote all ``out'' points to ``in'' points in a similar manner. 

\subsubsection{Finite-difference scheme: method of lines\label{subsec:FDmethod}}

To evolve Eq.~(\ref{u-eq}), we used a finite difference scheme based on the well-known Method of Lines (see Ref.~\cite{Schiesser} or Sec.~4.2 of Ref.~\cite{Rinne}). The first step is to rewrite the equations in first-order form, which is simply achieved by defining the auxiliary variables,
\beq
v_{\alp \bet}^{(m)} \equiv \partial_t u_{\alp \bet}^{(m)} .
\eeq
Hence, for each $m > 0$ we have a $40$-dimensional system (from 10 real and 10 imaginary parts of $u_{\alp \bet}$ and $v_{\alp \bet}$, respectively), which is first-order in time. (An exception is the $m=0$ mode, which is real and thus gives a 20-dimensional system.) The next step is to replace all spatial derivatives (in $r_\ast$ and $\theta$) with their finite-difference approximants. We chose to use fourth-order finite difference operators, i.e.
\begin{eqnarray}
\frac{\partial^2 X}{\partial r_\ast^2} &\rightarrow& \frac{1}{12 \Delr^2} \left( -X_{j+2} + 16X_{j+1} - 30X_{j}+ 16X_{j-1} - X_{j-2} \right) , \\
\frac{\partial X}{\partial r_\ast}  &\rightarrow& \frac{1}{12 \Delr}  \left( -X_{j+2} + 8X_{j+1} - 8X_{j-1} + X_{j-2} \right) ,
\end{eqnarray}
where $X \in \{ u^{(m)}_{\alp \bet}, u^{\mathcal{R}(m)}_{\alp \bet}  \}$, and $X_{j} = X(t, r_{0\ast} + j \Delr, \theta)$. Similar operators were used for the $\theta$ derivatives. 

To evolve the first-order equations (i.e., to progress forward in time $t$ by steps of $\Delt$), we applied a fourth-order Runge-Kutta step. For details, see Sec.~IVB in Paper II.
In vacuum (i.e., without sources), we would expect this scheme to be globally fourth-order accurate, 
so that doubling the grid resolution will reduce the finite-differencing error by a factor of $\sim 2^4$.
However, when the effective source is included, a naive application of the scheme (as here) will \emph{not} be globally fourth-order accurate, due to the logarithmic divergence of $S_{\alp \bet}^{\text{eff}(m)}$ on the worldline.  

\subsubsection{Boundary conditions}

The spatial boundaries of our grid are at $r_\ast=r_{\ast{\rm in}}$, $r_\ast=r_{\ast{\rm out}}$ and at  $\theta=0$, $\theta=\pi$ (but see below how reflection symmetry allows us to replace the latter boundary with one at $\theta=\pi/2$). Taking $r_{\ast{\rm in}}\ll -M$ and $r_{\ast{\rm out}}\gg M$ places the radial boundaries at the asymptotic domains where the usual ``ingoing'' and ``outgoing'' radiation conditions apply to the retarded perturbation. In practice, we do not actively impose these radial boundary conditions in our time-evolution scheme.  Rather, we set the radial boundaries far enough that a sufficient portion of the late-time solution in the neighborhood of the particle has no causal connection with these boundaries. 
For the modes $m\geq 2$ we confirm retrospectively that our solutions possess the correct asymptotic behavior. The situation is more delicate in the case of the modes $m=0,1$, in which the horizon regularity conditions (\ref{bc-h00})--(\ref{bc-h13}) need to be explicitly imposed in constructing the solutions, as we shall discuss in Sec.\  \ref{sec:low-multipoles}.

The physical boundary conditions at the poles were given in Eqs.\ (\ref{bc-poles}) and (\ref{bc-equator}). 
In our implementation we make use of so-called ``ghost points'', which lie in zones just outside the physical domain (at $\theta \le 0$ and $\theta \ge \pi/2$), whose values are artificially set to enforce the correct boundary conditions on the physical domain. This allows us to use the same finite-differencing molecule everywhere within (the physical part of) the grid. In addition, we halve the computational burden by evolving for grid points in the domain $0 \leq \theta \leq \pi/2$ only, and applying the symmetry condition (\ref{bc-equator}) on the equator.

\subsubsection{Initial conditions}
In Papers I and II we used very primitive initial data, and relied upon the radiative character of the equations to dissipate the ``junk radiation'' that is initially present. In this spirit, we will start with the trivial initial data set $u_{\alp \bet}^{(m)} = 0 = \partial_t u_{\alp \bet}^{(m)}$. We will show that this is sufficient for $m \ge 2$, but that more careful consideration is needed for the non-radiative modes contained in $m=0,1$.

\subsubsection{Decomposition in tensor spherical harmonics\label{subsec:lm-modes}}

The linearized equations governing Lorenz-gauge MPs on Schwarzschild spacetime may be separated using the set of ten tensor spherical harmonics. With this approach, the angular dependence of the equations is completely removed, leaving (for each $l,m$ where $l \ge m$) a set of ten coupled partial differential equations in independent variables $t$ and $r$. This set may be further decoupled into seven equations describing the even-parity part of the perturbation, and three equations describing the odd-parity part \cite{Regge:Wheeler:1957, Zerilli:1970}.
Thus far, we have spurned this decomposition, because our key motivation is to develop a method that can be generalized to axisymmetric spacetimes like Kerr's, where such a separation is not possible. However, we now give this decomposition for two reasons: firstly, to make possible comparisons with known results in the literature; and secondly, because this decomposition will later facilitate better understanding of the role of non-radiative modes in the Schwarzschild case.

Following Barack and Lousto \cite{Barack:Lousto:2005} [Eq.~(8) therein], one makes the decomposition
\beq
\hh_{\alp \bet} = \frac{\mu}{r} \sum_{l, m} \sum_{i=1}^{10} a^{(i)l} \hh^{(i)}_{lm}(t,r) Y_{\alp \bet}^{(i)lm}(\theta, \phi; r),   \label{hilm-def}
\eeq
where the tensorial harmonic basis functions $Y_{\alp \bet}^{(i)lm}(\theta,\phi;r)$ and coefficients $a^{(i)l}$ are defined in Sec.~IIB of \cite{Barack:Lousto:2005}. This should be compared with the $m$-mode expression, Eq.\ (\ref{h-ansatz}). It is straightforward to construct the $lm$ modes $\hh^{(i)}_{lm}(t,r)$ from the $m$-modes $u_{\alp \bet}^{(m)}(t,r,\theta)$ by integrating over $\theta$. In Appendix \ref{appendix:lmmodes} we give explicit expression for $\hh^{(i)}_{lm}(t,r)$ in terms of the $m$-mode variables.

\section{Numerical Results: Part I \label{sec:results:1}}

In this section we present a selection of numerical results. We show that the numerical simulation evolves towards equilibrium for modes $m \ge 2$, but that the lower modes ($m=0$ and $m=1$) are plagued by  instabilities which grow linearly with $t$. In Secs.~\ref{sec:low-multipoles} and \ref{sec:stabilization}, we investigate how these instabilities arise, and propose a way forward. In Sec.~\ref{sec:results:2} we present some results for the total GSF.

\subsection{$m$-modes\label{subsec:slices}}

Let us consider the results of a typical ``run'', i.e., a single simulation with a given $r_0$ and $m$, and with a particular set of numerical parameters $\{\text{num.} \} = \{ \Delt, \Delr, \Delq, \Gamma_{r_\ast}, \Gamma_{\theta}, t_{\max}\}$, where $t=t_{\max}$ is the physical evolution time.  We choose to present the results of the run by plotting data along three slices (as in Papers I \& II): (i) $t=t_\text{max}$, $\theta = \pi/2$, i.e.~the equatorial plane, (ii) $t=t_\text{max}$, $r=r_0$, i.e., from pole to pole and intersecting the worldline, and (iii) $r=r_0$, $\theta = \pi/2$, i.e., as a function of time along the worldline.

\begin{figure} 
 \begin{center}
  \includegraphics[width=12cm]{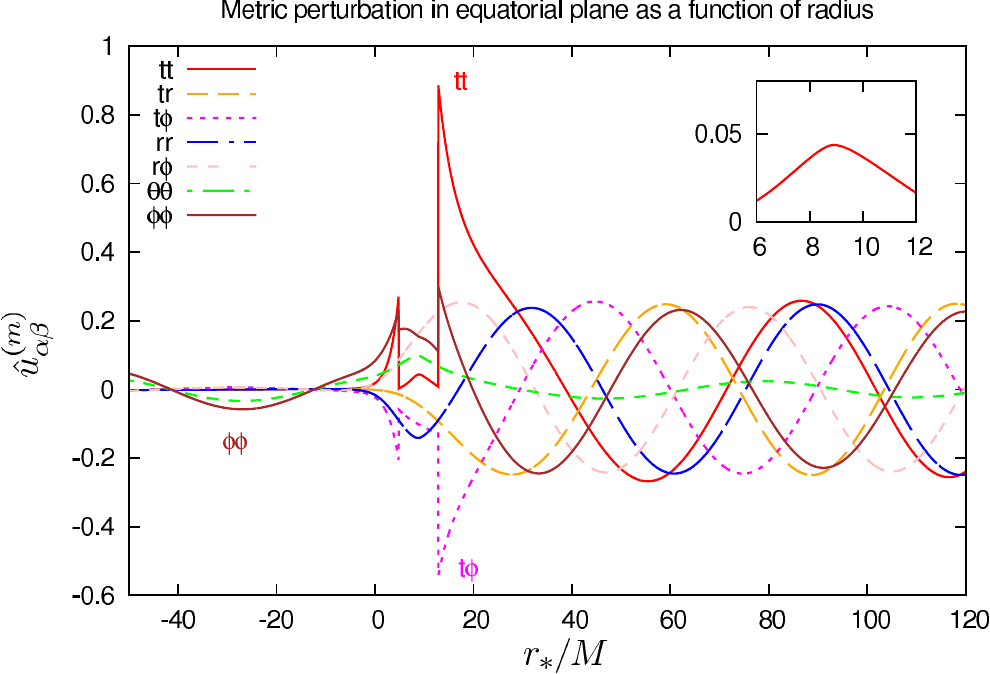}
 \end{center}
 \caption{Slice (i): The $m=2$ mode of the (trace-reversed) MP as a function of tortoise coordinate $r_\ast$, for $\theta = \pi/2$, $r_0 = 7M$ and $t=t_\text{max} = 300M$. The worldtube covers the region $r_{0\ast} - \Gamma_{r_\ast}/2 < r_\ast < r_{0\ast} + \Gamma_{r_\ast}/2$ where $r_{0\ast} \sim 8.8326 M$ and $\Gamma_{r_\ast} = 8M$. The curves show the components $\hat{u}_{\alp \bet}^{(m=2)}$ [defined in Eq.~(\ref{uhat-def}) and proportional to the trace-reverse perturbation, Eq.~(\ref{h-ansatz})] outside the worldtube, and the punctured version $\hat{u}_{\alp \bet}^{\mathcal{R}(m=2)}$ inside the worldtube. The $t\theta$, $r\theta$ and $\theta \phi$ components are not shown as they are zero on the equatorial plane by symmetry, Eq.~(\ref{bc-equator}).}
 \label{fig:slice1}
\end{figure}

\begin{figure} 
 \begin{center}
  \includegraphics[width=12cm]{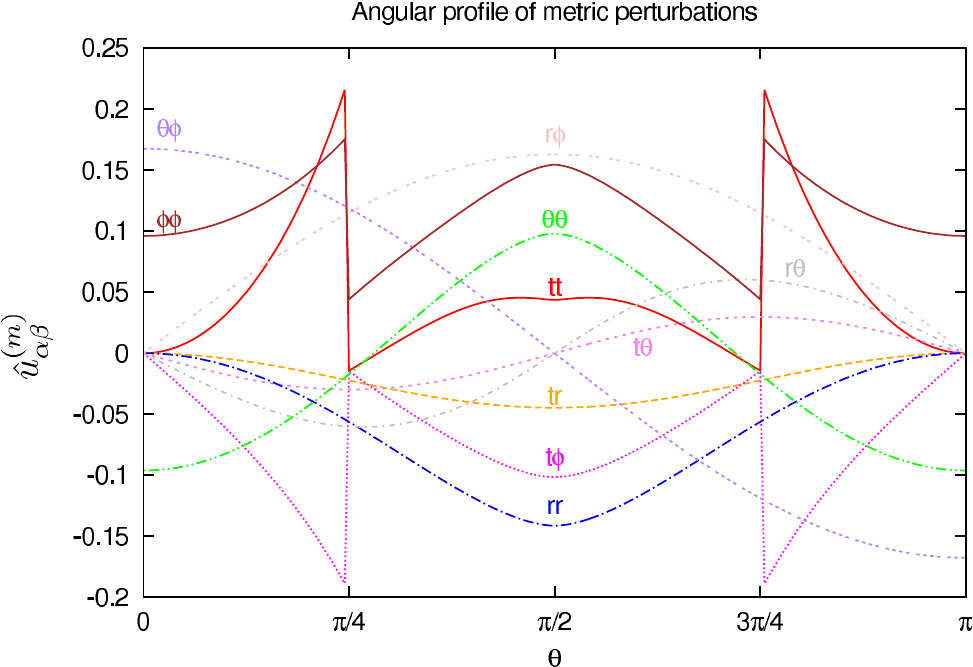}
 \end{center}
 \caption{Slice (ii): Showing the $m=2$ mode of the MP as a function of polar angle $\theta$, for $r = r_0 = 7M$, $t_\text{max} = 250M$. The worldtube covers the region $\pi/4 \le \theta \le 3 \pi/4$. The ten curves show the components $\hat{u}_{\alp \bet}^{(m=2)}$ [defined in Eq.~(\ref{uhat-def})] outside the worldtube, and the regularized version  $\hat{u}_{\alp \bet}^{\mathcal{R}(m=2)}$ inside the worldtube.  }
 \label{fig:slice2}
\end{figure}

\begin{figure} 
 \begin{center}
  \includegraphics[width=12cm]{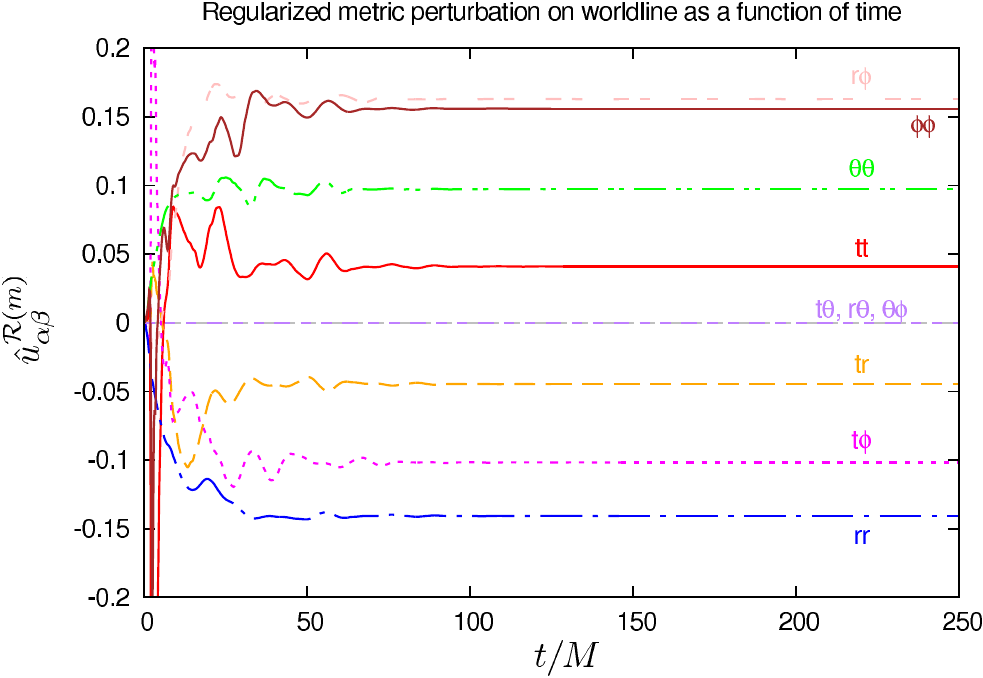}
 \end{center}
 \caption{Slice (iii): Showing the $m=2$ mode of the (regularized) MP on the worldline $r=r_0=7M$, $\theta=\pi/2$ as a function of simulation time $t$. }
 \label{fig:slice3}
\end{figure}

Figure \ref{fig:slice1} shows numerical data for the components of the MP as a function of $r_\ast$ at $\theta=\pi / 2$, $t = t_{\text{max}}$ [slice (i)]. Outside the worldtube we plot the real variables
\beq
\hat{u}^{(m)}_{\alp \bet}(t,r,\theta) \equiv \, \text{Re} \, \left[ u^{(m)}_{\alp\bet}(t,r,\theta) e^{i m \Omega t} \right] ,  \label{uhat-def}
\eeq
and inside we show the punctured version $\hat{u}^{\mathcal{R}(m)}_{\alp \bet}$.  The worldtube is visible as a ``trough'' around $r_0 = 7M$ ($r_{\ast0} \approx 8.832M$). In the asymptotic limits $r_\ast \rightarrow \pm \infty$ the original (complex) $m$-modes $u_{\alp\bet}^{(m)}$ go as $\sim \exp( \pm i m \Omega r_\ast )$. This behaviour is apparent as fixed-wavelength oscillations in the numerical data for $\hat{u}^{(m)}_{\alp \bet}$.  
Figure \ref{fig:slice2} shows numerical data as a function of polar angle, crossing the worldtube at a fixed time $t=t_{\rm max}$ [slice (ii)]. 
Figure \ref{fig:slice3} shows numerical data along the worldline as a function of time [slice (iii)]. The plot makes it clear that the components of the (regularized) MP on the worldline approach stationary values. In other words, the MP is ``co-rotating'' with the particle, and the modes have the expected time dependence. 

\subsection{Validation and tests\label{subsec:validation}}

\subsubsection{Gauge constraint violation}
\begin{figure} 
 \begin{center}
  \includegraphics[width=8cm]{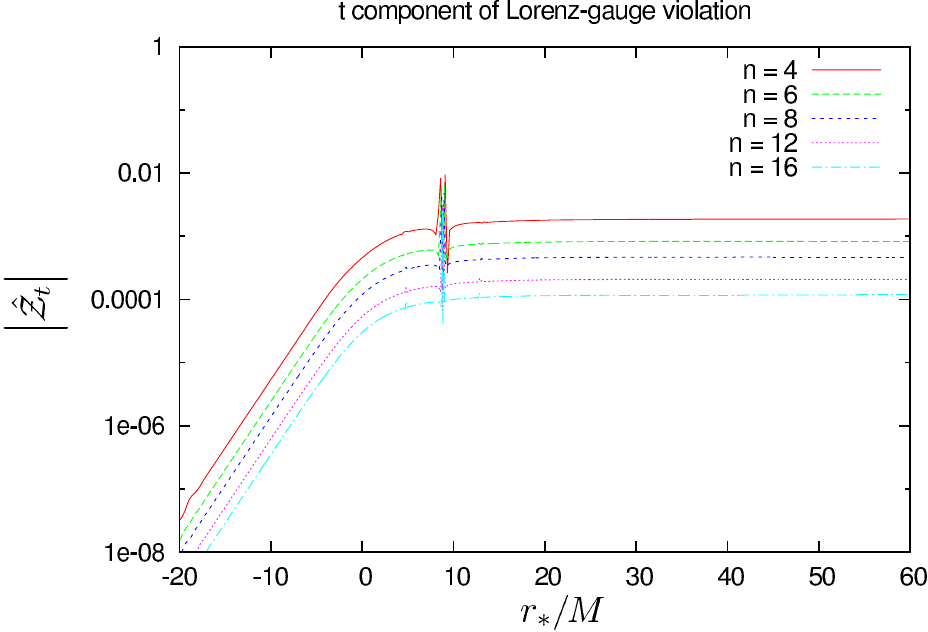}
  \includegraphics[width=8cm]{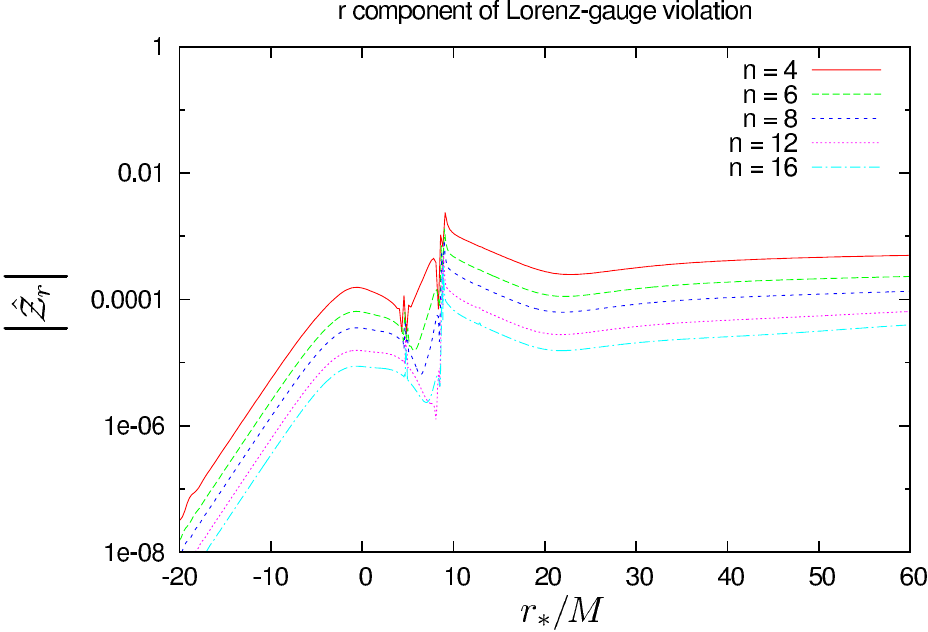}
  \includegraphics[width=8cm]{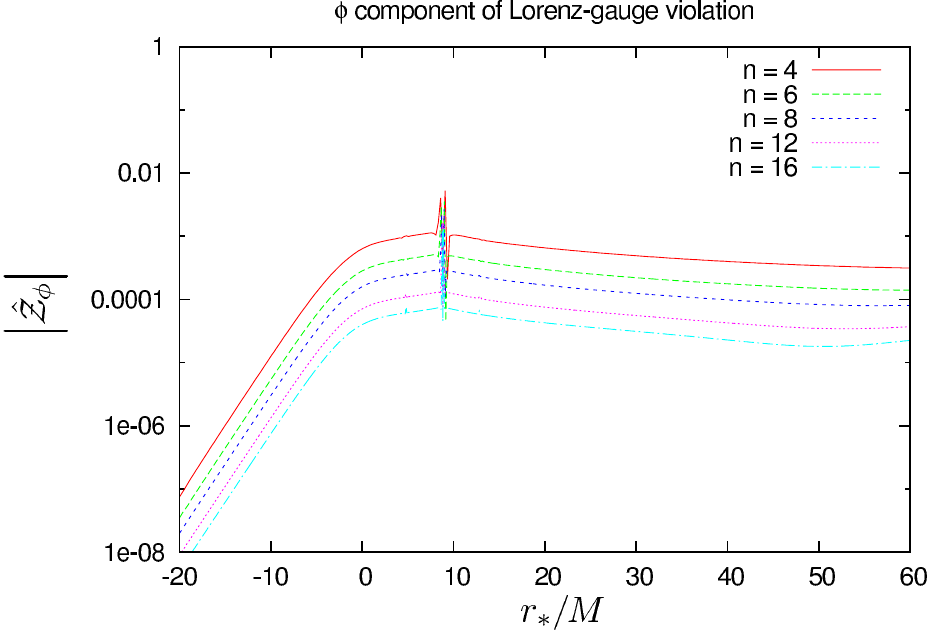}
 \end{center}
 \caption{Gauge constraint violation. The plots show the amplitude of $\ZZ^{(m)}_\alpha$ [defined in Eqs.~(\ref{gauge-t})--(\ref{gauge-p})], normalized by $\left|\partial_t h^{(m)}\right|=\left|m\Omega h^{(m)}\right|$, where $h^{(m)}$ is the trace of the $m$-mode MP---we denote this normalized quantity by $\hat\ZZ^{(m)}_\alpha\equiv \ZZ^{(m)}_\alpha/\left|m\Omega h^{(m)}\right|$. The 3 panels display the $t$, $r$ and $\phi$ components. Each panel shows runs at various resolutions $n$, where $\Delr = M / n = \Delt$ and $\Delq = \pi / (8 n)$.  The data shown is for the specific case $m=2$, $r_0 = 7M$, $t = 250M$, and given as a function of $r_\ast$ in the equatorial plane $\theta=\pi/2$. The spurious values very near $r_\ast = r_{0\ast}$ should be disregarded; they arise because the retarded field (and hence $Z_{\alpha}$) is not defined on the worldline.}
 \label{fig:gcd}
\end{figure}

The numerical solution only represents a physical solution of the linearized Einstein equations if the gauge constraints are satisfied. In practice, we do not expect the constraints to be identically zero, but rather to be ``small'' in some sense, and to approach zero as the resolution is improved. 
Fig.~\ref{fig:gcd} shows the profile of gauge constraint violation $\ZZ_{\alpha}^{(m)}$ [defined in Eqs.~(\ref{gauge-t})--(\ref{gauge-p})] as a function of $r_\ast$ on the equatorial plane $\theta=\pi/2$, in a run with $m=2$, $r_0=7M$ at $t=250M$. It makes sense to compare the magnitude of $\ZZ_{\alpha}^{(m)}$ (which is itself dimensionful) to some norm involving the MP's gradient, and in Fig.~\ref{fig:gcd} we have chosen as norm the quantity $\left|\partial_t h^{(m)}\right|=\left|m\Omega h^{(m)}\right|$, where $h^{(m)}$ is the trace of the $m$-mode MP. The plots show that the (normalized) constraint violation is maximal near the particle's position, but even there it remains quite small: $\sim10^{-4}$ for our highest resolution. More importantly, the constraint violation diminishes as the grid resolution is increased. 

\subsubsection{Decay of junk radiation}

\begin{figure} 
 \begin{center}
  \includegraphics[width=10cm]{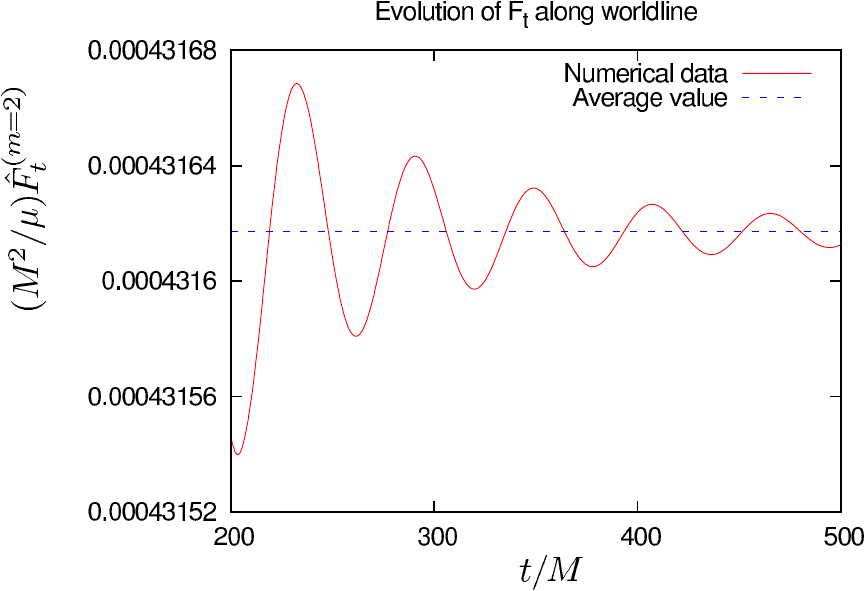}
 \end{center}
 \caption{Power-law decay of junk radiation. The plot shows numerical data for the self-force mode $\hat{F}_t^{(m)}$, for $m=2$ and $r_0 = 7M$, extracted from a time-domain simulation up to $t_{\text{max}} = 500M$ [the real quantity $\hat{F}_t^{(m)}$, defined in Eq.\ (\ref{Fm}), is formed by adding together the contributions from the two modes $\pm|m|$). The small damped oscillations (note the vertical scale varies by less than 0.04\%), with frequency $2 \Omega$ and an envelope scaling as $t^{-3}$, are due to the power-law decay of junk radiation generated by unphysical initial data.  The equilibrium value (dashed line) may be estimated by averaging over a complete wave cycle, as described in the text.}
 \label{fig:powerlaw-decay}
\end{figure}

Our trivial (unphysical) initial data create a burst of ``junk radiation'', which (we find, for $m \ge 2$) dissipates away with time as the numerical solution approaches a stationary state. An example is shown in Fig.\ \ref{fig:powerlaw-decay}. The figure shows the real quantity $\hat F_t^{(m=2)}$ defined in Eq.\ (\ref{Fm}) (recall this represents the combined contribution from the two modes $m=\pm 2$) as a function of $t$ at late time. This quantity is constant (non-oscillatory) in the physical solution. We see that junk radiation in the numerical solution superposes decaying oscillations of frequency $m\Omega$ on the physical $m$-mode solution. 
The amplitude of these oscillations is quite small in the example shown: note the vertical scale.
[The origin of oscillations in the junk radiation, which is not sourced by the particle, is simply the fact that $F_t^{(m)}$ is evaluated in a system co-rotating with the particle---note the factor $e^{im\phi}$ in Eq.\ (\ref{Fm}).]

If we were solving the actual perturbed Einstein equations, then we would expect the decay to equilibrium to follow the familiar power-law pattern characteristic of generic vacuum perturbations of Schwarzschild spacetime at late time: individual radiative multipoles die off as $\propto t^{-(2l+3)}$, and since the lowest radiative mode contained in a generic $m$-mode perturbation is $l=|m|$, the $|m|$-mode vacuum MP would be expected to exhibit a $\propto t^{-(2|m|+3)}$ decay tail at late time. Indeed, in the scalar-field case of papers I and II we have observed a relaxation pattern consistent with this rule. Here, however, crucially, we are not solving the perturbed Einstein equations but rather the system (\ref{EE}), which is not equivalent to the Einstein equations unless the constraint is satisfied. The junk radiation does not generally satisfy the constraint, and thus it does not represent a physical solution of the Einstein equations. The rate of decay of these junk solutions is a priori unclear. 

To understand the approach to equilibrium in our system, we experimented with vacuum runs starting with some Gaussian (non-Lorenz gauge, non-static) initial data. We found that, at late time, the system undergoes power-law decay in which (for $m \ge 2$) the BL metric components die off as $t^{-(2|m|-1)}$, i.e., significantly slower than the decay of a scalar field in vacuum. We also found that certain cancellations occur at late time, so that the gauge constraints $Z_\alp$ and the trace $h$ decay as $t^{-(2|m|+1)}$---faster than the metric components but still slower than a scalar field. We have confirmed the behaviour seen in the 2+1D simulations by also evolving our gauge-damped $Z_4$ system in 1+1D (as in Ref.~\cite{Barack:Lousto:2005}). Our vacuum 1+1D evolutions for specific multipoles $l\geq 2$ (and $m\leq l$) showed $t^{-(2l-1)}$ decay tails for the MP, and $t^{-(2l+1)}$ for $Z_\alp$ and $h$, consistent with our 2+1D results. 



The above (empirically deduced) relatively slow decay of junk radiation is an unfortunate feature of our constraint-damping formulation. It might be possible to improve this feature by controlling the form of the constraint-damping terms ($\propto\kappa$) at large $r$, which is where late-time decay tails form through backscattering. We leave this for future study. In the case of circular orbits considered here we may reduce the adverse effect of slow junk decay simply by averaging the late-time solutions over a complete wave-period 
$T_m \equiv 2\pi / (m\Omega)$. Assuming the junk radiation has the late-time form $\sim e^{im\Omega t}t^{-k}$, with $k$ as indicated above, such an averaging procedure effectively suppresses the amplitude of oscillations (e.g., in the real quantity $\hat{F}_\alp^{(m)}$) by a factor $\sim 2\pi^2[k(k+1)]^{-1}(t/T_m)^2\sim \frac{1}{2}\pi^2(t/T_{\rm orb})^2$ at leading order in $t/T_m$, where $T_{\rm orb}\equiv 2\pi/\Omega$ is the orbital period and where in the second expression we have approximated $k\sim 2|m|$. This is a significant gain for $t\gg T_{\rm orb}$. We have implemented this ad-hoc procedure in our analysis; the outcome is illustrated in Fig.\ \ref{fig:powerlaw-decay}.

\subsubsection{Convergence of results with grid resolution}


In vacuum, our finite-difference scheme is 4th-order accurate, in the sense that, for grid spacing $\lambda \Delt$, $\lambda \Delr$ and $\lambda \Delq$, the discretization error scales approximately as $\lambda^4$. In our case, however, the irregularity of $S^{(m)\text{eff}}_{\mu\nu}$ at the worldline disrupts the global convergence of the finite difference scheme. As in Paper I, we find that a 2nd-order puncture formulation leads to a global discretization error that scales with $\lambda^2 \ln \lambda$ at leading order. 

\begin{figure} 
 \begin{center}
  \includegraphics[width=8cm]{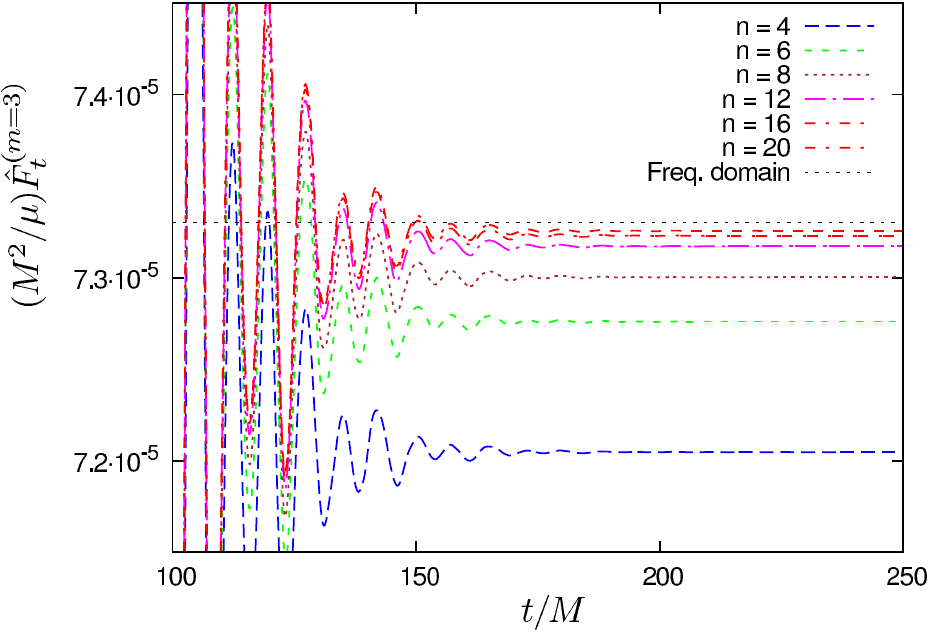}
  \includegraphics[width=8cm]{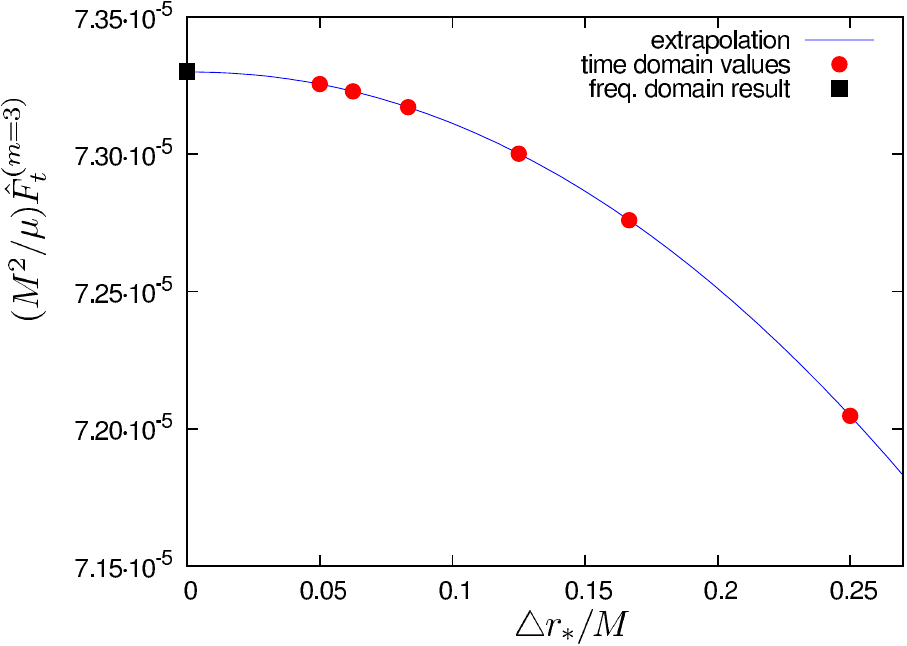}
 \end{center}
 \caption{Convergence of time-domain $m$-mode data with grid resolution. The left panel shows numerical data for $\hat F_t^{(m=3)}$ [cf.\ Eq.\ (\ref{Fm})], extracted on the worldline at $r_0 = 7M$ and plotted as a function of simulation time $t$, for a range of grid resolutions $n$, with $\Delr = M / n = (5/4) \Delt$ and $\Delq = \pi / (8 n)$, for $n \in \{4,6,8,12,16,20\}$.
The numerical data converges with finite-differencing error $\propto n^{-2} \ln n$.
The right panel illustrates our Richardson extrapolation procedure. Red circles show the values extracted at $t_{\text{max}} = 250M$ for grid resolutions corresponding to $n \in \{4,6,8,12,16,20\}$ from right to left.
The blue line shows the line of best fit, according to the model given in Eq.~(\ref{fit-model}). The black square in the right panel, and the straight (blue) dashed line in the left panel, mark the highly accurate value of $F_t^{(m)}$ obtained using a frequency-domain calculation \cite{Akcay:2011}. The relative difference between our extrapolated value and the accurate frequency-domain value is $\sim 2.5 \times 10^{-5}$.
}
 \label{fig:convergence}
\end{figure}

In order to improve our estimates of the physical results, we fitted an appropriate model to data extracted from simulations at various resolutions. The fitted model allows us to extrapolate to zero grid spacing, $\lambda \rightarrow 0$ (``Richardson's deferred approach to the limit'', \cite{NumericalRecipes}). We applied the model
\beq
X(\lambda) = X(0) + c_{\text{ln}} \lambda^2 \ln \lambda + c_2 \lambda^2 + c_3 \lambda^3 , \label{fit-model}
\eeq
where $c_{\text{ln}}$, $c_2$ and $c_3$ are numerical coefficients to be determined, and $X$ stands for $\hat{u}_{\alp\bet}^{\mathcal{R}(m)}$ or $\hat{F}_\alp^{(m)}$.
An example is shown in Fig.~\ref{fig:convergence}, for the mode $m=3$. We compare the extrapolated value with a highly accurate value which was obtained using a frequency-domain calculation by Akcay \cite{Akcay:2011} (summing over all frequency modes and all $l \ge 3$ for $m=3$). 
In this typical example, after extrapolation with the model above, Eq.~(\ref{fit-model}), we find agreement up to a fractional error of $\sim 2.5 \times 10^{-5}$. 

\subsubsection{Comparison with $lm$-mode results}

\begin{figure} 
 \begin{center}
  \includegraphics[width=14cm]{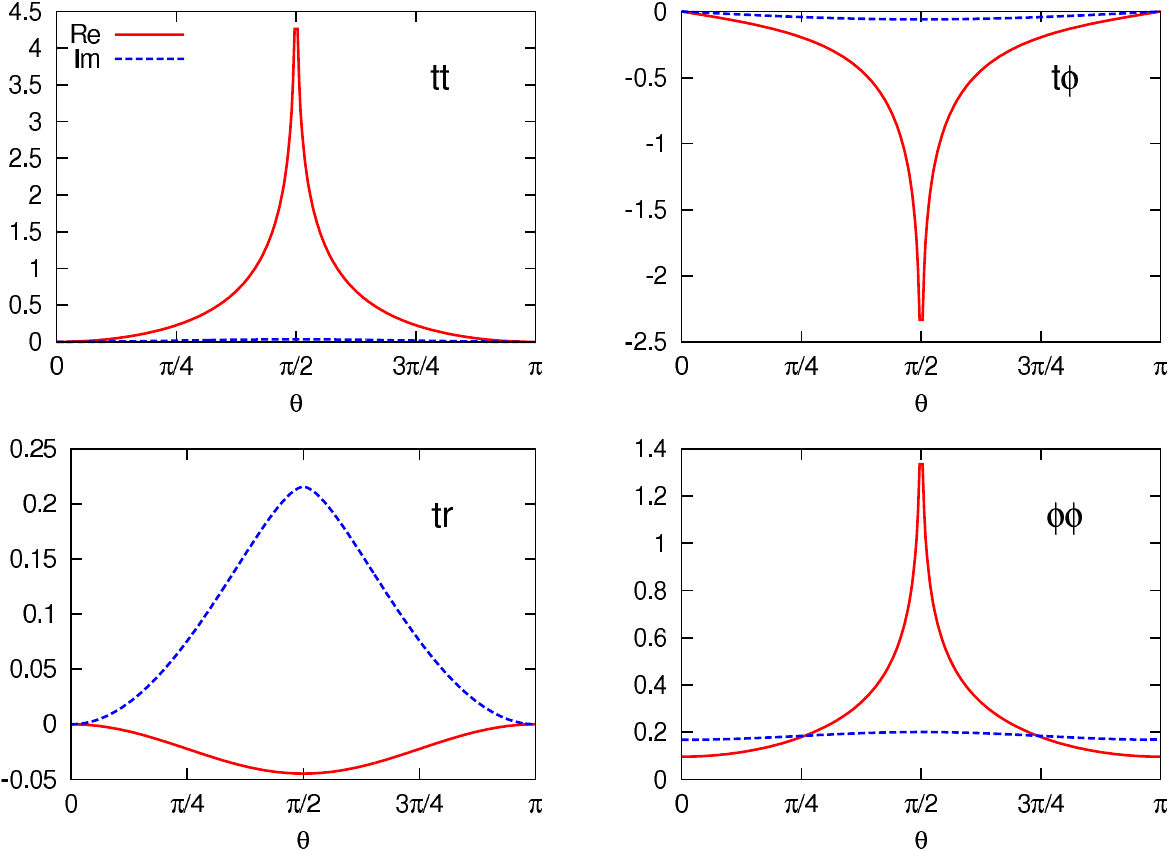}
 \end{center}
 \caption{The angular profile of the (complex) retarded field variables ${u}^{(m)}_{\alp\bet}e^{im\Omega t}$, for components $\alp\bet$ given by (clockwise from top left) $tt$, $t\phi$, $\phi\phi$ and $tr$. The real parts of the former three components are logarithmically divergent at the worldline. Here $m = 2$ and $r_0 = 7M$.}
 \label{fig:ang_profile}
\end{figure}

Figure \ref{fig:ang_profile} shows numerical data for the angular profile of the (unregularized) retarded MP. More precisely, it shows the real and imaginary parts of ${u}^{(m)}_{\alp\bet}e^{im\Omega t}$, for $m=2$ at $r=r_0=7M$. The plots show that the components $tt$, $t\phi$ and $\phi\phi$ diverge as the worldline is approached. It can be checked that this divergence is logarithmic, i.e.~$\sim \ln| \theta - \pi/2|$. The other  components (for example, $tr$ in Fig.~\ref{fig:ang_profile}) are regular across the worldline. 

We may extract numerical estimates for $lm$ modes from $m$-mode data by projecting the latter onto the tensor spherical harmonics, and using the relations given in Appendix~\ref{appendix:lmmodes}. With this approach, we have validated our results against previous frequency-domain studies \cite{Akcay:2011}. Table \ref{table:lm-validation} shows typical numerical results from this process. Here we have extracted the value of the $l=m=2$ even-parity ($i=1\ldots7$) modes on the worldline at $r=r_0=7M$. The small discrepancy between the $m$-mode estimate and the (highly-accurate) $lm$ mode results is at the expected level of discretization error of our time-domain scheme, $\sim 5\times 10^{-5}$ fractionally.

\begin{table}
\begin{tabular}{c | r r}
\hline\hline
$(i)$ & \multicolumn{2}{c}{$l=2$, $m = 2$} \\
\hline
\multirow{2}{*}{$1$} & $3.12437$ & $-\,0.26317 i$  \\ & $3.12457 $ & $-\, 0.26316i$ \\ \hline
\multirow{2}{*}{$2$} & $-0.23121$ & $+\,0.97576 i$  \\ & $-0.23121$ & $+\,0.97577i$ \\ \hline
\multirow{2}{*}{$3$} & $3.79704$ & $+\,0.44010 i$ \\ & $3.79727$ & $+\,0.44011i$ \\ \hline
\multirow{2}{*}{$4$} & $-0.92491$ & $+\,9.42906 i$ \\ & $-0.92491$ & $+\,9.42918i$ \\ \hline
\multirow{2}{*}{$5$} & $-2.33102$ & $-\,2.52793 i$ \\ & $-2.33103$ & $-\,2.52790i$ \\ \hline
\multirow{2}{*}{$6$} & $1.54687$ & $+\,0.60066 i$ \\ & $1.54684$ & $+\,0.60065i$ \\ \hline
\multirow{2}{*}{$7$} & $-5.33209$ & $-\,5.21910 i$  \\ & $-5.33189$ & $-\,5.21903i$ \\


%

\hline
\hline\hline
\end{tabular}
\caption{Example of numerical validation of $m$-mode results against $lm$-mode results for the case $l = m = 2$ and $r_0 = 7M$. The index $i$ labels individual tensor-harmonic modes $\hh^{(i)}_{lm}(r = r_0)$ [cf.\ Eq.\ (\ref{hilm-def})], with $i=1,\ldots,7$ corresponding to modes of even parity. In each entry, the upper value is extracted from our $m$-mode scheme, using Eqs.~(\ref{hilm-eq1})--(\ref{hilm-eq10}). The lower value is obtained from a highly accurate frequency-domain calculation \cite{Akcay:2011}. The small discrepancy is consistent with the expected level of error in the $m$-mode data. Similar accuracy is found for odd-parity modes (e.g.~$l=3$, $m=2$, $i=8\ldots10$) and for higher modes, $m > 2$.} 
\label{table:lm-validation}
\end{table}

\subsubsection{Convergence of the $m$-mode sum}

As established in Paper I (and in Ref.~\cite{Barack:Golbourn:Sago:2007}), with a 2nd-order puncture implementation we expect the modes of the residual field $\hat{u}_{\mu\nu}^{\mathcal{R}(m)}$, and the modes of the self-force $\hat{F}_r^{(m)}$, to fall away as $\sim m^{-2}$ in the large-$m$ limit. The dissipative components of the GSF ($F_t$ and $F_\phi$ for circular orbits), which we expect to be exponentially convergent with $m$, are exceptions to this rule. The individual modes of $F_t$ and $F_\phi$ have a gauge- and puncture-independent interpretation, in terms of the rate of loss of energy and angular momentum through gravitational wave emission. On the other hand, the individual modes of $F_r$ are somewhat arbitrary, since they depend on the analytic extension of the puncture function away from the worldline (which is implementation-dependent).

\begin{figure} 
 \begin{center}
  \includegraphics[width=8cm]{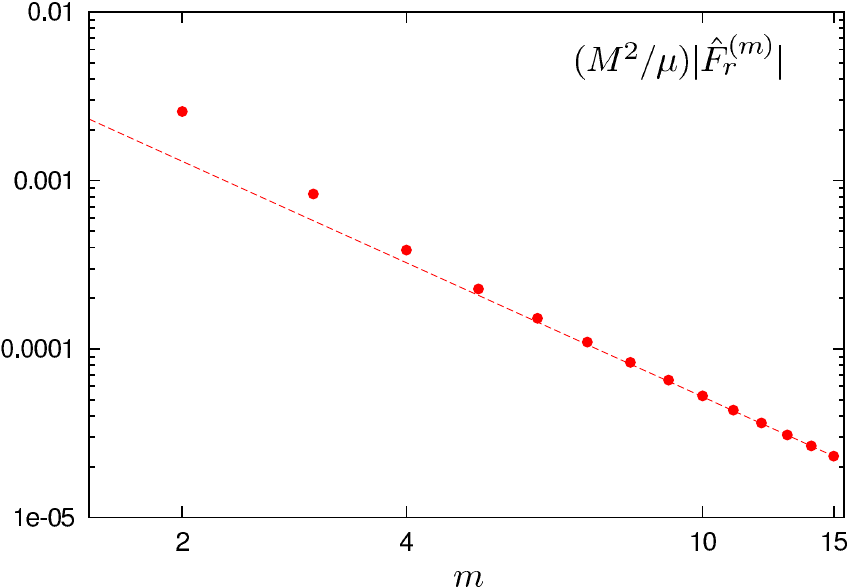}
  \includegraphics[width=8cm]{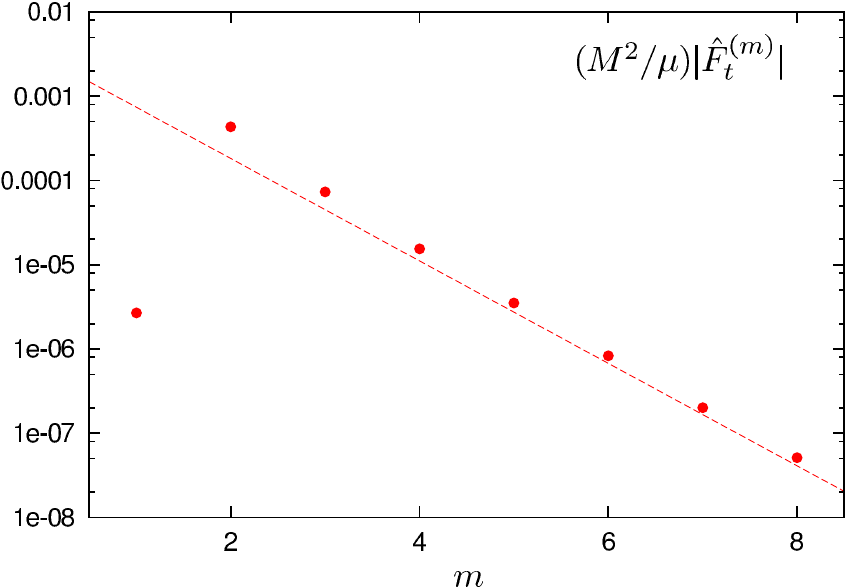}
 \end{center}
 \caption{Convergence of the $m$-mode sum. The left panel illustrates, on a log-log scale, the $m^{-2}$ power-law fall-off (at large $m$) of the $m$-mode contributions $\hat F_r^{(m)}$ for a circular orbit at $r_0 = 6M$. The dashed line is a reference $\sim 0.0052/m^2$. The right panel shows, on a semilog scale, the exponential decay exhibited by the $m$-modes $\hat{F}_t^{(m)}$ of the dissipative GSF component. The dashed line is a reference $0.003\times e^{-1.4m}$.
}
 \label{fig:m-mode-convergence}
\end{figure}

Figure \ref{fig:m-mode-convergence} confirms, in the example of $r_0=7M$, that our expectations for the large-$m$ behaviour of $\hat{F}_t^{(m)}$ and $\hat{F}_r^{(m)}$ are met.  

\subsection{The modes $m=0$ and $m=1$ \label{subsec:m01-numerical}}
In the preceding sections we have shown that, for the modes $m \ge 2$, the MP settles into an equilibrium configuration, and initial junk radiation dissipates with time. Unfortunately, this is not the case for the modes $m=0$ and $m=1$. 


In earlier 1+1D studies of the GSF, the modes $l=1$ and $l=0$ could not be obtained through the time-domain approach, and were obtained separately using frequency-domain analyses \cite{Barack:Ori:Sago:2008}. Indeed, a fully-time-domain Lorenz-gauge scheme has not yet been successfully developed. Barack and Sago \cite{Barack:Sago:2010} noted that ``experimentation suggests to us that the monopole and dipole cannot be evolved stably using this [time-domain] scheme. A naive application of the evolution scheme yields exponentially growing solutions, and, since our scheme gives us no handle on the boundary conditions, the occurrence of these unphysical solutions is difficult to control.'' 

\begin{figure} 
 \begin{center}
  \includegraphics[width=8cm]{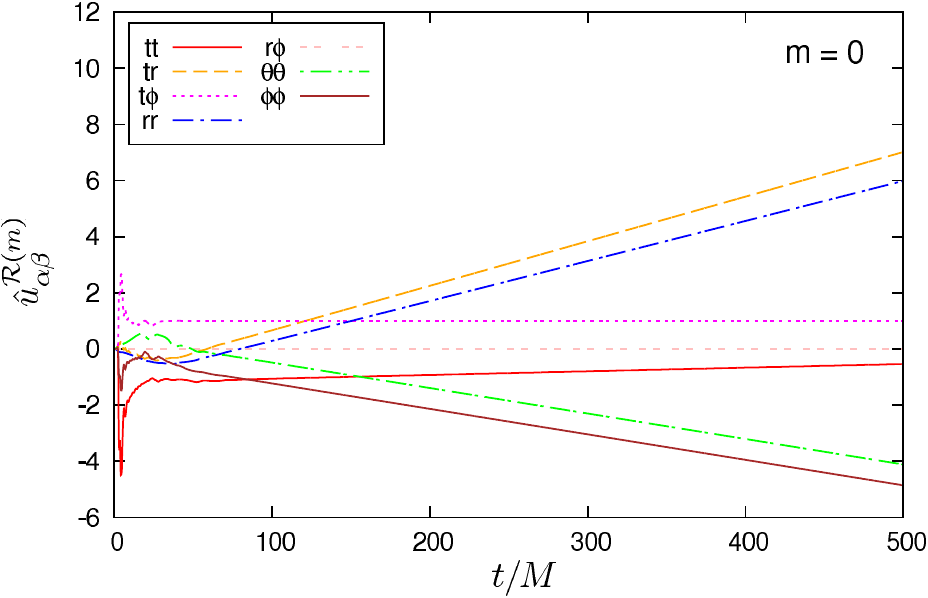}
  \includegraphics[width=8cm]{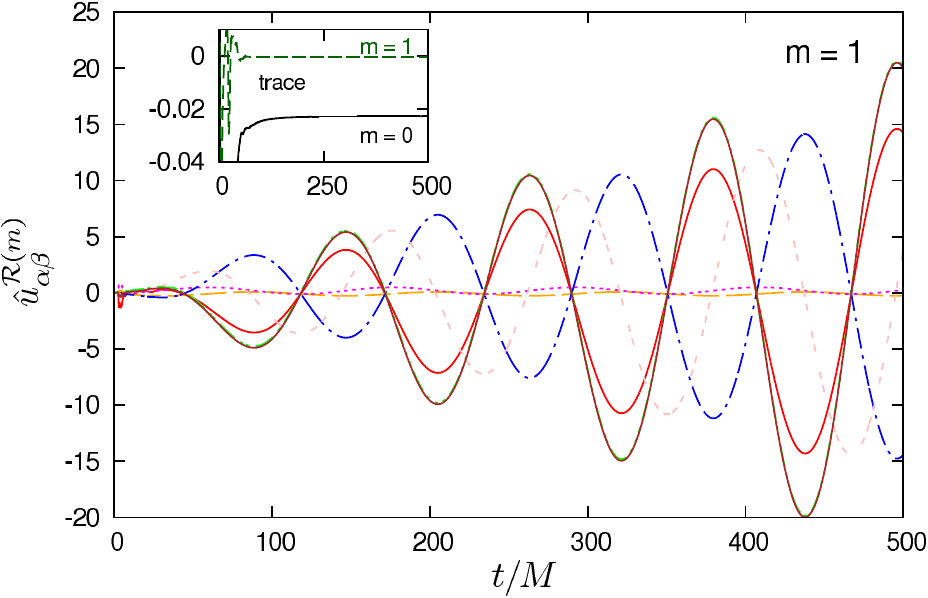}
 \end{center} 
 \caption{The non-zero components of the $m=0$ and $m=1$ modes of the (regularized) MP variables $\hat{u}^{(m)}_{\alp \bet}$ on the particle's worldline [cf.\ Eq.\ (\ref{uhat-def})] at $r_0 = 7M$, as a function of time $t$. Left panel: the $m=0$ mode, showing linear growth in the $tt$, $tr$, $rr$, $\theta\theta$ and $\phi\phi$ components. The $t\phi$ component approaches a stationary value. Right panel: the $m=1$ mode, evaluated on the worldline (at $\phi = \Omega t$). The linear-in-$t$ growth appears on the worldline as a growing oscillation of frequency $\Omega$ in components $tt$, $rr$, $r\theta$, $r\phi$, $\theta\theta$ and $\phi\phi$. For both $m=0$ and $m=1$ the trace approaches a stationary value (inset), which suggests that the linear-in-$t$ part is traceless.
}
 \label{fig:m01-t}
\end{figure}

With our choice of gauge-constraint damping (Sec.~\ref{subsec:gcd}) and a Cauchy ($t,r_\ast$) scheme we do \emph{not} observe exponentially growing solutions. On the other hand, we do observe solutions which grow linearly with time $t$. Numerical data from typical evolutions of the $m=0$ and $m=1$ modes are shown in Fig.~\ref{fig:m01-t}. In both cases, it seems that the instability can be attributed to a homogeneous (vacuum) mode, which is generically excited by our unphysical initial data. Such homogeneous modes appear in the co-rotating variables $\hat{u}_{\alp\bet}^{(m)}$ as oscillations with frequency $m \Omega$.

\begin{figure} 
 \begin{center}
  \includegraphics[width=10cm]{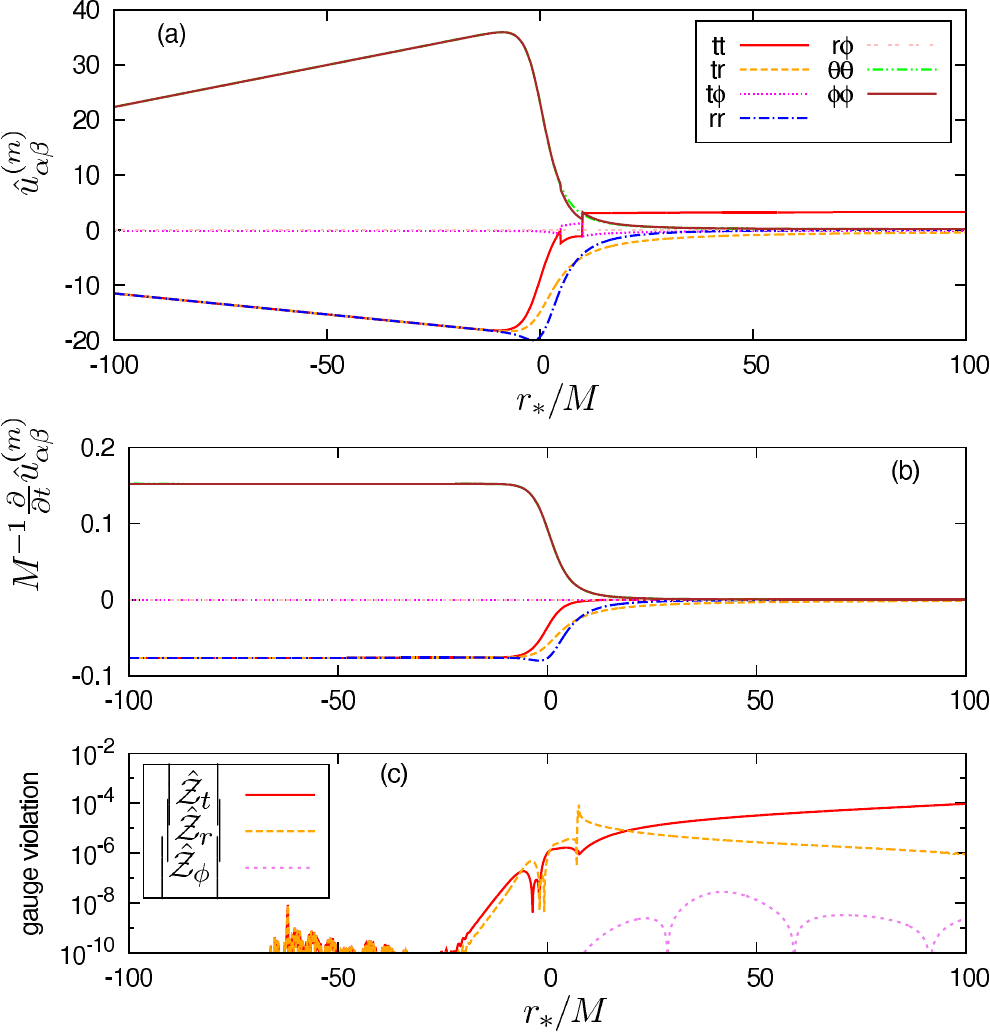}
 \end{center}
 \caption{Features of the linear-in-$t$ instability in the $m=0$ mode, illustrated here in the equatorial plane at $t=t_\text{max} = 250M$ for circular orbit at $r_0 = 6M$. 
 (a) Components of the MP $\hat{u}_{\alp \bet}^{(m=0)}$, showing that in the near-horizon regime ($r_\ast \lesssim -10M$) the components $tt$, $tr$, $rr$, $\theta\theta$ and $\phi\phi$ scale linearly with $r_\ast$. (b) The time derivative of the components, showing that in the near-horizon regime the same components scale linearly with $t$ as well; in the text we conclude that the growing mode is ``purely ingoing'' at the event horizon. (c) The (normalized) gauge-constraint violation, which appears to be small and which diminishes as grid resolution is improved. Here we show the amplitude of $\ZZ^{(m)}_\alpha$ [defined in Eq.~(\ref{gauge-t})--(\ref{gauge-p})], normalized by $\left|M h^{(m=0)}\right|$, where $h^{(m=0)}$ is the trace of the $m=0$ MP.
 }
 \label{fig:m0-mode}
\end{figure}

Figure \ref{fig:m0-mode} shows the radial profile of a typical evolution of the $m=0$ mode, at late times. The plots indicate that the components of the MP that grow linearly in $t$ also scale as $\sim r_\ast$ near the horizon, and we find that the near-horizon scaling is in fact proportional to the advanced time coordinate $v = t + r_\ast$. In the near horizon limit we find $\hh_{tt} \sim \hh_{tr} \sim \hh_{rr} \sim -\tfrac12 \hh_{\theta\theta} \sim-\tfrac12 \hh_{\phi\phi} = A v$ for some amplitude $A$ which depends on initial data. In other words, these troublesome modes demonstrate ``ingoing'' behaviour at $\Hplus$ [i.e., $(\partial_t - \partial_{r_\ast}) \hh_{\alp \bet} \rightarrow 0$]. Figure \ref{fig:m0-mode}(c) shows the components of the Lorenz-gauge violation $\ZZ_{\alp}$. We observe that the gauge violation remains small, suggesting that the growing modes are indeed valid Lorenz-gauge solutions. 

By examining snapshots of the pole-to-pole angular profile, we infer that only the low multipoles $l \le 2$ are implicated in the instability. This suspicion is further confirmed by evolving the equivalent $lm$ field equations in $1+1$D [Eqs.\ (\ref{monoeq00})--(\ref{monoeq22}) below].

In the next section, we derive explicit analytic expressions for the above linear-in-$t$ modes that are disrupting our naive implementation of the time-domain scheme. We show that these modes are pure (Lorenz)-gauge modes, which are closely related to ``scalar-type'' (traceless) gauge modes. In Sec.\ \ref{sec:stabilization} we then address the challenge of ``stabilizing'' the $m=0$ and $m=1$ evolutions, and in Sec.\ \ref{sec:results:2} we display some final results for the total GSF and compare with the literature.

\section{Low multipoles: Analytic considerations\label{sec:low-multipoles}}

In this section we consider the non-radiative multipoles of the MP, i.e.,~$l=0$ and $l=1$ in the decomposition of Sec.~\ref{subsec:lm-modes}. Physical solutions for these modes have been discussed before, for example in Ref.\ \cite{Detweiler:Poisson:2004,Ori:2004,Barack:Lousto:2005,Sago:Barack:Detweiler:2008}. Here, we give a further analysis which focuses on two issues: (i) the presence of conserved mass-energy and angular momentum in the low multipoles of the MP, associated with Killing vectors of the background spacetime (see the covariant derivation in Sec.~\ref{subsec:conservation}), and (ii) the existence of linear-in-$t$ (and Lorenz-gauge) gauge modes, which are regular on $\Hplus$ (in the sense of Sec.~\ref{subsec:bc}) as well as at infinity. A key point is that such gauge modes, which, at any finite time, are well-behaved everywhere, cannot be eliminated using boundary conditions alone. To eliminate these modes we must impose additional conditions; for example, the condition that the monopole perturbation is static (i.e.~$\partial_t h_{\mu\nu} = 0$ and $h_{tr} = 0$). 

Some parts of the analysis below, especially from Sec.~\ref{subsec:mono-homo} onwards, will take a more general form (less tied to the 1+1D context), with the idea of preparing the ground for a similar analysis in Kerr. 

\subsection{Monopole perturbation}

The $l=m=0$ monopole mode is spherically symmetric and of even parity; physically, it describes the mass perturbation due to the particle (up to a gauge).  In the Lorenz gauge, this mode is governed by four field equations and two gauge constraints. Analytic solutions for the monopole were previously obtained in Sec.~III of \cite{Detweiler:Poisson:2004} and in Sec.~IIID of \cite{Barack:Lousto:2005} for circular orbits. 
Here we present an alternative, but closely related, analysis.


\subsubsection{Monopole equations\label{subsec:mono-eq}}

Let us begin by stating the monopole equations in terms of the spherically-symmetric part of the relevant $m$-mode variables, i.e., $u_{\alp\beta} \equiv \tfrac{1}{2} \int_{0}^\pi \sin \theta u_{\alp\beta}^{(m=0)}(t,r,\theta) d\theta$. The nonvanishing components are related to the $lm$ modes $\hh^{(i)}_{lm}$ of Eq.~(\ref{hilm-def}) via
$u_{tt} = a (h^{(1)}_{lm} + h^{(3)}_{lm})$, $u_{tr} = a h^{(2)}_{lm}$, $u_{rr} = a (h^{(1)}_{lm} - h^{(3)}_{lm})$, and $u_{\theta\theta}=u_{\phi\phi} = a h^{(6)}_{lm}$, where $a = (16\pi)^{-1/2}$. The four Lorenz-gauge equations without constraint damping are
\begin{eqnarray}
D^2 u_{tt} &+& \frac{2(u_{tt} - u_{rr})}{r^4} + \frac{4(\dot{u}_{tr} - u_{tt}^\prime)}{r^2} + \frac{4f(u_{tt}-u_{rr})}{r^3} + \frac{4 f^2 u_{\theta\theta}}{r^3} =  \frac{4 \mu \mathcal{E}}{r_0} f_0 \delta(r_\ast - r_{0\ast}),  \label{monoeq00}  \\
D^2 u_{tr} &+& \frac{2(\dot{u}_{tt} + \dot{u}_{rr} - 2u_{tr}^\prime)}{r^2} - \frac{2 f^2 u_{tr}}{r^2}  = 0 ,   \label{monoeq01}  \\
D^2 u_{rr} &+& \frac{2(u_{rr} - u_{tt})}{r^4} + \frac{4(\dot{u}_{tr} - u_{rr}^\prime)}{r^2}  - \frac{4 f (u_{tt} - u_{rr})}{r^3} - \frac{4 f^2 u_{rr}}{r^2} - \frac{4 f^2 u_{\theta\theta}}{r^3} + \frac{4 f^3 u_{\theta\theta}}{r^2} = 0,   \label{monoeq11}  \\
D^2 u_{\theta\theta} &+& \frac{2(u_{tt} - u_{rr})}{r^3} + \frac{2 f (u_{rr} + u_{\theta\theta})}{r^2} - \frac{4f^2u_{\theta\theta}}{r^2} = \frac{2 \mu \mathcal{E}}{(r_0 - 2)^2}  f_0 \delta(r_\ast - r_{0\ast}) ,  \label{monoeq22} 
\end{eqnarray}
where $D^2  = -\partial_t^2 + \partial_{r_\ast}^2 - 2f/r^3$, and $\dot{u}$ and $u^\prime$ denote differentiation with respect to $t$ and $r_\ast$, respectively. Recall $r_0$ is the orbital radius and $f_0\equiv 1-2M/r_0$. The monopole gauge constraints may be written as
\begin{eqnarray}
\ZZ_t &\equiv& \frac{(u_{tr}^\prime - \dot{u}_{tt})}{r} + \frac{f u_{tr}}{r^2}  = 0 ,  \label{Xt-def}  \\
\ZZ_r &\equiv& \frac{(u_{rr}^\prime - \dot{u}_{tr})}{r} + \frac{(u_{tt} - u_{rr})}{r^3} + \frac{f u_{rr}}{r^2} - \frac{2 f^2 u_{\theta\theta}}{r^2} = 0 .  \label{Xr-def}
\end{eqnarray}
To obtain the dynamical equations with gauge constraint damping of the form given in Sec.~\ref{subsec:gcd}, one then takes the combinations Eq.~(\ref{monoeq00})$\,+\, 4\ZZ_t / r$, Eq.~(\ref{monoeq01})$\,+\, 2(\ZZ_t + \ZZ_r)/r$, Eq.~(\ref{monoeq11})$\,+\, 4 \ZZ_r / r$ and Eq.~(\ref{monoeq22}). 



\subsubsection{Mass-energy condition}

In Sec.~\ref{subsec:conservation} we obtained a relation between the particle's energy $\mu{\cal E}[=-Q(X^{\alpha}_{(t)})]$ and a closed 2-surface integral ${\cal F}$ over an antisymmetric tensor $F_{\alpha\beta}$ constructed from the MP, its derivatives, and the background timelike Killing vector $X^{\alpha}_{(t)}$ [Eq.\ (\ref{Fdef})]. We now specialize to a Schwrazschild background and to a circular orbit of radius $r=r_0$, choose our 2-surface to be a 2-sphere of $r$=const (and $t$=const), and define the ``energy'' functional $E[u_{\alp\beta};r] \equiv -{\cal F}(X^{\alpha}_{(t)},\partial \Sigma_r)$. We then have
\begin{eqnarray}\label{masscondition}
E[u_{\alp\beta}] =- r^2 \int F^{tr} d\Omega 
= \begin{cases} 
    \mu{\cal E}, & r> r_0, \\
    0, & r<r_0 . 
    \end{cases}  
\end{eqnarray}
An explicit expression for $E[u_{\alp\beta}]$ is obtained by substituting for $F^{tr}$ from Eq.~(\ref{Fdef}) with $X^{\alp} = \delta^{\alp}_t$, and noting that the surface integral picks out the monopole part of the MP:
\begin{eqnarray} \nonumber
 E [u_{\alp\beta}] &=&  -\tfrac{1}{4} r f^{-1} \left( u_{tt}^\prime - \dot{u}_{tr} - \frac{f}{r} u_{tt} - \frac{2 (u_{tt} - u_{rr})}{r^2} \right) ,  \\  &=& 
-\tfrac{1}{4} r f^{-1} \left( u_{tt}^\prime - u_{rr}^\prime - \frac{f (u_{tt} + u_{rr})}{r} - \frac{3(u_{tt} - u_{rr})}{r^2} + \frac{2 f^2 u_{\theta\theta}}{r}  \right) ,  \label{mass-energy}
\end{eqnarray}
where, in going from the first line to the second, we have used the Lorenz-gauge constraint (\ref{Xr-def}) to eliminate $\dot{u}_{tr}$. 
It is straightforward to use the vacuum version of Eqs.~(\ref{monoeq00})--(\ref{Xr-def}) to confirm explicitly that $\partial_t E [u_{\alp\beta}]=0=\partial_r E [u_{\alp\beta}]$, i.e., that the MP combination on the right-hand side of (\ref{mass-energy}) is indeed constant for $r\ne r_0$, as expected.

Eq.\ (\ref{masscondition}) gives a necessary condition for the MP $u_{\alpha\beta}$ to represent a physical monopole solution. The monopole piece of our $m=0$ numerical solution must satisfy this condition.



\subsubsection{Static homogeneous Lorenz-gauge solutions\label{subsec:mono-homo}}
Now let us consider the family of solutions to the homogeneous part of the monopole equations (\ref{monoeq00})--(\ref{Xr-def}). With the staticity condition $u_{tr}=0$, we have 3 independent second-order field equations [Eqs.\ (\ref{monoeq00}), (\ref{monoeq11}) and (\ref{monoeq22})], and a single nontrivial gauge condition [Eq.\ (\ref{Xr-def})]. Since the gauge condition (\ref{Xr-def}) gives $u_{\theta\theta}$ algebraically in terms of $u_{tt}$ and $u_{rr}$ (and the first derivative $u'_{rr}$), the system effectively reduces to a set of two second-order ordinary differential equations. Hence we expect the complete basis of homogeneous solutions to be 4-dimensional. Below we will construct a complete basis of four homogeneous solutions. We will characterize each basis solution by its mass content, by whether it is traceless or tracefull, and by whether or not it is regular at the future horizon $\Hplus$ and at spatial infinity. Of our four solutions, only one will possess a non-zero mass-energy; the rest will be pure-gauge solutions.
For notational simplicity, in the rest of this section we usually adopt the convention $M=1$.

{\it Solution A:---} Our first solution, naturally called the ``conformal'' solution, is given by 
\beq \label{SolA}
h_{\mu \nu}^{(A)} = \mu g_{\mu \nu}.
\eeq 
This is clearly a valid Lorenz-gauge solution because the covariant derivative of $g_{\mu \nu}$ is zero (hence ${g_{\mu \nu}}^{;\nu} = 0$ and also $\Box g_{\mu \nu} = 0$), and the background is Ricci-flat (hence $2 {{{R^\lambda}_\mu}^\sigma}_\nu g_{\lambda \sigma} = 2 R_{\mu \nu} = 0$). This solution has a (constant) nonzero trace ($h^{(A)} = 4$), and a nonzero mass energy $E[u_{\alp\bet}]$. It is regular at the horizon [in the sense of Eqs.~(\ref{bc-h00})--(\ref{bc-h11})], but note that it does not fall off to zero at infinity (but approaches a constant value there). 

{\it Solutions B and C:---} Next we consider the ``scalar'' pure-gauge Lorenz-gauge solutions, given by $h_{\alp \bet} = \xi_{\alp ; \bet} + \xi_{\bet ; \alp}$, where $\xi_\alp = \Phi_{;\alp}$ for some scalar field $\Phi$. Note the MP trace is $h=2\Box\Phi$. The Lorenz-gauge condition, together with Ricci-flatness, implies that
\beq \label{consttrace}
h_{, \alp} = 2\left( \Box \Phi \right)_{, \alp}= 2\Box \xi_{\alp} = 0,
\eeq
hence such solutions have a constant trace. The ansatz $\Phi = \Phi(r)$ leads to two independent scalar pure-gauge solutions given by
\beq
 \xi^{(B)}_{r} = \mu (r^3 - 8) / (f r^2) , 
\eeq
and
\beq
 \xi^{(C)}_{r} =  \mu/(f r^2)
\eeq
(with all other components vanishing),
of which the first is tracefull ($h^{(B)}=6$) and the second traceless ($h^{(C)}=0$).
The corresponding MPs are given explicitly in Eqs.\ (\ref{mono-hB}) and (\ref{mono-hC}) below. These solutions are massless. We will consider their regularity properties in the next subsection.

{\it Solution D:---} Finally, let us consider a more general pure-gauge homogeneous solution, derived from a gauge displacement 
\beq
\xi^\alp_{(D)} = \mu [t , \xi_{(D)}^{r}(r), 0, 0 ] ,
\eeq
which is (up to a multiplicative constant) the most general form of $\xi^\alp$ consistent with a static MP. The Lorenz-gauge condition gives an inhomogeneous second-order ordinary differential equation for $\xi_{(D)}^{r}(r)$. The above scalar modes $\xi^{(B)}_{r}$ and $\xi^{(C)}_{r}$ are two independent solutions of the homogeneous part of this equation, and as a particular solution to the inhomegenous equation we take
\beq
\xi_{(D)}^{r} \equiv  -\frac{\mu }{3r^2} \left[ (r^2 + r + 4) r - (r^3-8) \ln f + 8 \ln r  \right] .
\eeq
The corresponding MP in given in Eq.\ (\ref{mono-hD}) below. This solution is tracefull ($h^{(D)}=2\ln f$) and massless. We will consider its regularity properties in the next subsection.

To summarize, we have constructed a complete basis of independent homogeneous solutions A--D to the static monopole equations. Writing $\vec{h} = \{ h, E, h_{tt}, h_{rr}, r^{-2} h_{\theta\theta} = (r \sin\theta)^{-2} h_{\phi\phi} \}$ for any of the solutions A--D, we have, explicitly,
\begin{eqnarray}
\vec{h}_{A} &=& \mu \{ 4, 1/2 ,  -f, f^{-1}, 1 \} ,  \label{mono-hA} \\
\vec{h}_{B} &=& \mu \{ 6, 0, -2 f P(r) /r^3, 2 f^{-1} Q(r)/r^3, 2 f P(r) / r^2 \} ,  \label{mono-hB} \\
\vec{h}_{C} &=& \mu \{ 0, 0, -2/r^4, -2(2r - 3)f^{-2}/r^4, 2 / r^3 \} , \label{mono-hC}  \\
\vec{h}_{D} &=& \mu \{ 2 \ln f, 0, -\tfrac{2}{3r^4}[rW(r) + r f P(r)\ln f - 8 \ln r] , \nn \\ 
 && \quad \tfrac{2}{3f^2 r^4} [ -r K(r) + L(r) \ln f + 8(2r-3)\ln r ] , \nn \\
 && \quad\tfrac{2}{3 r^3} [ -r (P(r) - r) + (r^3 - 8)\ln f - 8 \ln r ] \} . \label{mono-hD} 
\end{eqnarray}
In these expressions
\begin{eqnarray}
P(r) &=&  r^2 + 2r + 4  , \\
Q(r) &=& r^3 - r^2 - 2r + 12 , \\
W(r) &=& 3r^3 - 7r^2 - r - 4 , \\
K(r) &=& r^3 - 5r^2 - 5r + 12 , \\
L(r) &=& r^4 - 3r^3 + 16r - 24 .
\end{eqnarray}

\subsubsection{Static inhomogeneous Lorenz-gauge solution\label{subsec:static-mono}} 

We now attempt to construct a unique physical monopole solution, fulfilling the following list of requirements. 
(i) It is a solution to the inhomogeneous equations (\ref{monoeq00})--(\ref{monoeq22}).
(ii) It is a Lorenz-gauge solutions, i.e., Eqs.\ (\ref{Xt-def}) and (\ref{Xr-def}) are satisfied.
(iii) The MP is static in the sense that $\partial_t h_{\alp \bet} = 0$ \emph{and} $h_{ti} = 0$ for $i=r,\theta,\phi$.
(iv) The MP is continuous across $r=r_0$.
(v) The mass-energy condition (\ref{masscondition}) is satisfied.
(vi) The MP is regular across $\Hplus$ (Sec.~\ref{subsec:bc}).
(vii) The MP is regular at $r\to\infty$, in the sense that $h_{\mu\nu}/\hat\gamma_{\mu\nu}$ (no summation) falls off to zero at least as $\sim 1/r$. We will find (in confirmation of \cite{Barack:Lousto:2005,Barack:Sago:2007, Sago:Barack:Detweiler:2008}) that it is not possible to construct a solution that meets all such requirements. 

Let us first consider horizon regularity. It is easy to see, using the the conditions of Eqs.~(\ref{bc-h00})--(\ref{bc-h11}), that solutions A and B are regular at the horizon, whereas C and D are not. However, the conformal solution A has mass-energy (and it is the only solution with this property), so by above requirement (v) is it not allowed within the interior of the orbit. Thus we may immediately write
\beq
\vec{h}_{\text{mono}}^{\text{Lor}} = \begin{cases} 
\vec{h}_{\text{int}}  = a_{B} \vec{h}_{B} , & r < r_0 , \\
\vec{h}_{\text{ext}} = 2 \mathcal{E} \vec{h}_{A} + b_{B} \vec{h}_{B} + b_{C} \vec{h}_{C} + b_{D} \vec{h}_{D} , & r > r_0 , \end{cases}  \label{h-mono-static}
\eeq
with four coefficients $a_{B}$, $b_{B}$, $b_{C}$ and $b_{D}$  to be determined. Here the coefficient of $\vec{h}_{A}$ in the exterior is determined by the mass condition (\ref{masscondition}), recalling, again, that $\vec{h}_{A}$ is the only massfull solution, with $E=\mu/2$. 
The continuity requirement (iv) imposes 3 additional conditions, giving 
\begin{eqnarray}
a_{B} - b_{B} &=& -\frac{\alp}{3} \left[ (r_0 - 3)\ln f_0 - 4 r_0 f_0 \right]  \label{mono-aB} \\
b_{C} &=& -\frac{\alp}{3} \left[ r_0^2 + 20r_0 - 64 + 8(r_0 - 3) \ln r_0 \right] \label{mono-bC} \\
b_{D} &=& -\alp (r_0 - 3) , \label{mono-bD}
\end{eqnarray}
where $f_0 = 1 - 2 / r_0$, and
\beq
\alp \equiv \frac{\mathcal{E}}{r_0 f_0} . 
\eeq

Now let us consider the asymptotic behaviour at $r\to\infty$. We find
\begin{eqnarray} \label{htt}
h_{tt} &\sim&
 - 2\mu \alp + \mathcal{O}(r^{-1}), \\
h_{rr} , \, r^{-2}h_{\theta\theta} , \, (r\sin\theta)^{-2} h_{\phi\phi}  &\sim& \frac{2 \mu\alp (4r_0 - 9)}{3} + 2 \mu b_B + \mathcal{O}(r^{-1}) .
\end{eqnarray}
The leading (constant) term in the spatial components ($rr$,$\theta\theta$,$\phi\phi$) can be removed by fixing the remaining degree of freedom to be 
\beq
b_B = -\alp (4r_0 - 9)/3, \quad \Rightarrow \quad a_B =  \alp [1 - (r_0-3) \ln f_0 ] / 3 . \label{mono-bB}
\eeq
However, the $tt$ component of the perturbation remains asymptotically nonzero, which cannot be remedied without violating at least one of the other requirements (i)-(vi).

It is straightforward to verify that the static solution given above is identical to the solution presented in Sec.~IIID of Ref.~\cite{Barack:Lousto:2005} (or the solution inferred from \cite{Detweiler:Poisson:2004}), as expected. Here, for the first time, we derived this solution from a complete basis of (static) solutions to the Lorenz-gauge  monopole equations. Having such a complete basis at hand is useful in general, because it allows the construction of a physical mass-perturbation solution also for non-circular orbits. [Indeed, the above complete basis was previously used by one of the authors, in \cite{Barack:Sago:2010}, to construct a physical monopole solution for generic bound (eccentric) orbits; however, Ref.\ \cite{Barack:Sago:2010} did not give the complete basis explicitly, as we do here.] Note also that our construction avoids the explicit imposition of jump conditions for the MP derivatives at $r_0$, but it can be checked that these are satisfied. In our construction we have replaced the jump conditions with a ``mass condition'', with the foresight that in the Kerr problem one can still impose the latter, but one cannot impose explicit monopole junction conditions (because the ``monopole'' piece cannot be separated out in Kerr). Indeed, many of the details of the above analysis are transferable to the Kerr case, as we shall show in paper IV.

To obtain an asymptotically regular mass-perturbation solution, it is most natural to relax the Lorenz-gauge condition.
This approach was taken in Ref.~\cite{Sago:Barack:Detweiler:2008}, via the introduction of the simple non-Lorenz gauge transformation 
\beq
\delta h^{\text{NL}}_{\mu \nu} = \xi^{\text{NL}}_{\mu ; \nu} +  \xi^{\text{NL}}_{\nu ; \mu} ,
\eeq
with $\xi_{\text{NL}}^\mu = -\mu\alp t \delta_{t}^{\mu}$, so that
\beq
\delta h^{\text{NL}}_{tt} = 2 \mu\alp f ,
\eeq
and the other components are zero. Recalling Eq.\ (\ref{htt}), we see that the new perturbation $h_{\mu\nu}+\delta h^{\text{NL}}_{tt}$ is $\mathcal{O}(r^{-1})$ at $r\to\infty$ as desired. 
We note, however, that this new perturbation now {\it fails to be regular at $\Hplus$}: From Eq.\ (\ref{bc-h11}) we see that its Eddington--Finkelstein RR component diverges there as $\sim 2\mu\alpha f^{-1}$. 
An alternative choice, $\tilde \xi_{\text{NL}}^\mu = -\mu\alp (t + r_\ast - r) \delta_{t}^{\mu}$, leads to
\beq
\widetilde{\delta h}^{\text{NL}}_{tt} = 2\mu \alp f , \quad \quad \widetilde{\delta h}^{\text{NL}}_{tr} = 2\mu \alp M/ r ,
\eeq
with all other components unchanged. This MP \emph{is} regular at $\Hplus$, but it is not static in the sense of condition (v). We have not been able to find a simple transformation away from Lorenz gauge leading to a monopole solution that is static, continuous, globally regular, and has the correct mass-energy content.

Staying within the Lorenz gauge, it is in fact possible to write down a Lorenz-gauge monopole solution that is static, continuous and globally regular -- but has the wrong mass-energy $E$. This solution, first obtained by Berndtson \cite{Berndtson:2007}, is given by
\beq
\hat{\vec{h}}^{\text{Lor}}_{\text{mono}} = \vec{h}_{\text{mono}}^{\text{Lor}} - 2 \alp \vec{h}_{A} + \alp \vec{h}_{B} .  \label{incorrect-correct}
\eeq
While this solution is nonphysical [because it corresponds to a mass perturbation from a nongeodesic particle with energy $\mu \mathcal{E} - 2 \mu\alp = \mu \mathcal{E} (r_0 - 3M)/(r_0-2M)$], it is a useful solution nonetheless, because it can arise in time-domain simulations (which usually seek globally regular solutions). If we identify this nonphysical solution in the numerical data, we can easily ``correct'' the mass-energy by applying Eq.~(\ref{incorrect-correct}) in reverse. We will come back to this point later. For future reference, we shall refer to $\hat{\vec{h}}^{\text{Lor}}_{\text{mono}}$ as the ``asymptotically-flat'' Lorenz-gauge solution.

\subsubsection{Nonstatic and nonstationary (linearly growing) monopole solutions\label{subsec:nonstatic}}

In Sec.~\ref{subsec:m01-numerical} we saw that time-domain evolutions of the Lorenz-gauge monopole equations can excite pure-gauge modes that grow linearly with time $t$. In what follows we give an analytic  description of such modes. For notational brevity we set $\mu=1=M$ throughout this subsection.

First, observe that if the above staticity condition (v) is relaxed, then the monopole MP is no longer unique, even if all other conditions are satisfied. Consider, for example, the Lorenz-gauge scalar perturbation $\Phi^{\rm nonstat} = \tfrac12 t + \ln f$, for which $\xi^{\rm nonstat}_{\mu} = \Phi^{\rm nonstat}_{;\mu} = \left[\tfrac12, 2/(r^2 f), 0, 0\right]$. The nonzero components of the corresponding MP are
\begin{eqnarray}
h^{\rm nonstat}_{tt} &=& -\frac{4}{r^4},  \nonumber \\
h^{\rm nonstat}_{tr} &=& -\frac{1}{r^2 f},  \nonumber \\
h^{\rm nonstat}_{rr} &=& - \frac{4 (2r - 3)}{r^4 f^2} , \nonumber \\
h^{\rm nonstat}_{\theta\theta}&=&h_{\phi\phi}/\sin^2\theta = \frac{4}{r} .
\label{hnonstat}
\end{eqnarray}
This homogeneous, Lorenz-gauge MP is regular on $\Hplus$ according to criteria (\ref{bc-h00})--(\ref{bc-h11}), and also asymptotically flat. It is stationary in the sense that $\partial_t h^{\rm nonstat}_{\mu \nu} = 0$, but it is {\em nonstatic} since $h_{tr} \neq 0$. This MP is  traceless, $h^{\rm nonstat} = 0$. As we demonstrate below in Fig.\ \ref{fig:linear}, this nonstatic mode features in our 2+1D numerical $m=0$ solutions.

We seek a solution that satisfies all criteria (i)--(vii), except staticity, and for which the metric components grow linearly in time. We are guided by the observation that our linear-in-$t$ numerical solutions seemed to have perfectly stationary ($t$-independent) mass-energy $E$ and trace $h$. This strongly suggests that the mode implicated in the linear growth is pure-gauge and of a constant trace (or traceless). We will now construct analytically such a linearly-growing Lorenz gauge mode, which is globally regular (at any finite time). 

Let us first consider the scalar gauge mode $\Phi = \tfrac12 t \ln f$, which is a simple monopole solution of $\Box \Phi = 0$, and hence it generates a traceless Lorenz-gauge solution. This leads to $h_{\alp \alp} = t \times h_{\alp \alp}^{C}$ (no sum), where $h_{\alp \alp}^{C}$ is the homogeneous solution C given in Eq.~(\ref{mono-hC}), and $h_{tr} = (2 - \ln f)/(r^2 f)$. The diagonal components of this solution grow linearly in time, whereas the $tr$ component is stationary. Now consider a second scalar mode, constructed from $\Phi = t^2 + R(r)$ where $R(r)$ is chosen so that $\Box \Phi = \text{const}$. This generates a MP of a constant trace [recall Eq.\ (\ref{consttrace})], which has stationary diagonal components, and a $tr$ component that grows linearly in $t$. By taking a certain linear combination of these two modes we construct the gauge vector $\xi^{\rm lin}_{\alpha}=(\xi_{t}^{\rm lin},\xi_{r}^{\rm lin},0,0)$, with 
\beq
\xi_{t}^{\rm lin} = \ln (2 f) + t / 2 + \frac{13}{6}, \quad \quad
\xi_{r}^{\rm lin} = \frac{2t}{r^2 f} + \frac{r^3 + 3r^2 + 12r + 24 \ln(fr)}{6 r^2 f} - \frac{r}{6f} .
\eeq
This  generates a homogeneous Lorenz-gauge monopole MP $h^{\rm lin}_{\alp\beta}$, whose nonzero components read 
\begin{eqnarray}
h^{\rm lin}_{tt} &=& -\frac{-r^4 + 4t + r^2 + 4 r + 8\ln(rf)}{r^4}, \nonumber \\
h^{\rm lin}_{tr} &=& -\frac{t + \tfrac13 + 2\ln(2 f)}{r^2 f}, \nonumber \\
h^{\rm lin}_{rr} &=& -\frac{4t (2r - 3) + 5 r^2 - 12 r + 8 (2 r - 3) \ln(r f)}{r^4 f^2}, \nonumber \\
r^{-2} h^{\rm lin}_{\theta\theta} &=& \frac{4 t + r^2 + 4 r + 8 \ln(r f)}{r^3} = (r \sin \theta)^{-2} h_{\phi\phi}. \label{hlin}
\end{eqnarray}
We note that the origin of time is essentially arbitrary here, so $t \rightarrow t + c$ also yields a solution, which is equivalent to adding a multiple of the stationary traceless solution (\ref{hnonstat}). 

It can be checked that the pure-gauge solution (\ref{hlin}) is regular at $\Hplus$ according to the criteria of Sec.~\ref{subsec:bc}.
It is also regular at $r\to\infty$ in all components except $tt$, for which $h_{tt}^{\rm lin} \sim 1 + \mathcal{O}(r^{-1})$. It is a pure-gauge solution with zero mass-energy, $E = 0$, and a constant trace $h = 1$. By combining $h^{\rm lin}_{\alp\beta}$ with the inhomogeneous static Lorenz-gauge solution ${h}_{\text{mono}}^{\text{Lor}} $ of Sec.~\ref{subsec:static-mono}, it is possible to construct a Lorenz-gauge monopole solution with the correct mass-energy, which is both horizon-regular and asymptotically flat. The metric components of this solution grow linearly with $t$, but its trace is stationary. These features are all in common with those of the linearly-growing mode observed in our $m=0$ numerical simulations.  
Hence, the solution $h^{\rm lin}_{\alp\beta}$ a good candidate for being the one implicated in the linear instability. 

The above suspicion is indeed confirmed by examining the numerical data. An example is shown in Fig.\ \ref{fig:linear}. In Sec.\ \ref{sec:stabilization} we shall take advantage of the analytic insight thus obtained into the origin of the linear instability, in order to find a cure to it. 

\begin{figure} 
 \begin{center}
\includegraphics[width=8cm]{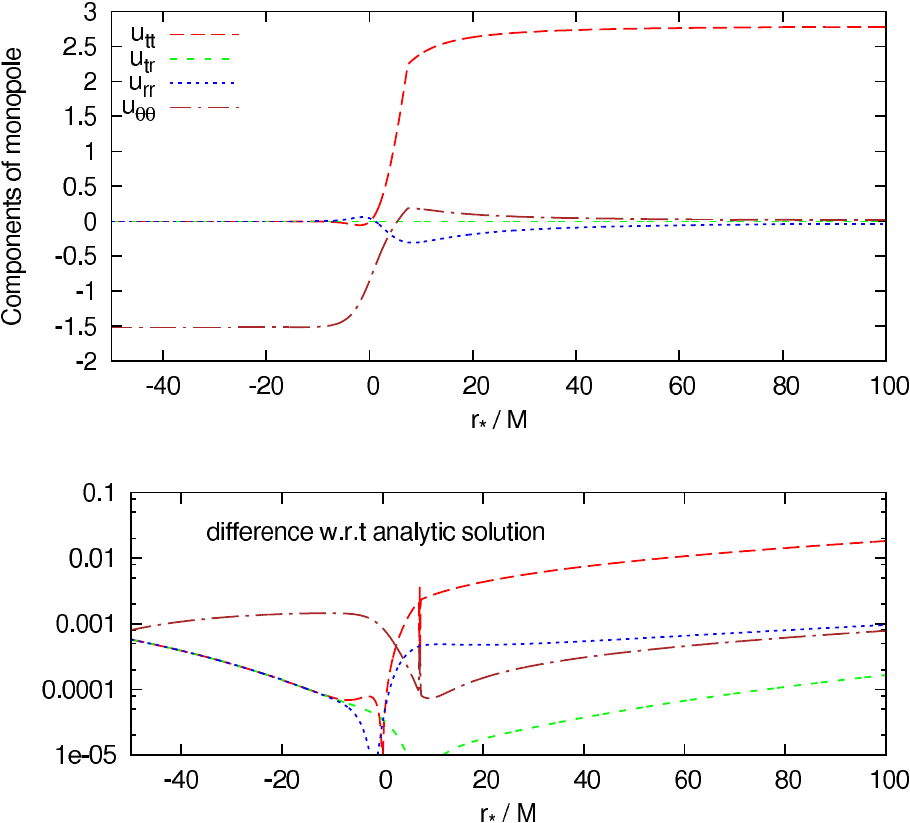}
\includegraphics[width=8cm]{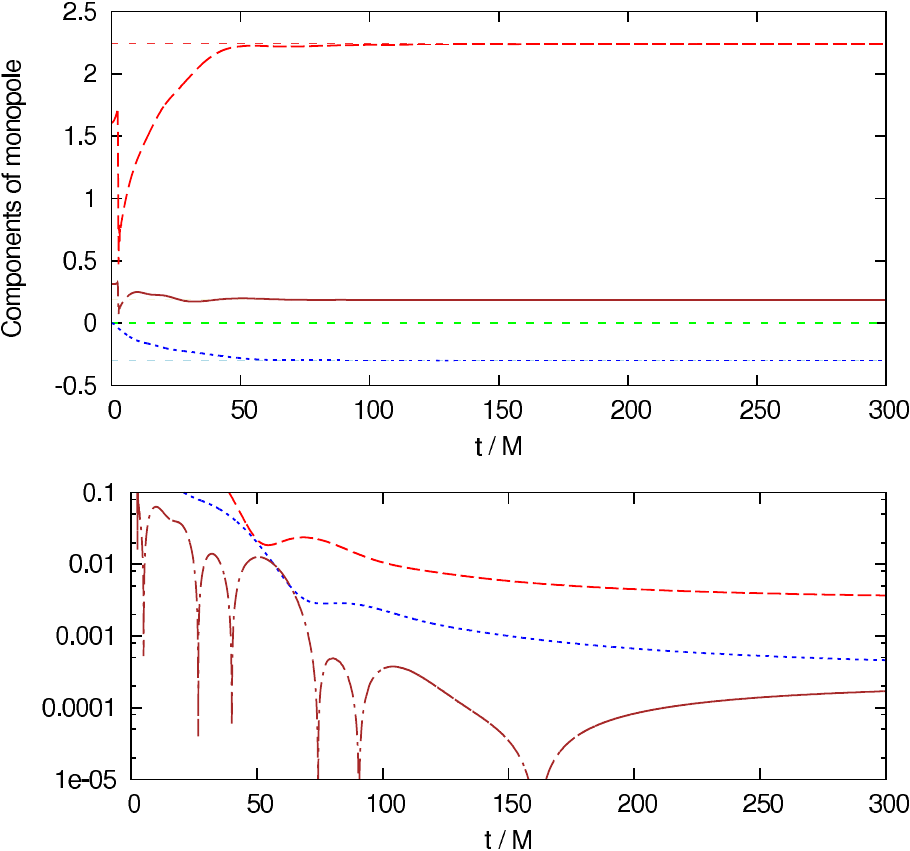}
 \end{center} 
 \caption
 {
Removal of the linear-in-t gauge mode from the m = 0 numerical data. The implicated mode is a certain homogeneous Lorenz-gauge mode of a monopolar angular dependence, which we identified analytically in Eq.~(\ref{hlin}). Here we remove the problematic mode (for the case shown in Fig.~\ref{fig:m0-mode}, with $r_0=6M$) by subtracting a suitable amplitude of this analytic solution. We fix the amplitude (and ``origin of time'') of the gauge mode to be subtracted by demanding that $h_{tr} = \dot{h}_{tr} = 0$ on the worldline (after subtraction). Above, the upper plots show the monopole mode for $r_0 = 6M$, obtained from $m=0$ data by projection [via Eqs.~(\ref{hilm-eq1})--(\ref{hilm-eq10})], \emph{after} the removal is attempted. The left plots show the radial profile, and the right plots show the solution on the worldline as function of time. The data shows that the ``cleaned'' monopole solution is both stationary ($\partial_t u_{\alpha \beta}$) \emph{and} static ($u_{tr} = 0$), up to numerical error. The lower plots show the difference between the numerical data and the analytic monopole solution $\hat{\vec{h}}^{\text{Lor}}_{\text{mono}}$ [Eq.~(\ref{incorrect-correct})] obtained by Berndtson \cite{Berndtson:2007} which, as argued in Sec.~\ref{subsec:static-mono}, is asymptotically-regular but has incorrect mass-energy. The difference between the numerical and analytical solutions is small, and diminishes as grid resolution is improved. Nevertheless, the numerical accuracy remains unsatisfactory and so in Sec.~\ref{sec:stabilization} we develop alternative methods to obtain stable evolutions (Fig.~\ref{fig:m0-stable}).
 }
 \label{fig:linear}
\end{figure}

\subsection{Dipole perturbation}

A particle in a circular equatorial orbit generates a dipolar $l=1$ MP with two parts: an odd-parity perturbation in the axisymmetric $l=1, m=0$ mode, and an even-parity perturbation in  the $l=1$, $m=\pm 1$ modes. The former part is attributed to the angular momentum perturbation due to the particle, and the latter is associated with the recoil motion of the large black hole at $O(\mu)$. We consider each of these two parts in turn.

\subsubsection{Odd-parity mode ($l=1, m=0$)}

The odd-parity axisymmetric Lorenz-gauge dipole perturbation is governed by a single dynamical equation, with two independent static homogeneous vacuum solutions given (up to multiplicative constants) by \cite{Barack:Lousto:2005} $h_{t \phi} = \mu r^2 \sin^2 \theta$ and   $h_{t \phi}=-2\mu  r^{-1} \sin^2 \theta$ (other components vanish).
The former solution is pure-gauge, and is generated by the vector $\xi^{a} = [0,0,0,t]$ (which satisfies the Lorenz gauge condition, $\Box \xi^a = 0$). The latter solution contains ``angular momentum'' $L[u_{\alp\beta};r] \equiv -{\cal F}(X^{\alpha}_{(\phi)},\partial \Sigma_r)=1$. Note that the latter solution may be obtained through a direct linear variation of the Kerr metric with respect to the background angular momentum $aM$ (with fixed Boyer-Lindquist coordinates), at the limit $a\to 0$. (This procedure yields a Lorenz-gauge perturbation only in the $a\to 0$ limit.) 

Following the argument of Sec.~\ref{subsec:conservation}, the physical solution should have an angular momentum of $L=\mu \mathcal{L}$ for $r>r_0$, and of $L=0$ for $r<r_0$. Therefore,
\beq
h_{t \phi}^{(l=1,m=0)} = \begin{cases}
-2 \mu \mathcal{L} (r^2 / r_0^3) \sin^2 \theta, & r < r_0,  \\
-2 \mu \mathcal{L} r^{-1} \sin^2 \theta, & r \ge r_0.
\end{cases}  \label{m0l1-soln}
\eeq
This matches the solution originally obtained by Zerilli \cite{Zerilli:1970} (and reproduced in Refs.~\cite{Barack:Lousto:2005,Detweiler:Poisson:2004}).

Within the $m$-mode scheme in 2+1D we do not decompose into $l$-modes, and we start with trivial initial data. Then it is not guaranteed, a priori, that the numerical solution would evolve towards one that has the correct angular momentum content.  However, by computing the surface integral $L[u_{\alp\beta};r]$ we can ``measure'' the angular momentum in the numerical $m=0$ solution, and, if necessary, simply add a multiple of the  homogeneous solution $h_{t \phi}=-2 r^{-1} \sin^2 \theta$ (which is regular on $\Hplus$), in order to ``adjust'' the angular momentum. In practice, we indeed find in our implementation that the $m=0$ numerical solution does not carry the correct angular momentum, and we apply the above simple cure.  

The odd-parity dipole does not contribute to the linear instability observed in our $m=0$ solutions. This is clear from the fact that the unstable mode has a monopolar ($\theta$-independent) angular profile.

\subsubsection{Even-parity modes ($l=1, m=\pm 1$)\label{sec:l1m1}}



The time-domain simulations of Sec.~\ref{subsec:m01-numerical} (see Fig.~\ref{fig:m01-t}) suggest that the $l=1$, $m=\pm1$ even-parity modes are susceptible to a linear gauge instability. Let us explore the origin of this instability. 
As in Sec.~\ref{subsec:nonstatic}, let us begin by considering the static traceless scalar (Lorenz-)gauge mode $\xi_{\alpha} = \Phi_{,\alpha}$, with $\Box \Phi = 0$. This admits separable $l=1$, $m=\pm 1$ static solutions of the form $\Phi_{\pm} = \psi(r)\mathcal{Y}_{\pm}$, where $\mathcal{Y}_{\pm}\equiv \sin \theta e^{\pm i \phi}$ and hereafter the label $\pm$ corresponds to $m=\pm 1$. 
 Two independent radial solutions are 
\beq\label{psi12}
\psi_1 = r - M, \quad \quad \text{and} \quad \quad \psi_2 = 2M + (r-M)\ln f .
\eeq

\begin{figure} 
 \begin{center}
  \includegraphics[width=9cm]{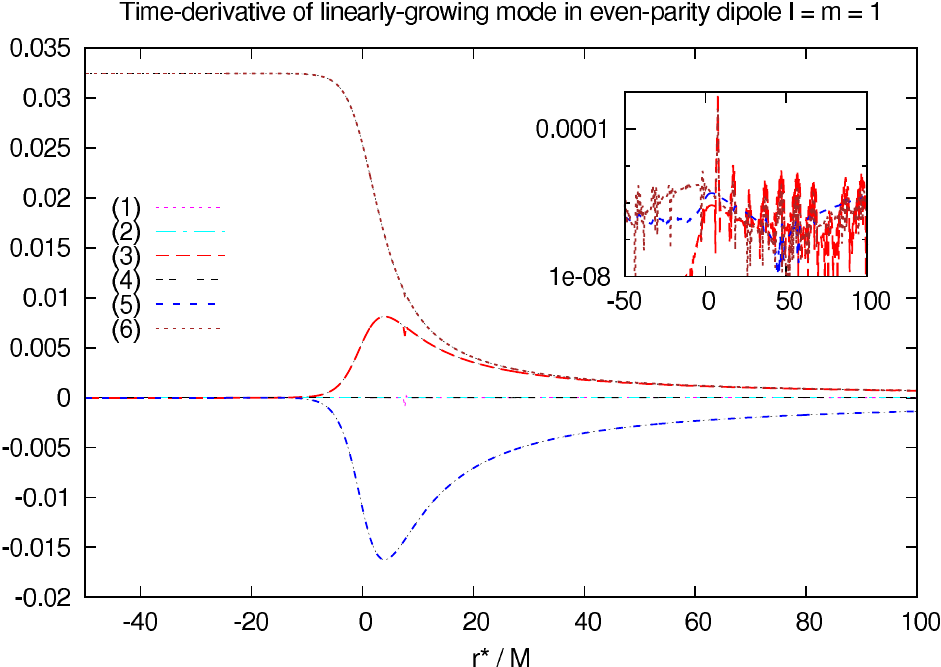}
 \end{center} 
 \caption{
 Numerical check on the relevance of the analytic solution for the linear-in-$t$ Lorenz-gauge mode in the even-parity dipole sector. In Sec.~\ref{sec:l1m1} we asserted that the numerical solution obtained in the $m=1$ case is a superposition of the physical solution, with time dependence $\exp(-i \Omega t)$, and a pure Lorenz-gauge dipole ($l=1$) mode, with linear-in-$t$ behaviour [Eq.~(\ref{culprit})]. 
 The linear growth in the numerical solution may be estimated by applying a simple filter. The plot above shows $\text{Re}\left( [1 - i \Omega^{-1} \partial_t ] \partial_t \hh^{(i)}_{11} \right)$ for $i = 1\ldots6$, where $\hh^{(i)}_{lm}$ denotes the radial variables used in Ref.~\cite{Barack:Lousto:2005}, and $r_0 = 6M$. The analytic solution (\ref{culprit}) has a very simple linear-in-$t$ profile given by $2 \dot{h}^{(3)} = - \dot{h}^{(5)} = 2 f \dot{h}^{(6)} \propto f / r$ with $\dot{h}^{(1)} = \dot{h}^{(2)} = \dot{h}^{(4)} = \dot{h}^{(7)} = 0$ where here $\dot{h}^{(i)} \equiv \partial_t \hh^{(i)}_{11}$. The plot above shows that the time-derivative of the linearly-growing part of the numerical solution is in excellent agreement with the analytic solution (with the difference between the two shown in the inset).
}
 \label{fig:linear_dipole}
\end{figure}

Now let us consider a solution of the form $\Phi_{\pm} =  t \psi_1 \mathcal{Y}_{\pm}$. This is also a traceless scalar Lorenz-gauge solution (satisfying $\Box \Phi = 0$), and it gives rise to a linear-in-$t$ perturbation. This perturbation is not regular on $\Hplus$, and nor is it regular as $r\to \infty$. However, we can construct a globally regular, time-growing solution by combining it with the static scalar mode $\psi_2\mathcal{Y}_{\pm}$, and also with an ``electric''-type gauge mode $\xi_\mu\propto \delta^t_{\mu}\mathcal{Y}_{\pm}$ (we use here the language of Ref.~\cite{Ori:2004}). The result is
\beq \label{culprit}
\xi_{\mu}^{\pm} = -2 (r-2) \delta_\mu^t \mathcal{Y}_{\pm} + \Phi^{\pm}_{,\mu}, \quad \quad \text{where} \quad \quad \Phi^{\pm} = \left[t (r-1) + 2\left(2+(r-1)\ln f\right) \right] \mathcal{Y}_{\pm} .
\eeq 
This gauge vector gives rise to a Lorenz-gauge MP which is globally regular and whose components $tt$, $rr$, $r\theta$, $r\phi$, $\theta\theta$ and $\phi\phi$ grow linearly in time, although its trace remains static, $h = 0$. It therefore has all the characteristics observed in the linearly-growing mode of the numerical $m=1$ solution. In Fig.~\ref{fig:linear_dipole} we show that the linearly-growing part of the numerical data (obtained by applying a filter) is indeed that given by the analytic solution in Eq.\ (\ref{culprit}). 

Note that, since the problematic growing gauge mode satisfies physical boundary conditions at any finite time, it cannot be eliminated using boundary conditions alone. Regularity at $\Hplus$ also means that schemes employing alternative time slicing (e.g., hyperboloidal \cite{Zenginoglu:2008, Zenginoglu:2011} or double-null constructions) would not eliminate such instabilities,  although these routes remain to be explored.

\section{Low Multipoles: Generalized Lorenz gauge and stabilization\label{sec:stabilization}}

In Sec.~\ref{subsec:gauge-choice} we discussed the formulation of the linearized Einstein equations in a generalized Lorenz gauge (GLG), which is defined by a gauge driver function $H_{\mu}$ ($=0$ in the Lorenz-gauge case). In this section, we make use of the freedom to choose the gauge driver $H_{\mu}$ to seek a formulation which is \emph{stable}, in the sense that time-domain evolutions of generic initial data are free from linearly (or exponentially) growing modes, and which lead to stationary values for $\hat{u}_{\alp\bet}^{(m)}$ on the worldline. We then describe a method for recovering the Lorenz gauge solution from the numerical solution in the generalized gauge.

\subsection{The $m=0$ perturbation}

The gauge-mode instability seen in numerical results for $m=0,1$ (Sec.~\ref{subsec:m01-numerical}) arises even in vacuum, for generic (non-Lorenz-gauge) initial data. In the previous section, we argued that the instability is entirely in the monopole and even-parity dipole sectors. We may therefore use the simpler framework of the vacuum 1+1D monopole equations [formulated in Eqs.~(\ref{monoeq00})--(\ref{Xr-def}), and with suitable GLG corrections] to develop and test ideas for stabilizing the evolution. Once this is achieved, we will implement the same ideas in our $m$-mode scheme.



First, we have tried out various choices of a generalized Lorenz gauge driver $H_{\mu}$, with the aim of eliminating the linear-in-$t$ modes without exciting other (e.g., exponentially-growing) instabilities. The goal was to obtain a stationary monopole MP that is regular at $\Hplus$ and approaches $H_{\mu}=0$ (i.e., Lorenz gauge) at late time. After some experimentation, we focused on the following general class of gauge drivers, which gave good results:
\beq
H_{\mu} =  \chi \, n_\mu , \quad \quad \text{where} \quad \chi \equiv \frac{A}{r^n} U . \label{H-choice}
\eeq
Here $n_{\mu} = [1, f^{-1}, 0, 0]$ is an ingoing null vector, $U$ is chosen to be one of the metric components $U \in \{ u^{(m=0)}_{tt}, u^{(m=0)}_{tr}, u^{(m=0)}_{rr} \}$, $A$ is a constant, and $n$ is a positive integer. We settled on the specific choice $U = u_{tr}^{(m=0)}$, $A = -1/4$ and $n=3$. The choice $U = u_{tr}^{(m=0)}$ guarantees that, if the eventual monopole solution is static (i.e., $u_{tr}^{(m=0)} = 0$ up to numerical error), then the solution will also be in Lorenz gauge, $Z_{\mu} = H_{\mu} = 0$.

With the above choice of a GLG, we established experimentally that the {\em sourced} 1+1D monopole equations also evolve stably, and that the solution is regular at the horizon. In the next step we specified as initial data the analytic solution $\hat{\vec{h}}^{\text{Lor}}_{\text{mono}}$ given in Eq.~(\ref{incorrect-correct}), which, recall, is a globally-regular Lorenz-gauge monopole (which, however, does not have the correct mass). With these initial data, we found that, as the numerical resolution increases, the system approaches staticity, i.e., $\partial_t u_{\alp \bet} \rightarrow 0$ and $u_{tr} \rightarrow 0$. As discussed in Sec.~\ref{subsec:static-mono}, it is then easy to ``correct the mass'' of the monopole solution, by applying the the analytic transformation (\ref{incorrect-correct}) in reverse. 

With the above preliminary tests passed, the next challenge was to implement the GLG approach in the 2+1D setting. Moving to a GLG leads to modifications of the constraint-damped Z4 system, Eq.~(\ref{ABC-eq}). Specifically, a new term $\mathcal{\hat{B}}_{\alp \bet}$ is introduced [see Eq.~(\ref{B-eq-alt})], the constraint damping terms $\mathcal{C}_{\alp \bet}$ are changed, and the effective source may be modified. However, with the simple choice of $H_{\alp} \propto u_{tr}$ above, the required modifications are rather simple, because the $tr$ component of the $m=0$ mode of the puncture function happens to vanish (for circular orbits). This implies that the effective source is not modified, and we only have to include extra terms on the left-hand side of the evolution equations. The extra terms are generated by $\mathcal{\hat{B}}_{\alp \bet}$ in Eq.~(\ref{B-eq-alt}), and $\mathcal{C}_{\alp \bet} \rightarrow \mathcal{C}_{\alp\bet} - \kappa (n_\alp H_\bet + n_\bet H_\alp)$.  At the level of the $m$-mode equations we evolve Eq.~(\ref{u-eq}) with $\mathcal{\tilde{M}}^{(m=0)}_{\alp \bet} \rightarrow \mathcal{\tilde{M}}^{(m=0)}_{\alp \bet} + \Delta \mathcal{\tilde{M}}^{(m=0)}_{\alp \bet}$, where 
\begin{eqnarray}
\Delta \mathcal{\tilde{M}}^{(m=0)}_{tt} &=& - f r \left(\dot{\chi} + \chi^\prime + \frac{2 \chi}{r}\left( 1 - 3 / r \right) \right) + C, \\
\Delta \mathcal{\tilde{M}}^{(m=0)}_{tr} &=& - f r \left(\dot{\chi} + \chi^\prime - \frac{2 \chi}{r^2} \right) + C, \\
\Delta \mathcal{\tilde{M}}^{(m=0)}_{t\theta} &=&   \Delta \mathcal{\tilde{M}}^{(m=0)}_{r\theta}  = - f \partial_\theta \chi , \\
\Delta \mathcal{\tilde{M}}^{(m=0)}_{rr} &=& - f r \left(\dot{\chi} + \chi^\prime - \frac{2 \chi}{r} (1 - 1/r) \right) + C, \\
\Delta \mathcal{\tilde{M}}^{(m=0)}_{\theta\theta} &=& \Delta \mathcal{\tilde{M}}^{(m=0)}_{\phi\phi} = -r (\dot{\chi} - \chi^\prime)   ,
\end{eqnarray}
and $C = -4 f \chi / r$. All other components of $\Delta \mathcal{\tilde{M}}^{(m=0)}_{\alp \bet}$ are zero.

\begin{figure} 
 \begin{center}
  \includegraphics[width=8cm]{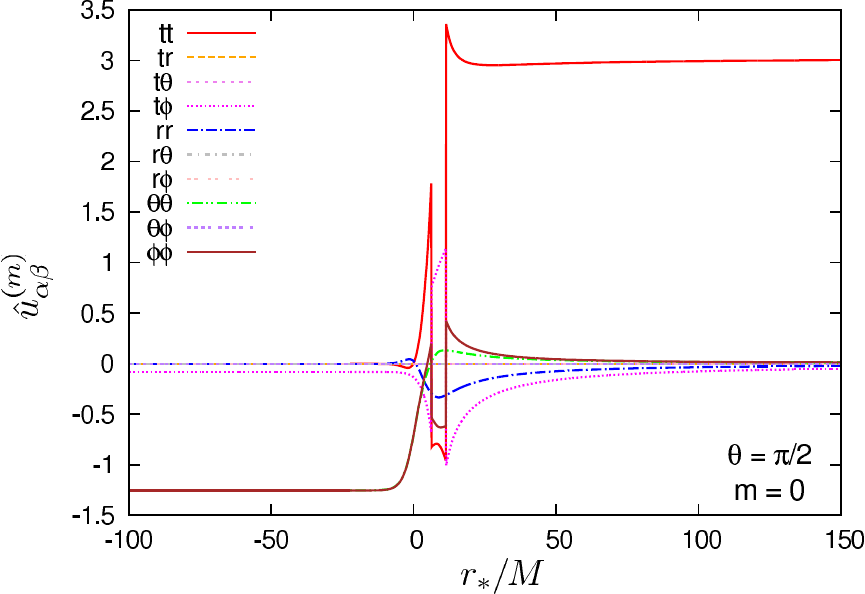}
  \includegraphics[width=8cm]{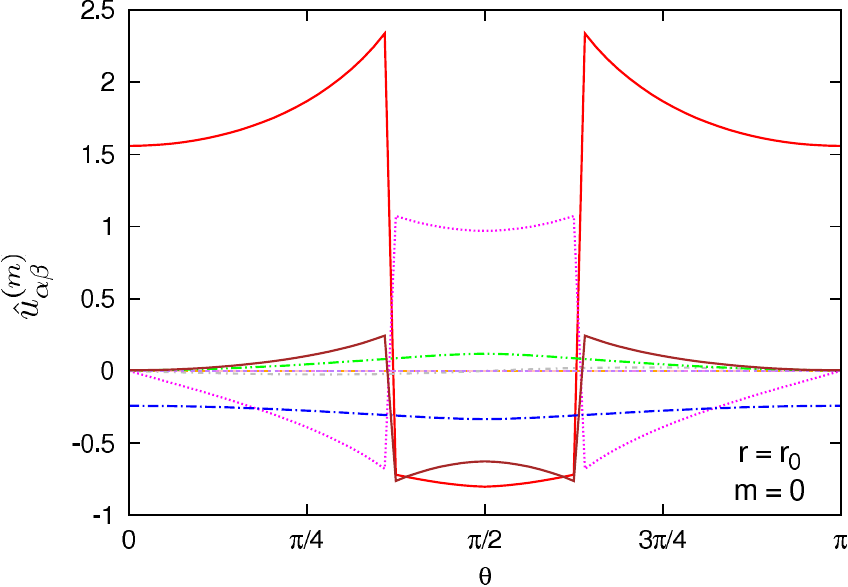}
 \end{center}
 \caption{Stable GLG evolution of the $m=0$ mode. These plots show numerical data (at a suitably late time $t_{\text{max}} = 250M$) for the $m=0$ mode from the ``stabilized'' evolution scheme using a generalized Lorenz gauge and analytic initial conditions for $l=0$ and $l=1$ modes (see text). The left plot shows the radial profile in the equatorial plane, and the right plot shows the angular profile at $r=r_0=7M$. In both plots, the central worldtube is visible as a trough, inside which we show $u_{\alp\bet}^{\mathcal{R}(m)}$ (rather than $u_{\alp\bet}^{(m)}$). Note that, since $u_{tr}$ vanishes with increasing resolution (see Fig.~\ref{fig:m0-u01} below), the data is both static and in Lorenz gauge (up to numerical error), as desired. The numerical solution is globally-regular but has the wrong mass-energy; this is easily rectified by adding a specific analytically-given monopole solution as described in the text.}
 \label{fig:m0-stable}
\end{figure}

\begin{figure} 
 \begin{center}
  \includegraphics[width=8cm]{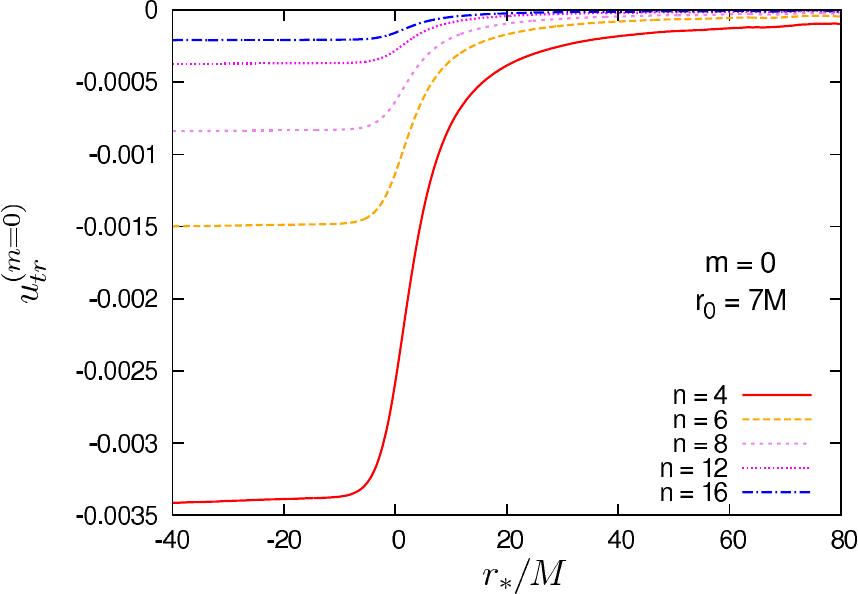}
 \end{center}
 \caption{Numerical data for the $tr$ component of the $m=0$ MP, for various grid resolutions $\Delr = M / n$. Referring to the evolution shown in Fig.\ \ref{fig:m0-stable}, the plot shows the residual non-staticity left over in the $m=0$ mode. It illustrates that, as the grid resolution is improved, the residual non-staticity and the deviation from Lorenz gauge both vanish: $H_{\alp} \propto u_{tr} \rightarrow 0$ as $n\to\infty$.}
 \label{fig:m0-u01}
\end{figure}

Figure \ref{fig:m0-stable} shows sample numerical results for the $m=0$ mode, using the scheme outlined above. The evolution starts from analytic initial data given by the combination of the monopole solution (\ref{incorrect-correct}) and the dipole solution (\ref{m0l1-soln}). 
Initially, the $tr$ component is zero everywhere, but, for a finite resolution, it evolves to a stationary solution, which is small but non-zero. Figure \ref{fig:m0-u01} shows that, as the grid resolution is improved, the stationary value of $u_{tr}$ converges to zero. Hence we are able to recover the static $m=0$ MP in Lorenz gauge, although with a monopole part $\hat{\vec{h}}^{\text{Lor}}_{\text{mono}}$ having the ``wrong mass''. This, however, is easily rectified by applying the the analytic transformation (\ref{incorrect-correct}) in reverse. 

The above demonstrates that, when appropriate initial data is known analytically, the correct $m=0$ Lorenz-gauge solution can be obtained through a time-domain evolution. But a natural question arises: how can we proceed in the Kerr case, where monopole initial data are not available in the Lorenz-gauge, and where the equations do not separate into 1+1D? In other words, if we are completely ignorant of the correct initial data, can we still devise a scheme that will evolve towards the static ($h_{tr} = 0$) Lorenz-gauge solution? 

We tested a simple but general approach. First, we evolved the sourced equations starting with trivial ($u_{\alp\bet} = 0$) initial data. Once a (nearly) stationary solution was reached, we read off the value ($y_1$) of the $u_{tr}$ component in the near-horizon limit ($r_\ast \ll 0$). Next, we evolved the vacuum equations again, this time with monopolar Gaussian initial data with amplitude $c$ (and some width), and read off a different value ($y_{2}$) of $u_{tr}$ in the same limit. Finally, exploiting the linearity of the equations, we  ``tuned'' the amplitude of the Gaussian initial data to $- c y_1 / (y_2 - y_1)$ (keeping the Gaussian width fixed), to obtain a stationary solution in which $u_{tr} = 0$ at the horizon. Rather than performing a third evolution, we can instead make an appropriate linear combination of the first and second stationary solutions. It turns out that the resulting solution is in fact static (up to numerical error), in the sense that $u_{tr} = 0$ \text{everywhere} and not just at the horizon. To see why, consider the $t$ component of the monopole part of the gauge constraint, which reads
\beq \label{gaugeEH}
u_{tr}^\prime - \dot{u}_{tt} + \hat f(r) u_{tr} = 0,
\eeq
where $\hat f(r)$ is a certain function whose specific form is unimportant here, except the feature that its value is bounded at the horizon. In the stationary limit we have $\dot{u}_{tt} = 0$, and it is easily seen that, with the boundary condition $u_{tr}=0$ at the horizon, the only solution to Eq.\ (\ref{gaugeEH}) is the trivial one,  $u_{tr}\equiv 0$. In Sec.~\ref{subsec:static-mono} we argued that the static, globally-regular Lorenz-gauge monopole solution is unique; hence it seems we have a numerical approach which recovers the desired solution (up to a trivial mass correction). It remains to be shown that this method can be applied to the Kerr problem.


\subsection{The $m=1$ perturbation}

Despite much experimentation with choices of $H_{\alp}$, we have not yet found a GLG that can lead to a stable evolution in the $l=1$, $m=1$ even-parity sector. Finding such a gauge remains a high priority for future work.

\begin{figure} 
 \begin{center}
  \includegraphics[width=8cm]{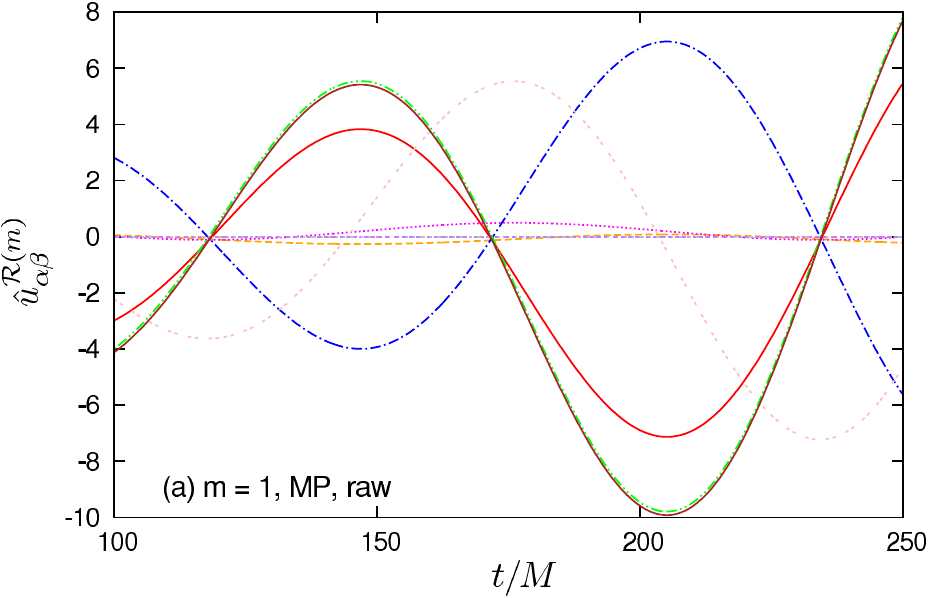}
  \includegraphics[width=8cm]{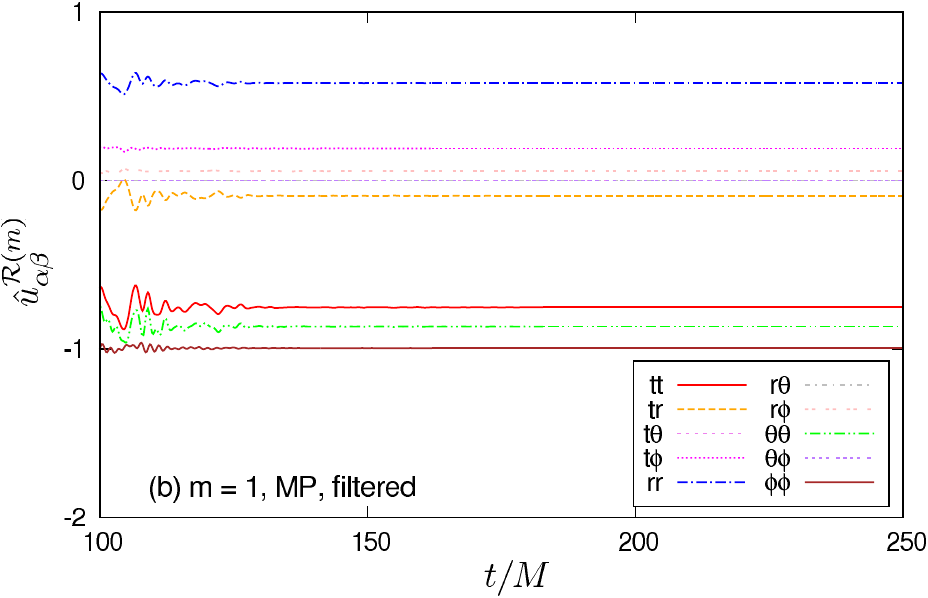}
  \includegraphics[width=8cm]{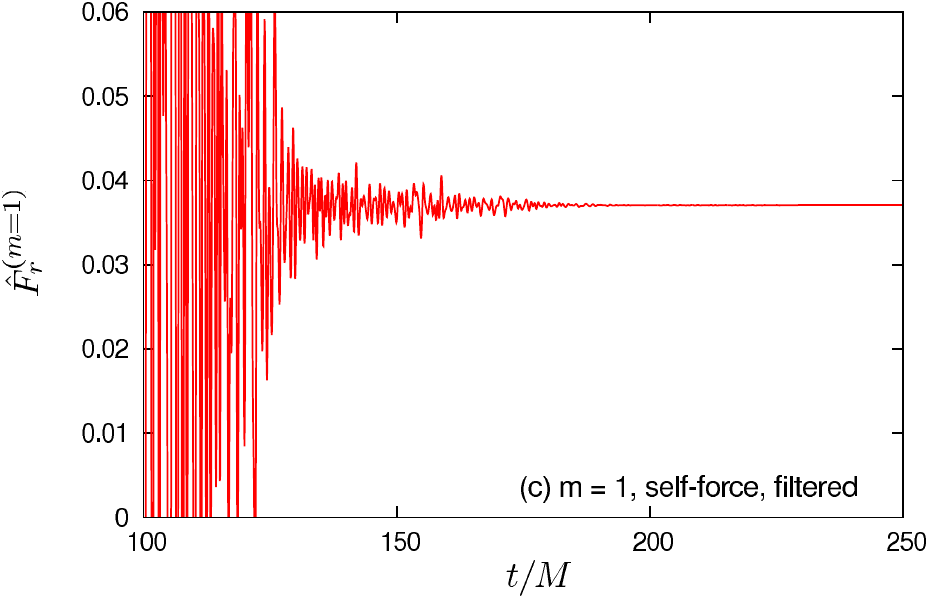}
 \end{center}
 \caption{Application of a frequency filter, Eq.~(\ref{freq-filter}), to the $m=1$ mode. Here we show numerical data on the worldline for $r_0=7M$, for (a) the unfiltered field components $\hat{u}^{\mathcal{R}}_{\alp \bet}$, whose envelops exhibit a linear-in-$t$ growth; (b) the components after filtering, which appear stationary; and (c) the radial self-force mode $\hat{F}_r^{(m=1)}$, after the filter has been applied. The $m=1$ mode contribution overwhelmingly dominates the total radial GSF, whose value is $\sim 0.0301 (\mu/M)^2$ (cf.\ Table \ref{tbl:Force}).
 }
 \label{fig:m1-filtered}
\end{figure}

Recall the $m=1$ numerical instability takes the form of a linearly growing (Lorenz-)gauge mode. For circular orbits, we can remedy the situation in an ad-hoc manner, simply by applying a ``frequency filter''
\beq
u_{\alp \bet}^{(m=1)} \rightarrow - \frac{1}{\Omega^2} \frac{\partial^2}{\partial t^2} u_{\alp \bet}^{(m=1)}  \label{freq-filter}
\eeq
to our numerical solutions. This eliminates any linearly growing gauge modes, while retaining the $\propto \exp(\pm i \Omega t)$ part of the dipole perturbation, which has the physically desired time dependence. 
The main drawback of this procedure is that we must compute the derivatives numerically, and this has the potential to amplify numerical error. In practice, we find that the trick works reasonably well for orbits in the strong field, but decreases in effectiveness at larger radii, as the orbital frequency decreases. Figure \ref{fig:m1-filtered} shows the effect of applying this filter in the case $r_0 = 7M$. An advantage of this technique is that it can also be applied to circular orbits in Kerr spacetime. However, it should be viewed as a stopgap until one can find a stable gauge for $m=1$ evolutions.

The limitations of the frequency-filter method are revealed when we attempt to compute the dissipative component of the GSF. 
The mode $F_t^{(m=1)}$ arises primarily from an $l=3$ contribution, since $l=1$ is nonradiative. However, we find that an $l=1$ contribution does arise generically in finite-resolution simulations in 2+1D (its amplitude reduces to zero only in the limit of infinite resolution). Unfortunately,  for realistic resolutions we find that this spurious contribution in fact dominates the numerical value of $F_t^{(m=1)}$.  

\begin{figure}
 \begin{center}
   \includegraphics[width=8cm]{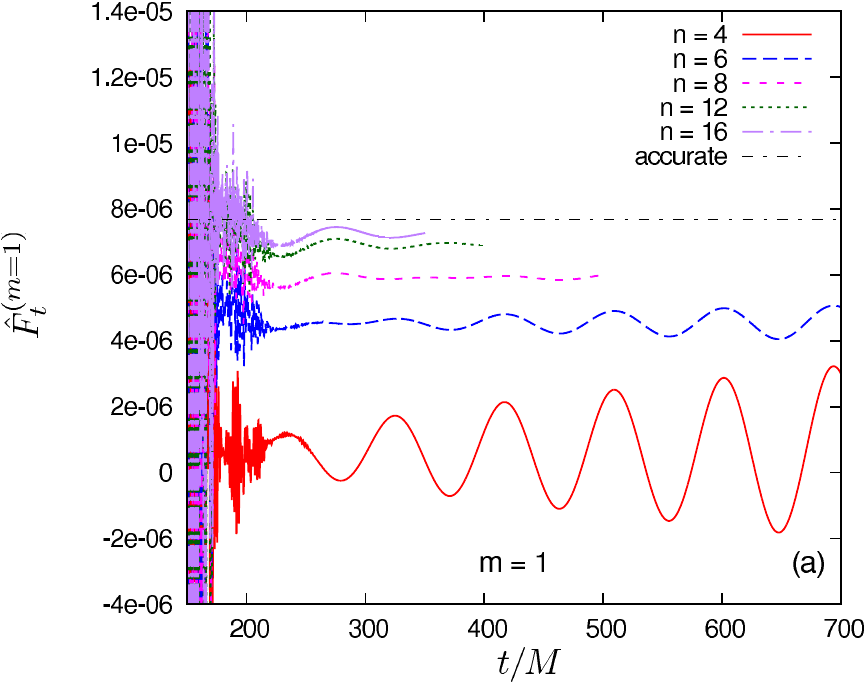}
   \includegraphics[width=8cm]{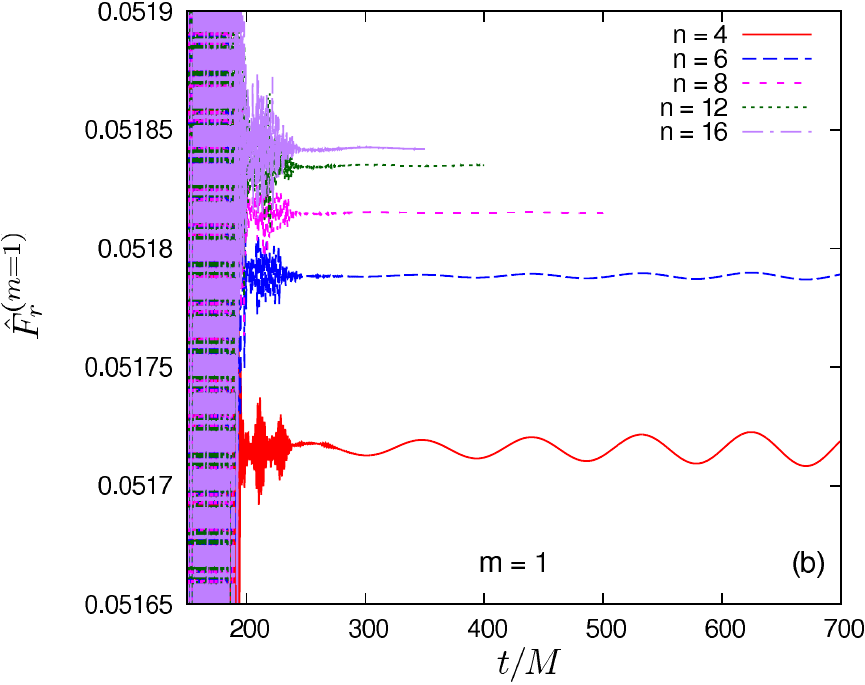}
 \end{center}
 \caption{Filtered numerical data for $F_{\mu}^{(m=1)}$. The plots show the temporal and radial components of the GSF, $\hat{F}_{t}^{(m=1)}$ and $\hat{F}_{r}^{(m=1)}$, extracted on the worldline as a function of time, for $r_0=6M$ and with various resolutions $\Delr = M / n$, $\Delq = \pi / (8n)$, after the application of the frequency filter (\ref{freq-filter}). Both plots exhibit non-decaying oscillations that persist after filtering (which removes the linear-in-$t$ instability). In the case of $F_t$ (left panel) the oscillations are large in comparison with the accurate value, which is quoted here from Ref.\ \cite{AWB} and shown as a dash-dot black line. In the case of $F_r$ (right panel) the oscillations lead to a much smaller relative error.}
\label{fig:F-m1}
\end{figure}

The problem is illustrated in Fig.~\ref{fig:F-m1}. The figure shows filtered data for the $m=1$ temporal and radial components of the GSF, as a function of simulation time $t$, for various grid resolutions $\Delr = M / n$, $\Delq = \pi / (8n)$. The plot for $F_t^{(m=1)}$ shows the imprint of residual oscillations, at frequency $\Omega$. At low resolutions (e.g., $n=4$), it is clear that the oscillations begin to grow at late times. At higher resolutions, the amplitude of the oscillations is reduced. Nevertheless, at achievable resolutions, the oscillations are still large in comparison with the magnitude of $F_t^{(m=1)}$, which makes the extrapolation to $n \rightarrow \infty$ rather fraught (even after a suitable averaging procedure), and leads to substantial loss of accuracy. The plot for $F_r$ shows similar oscillations, with a similar absolute amplitude. In this case, however, the oscillations are very small in comparison with the magnitude of $F_r^{(m=1)}$. 

The above limitations emphasize the need for a more systematic stabilization method for the $m=1$ mode. We will revisit the problem in Paper IV of the series, in the Kerr context.

\section{Numerical results: Part II\label{sec:results:2}}

Let us now present a selection of numerical results for the total GSF, and compare with previous results available in the literature \cite{Barack:Sago:2007, Sago:Barack:Detweiler:2008}. 

The accuracy of any numerical calculation is naturally limited by various sources of error. In Sec.~IV of Paper I, and Sec.~IVF of Paper II, we described three important sources: discretization error, due to the use of a grid of finite resolution; relaxation error, due to the residual effect of imperfect initial data; and $m$-mode summation error, due to approximating an infinite sum with a finite number of terms. In principle, these errors can be reduced by increasing the resolution (decreasing $\Delr, \Delq$); running for longer (increasing $t_{\text{max}}$); and computing more $m$-modes (increasing $m_{\text{max}}$), respectively. In practice, the situation is more subtle, since the sources of error are co-dependent. For example, computing more modes requires additional resolution to resolve the sharper physical features and faster oscillation ($\sim \exp(-i m \Omega t)$) of the high-$m$ modes. 
We applied the methods of Papers I \& II to reduce these errors as far as possible (i.e.~Richardson extrapolation; power-law fitting for late-time tails; and fitting the large-$m$ to an analytic model). The residual errors were then estimated in the standard way, by comparing results from different resolutions $n_{\text{max}}$, times $t_{\text{max}}$ and high-$m$ cut-offs $m_{\text{max}}$. 

To be specific, to obtain the total GSF for a given orbital radius $r_0$, we computed all modes up to $m_{\rm max}=15$ for resolutions $n = 4, 6, 8, 12$ and $16$, with $\Delr = M / n$ and $\Delq = \pi / (8n)$. Since the $m$-modes $\hat F^{r(m)}_{\rm self}$ fall off rather slowly, as $m^{-2}$, at large $m$, it is important to estimate the contribution from the remaining $m>m_{\rm max}$ modes by fitting to an analytic model, $\hat F^{r(m)}_{\rm self} \approx A/m^2 + B/m^3 + C/m^4$. On the other hand, the $m$-modes $\hat F^{t(m)}_{\rm self}$ fall off exponentially fast with $m$, so for this component the large-$m$ contribution is negligible---cf.\ Figure \ref{fig:m-mode-convergence}. For the modes $m=0$ and $m=1$ we used the method described in the preceding section. As discussed above, the $m=1$ contribution to $\hat F^{t}_{\rm self}$ suffers from a large numerical error, especially at large $r_0$. To obtain a final value for $\hat F^{t}_{\rm self}(r\geq 10M)$, we allowed ourselves here to use the crude estimate $F_t^{(m=1)} \approx \frac{25M}{896r_0} F_t^{(m=2)}$ from Ref.\ \cite{Finn:Thorne:2000}. 
The error in $m=1$ (due to the limitation of the frequency-filter method for eliminating gauge modes) is more perfidious than a relaxation error, because it does not necessarily diminish as $t_{\text{max}}$ increases (see Fig.~\ref{fig:F-m1}). We found that it dominates the error budget for our calculation of $F_t^{\rm self}$. However, for calculations of $F_r^{\rm self}$ and $\tilde{H}$ the discretization and $m$-mode summation errors are dominant, and broadly comparable in magnitude at these resolutions.

Table \ref{tbl:Force} displays an indicative sample of results for the total GSF, as well as results for the Detweiler MP invariant $\tilde{H}$ defined in Eq.~(\ref{Htil-def}). The error estimate, shown as a digit in parantheses, is found by combining the various error estimates in quadrature. Comparison with the accurate results of Refs.\  \cite{Barack:Sago:2007, Sago:Barack:Detweiler:2008} (obtained via a 1+1D implementation) shows that our method recovers the GSF and $\tilde{H}$ to within a few parts in $10^{4}$.  In all cases, our results agree with those of \cite{Barack:Sago:2007, Sago:Barack:Detweiler:2008} to within the estimated error in our data.

\begin{table}
\begin{tabular}{| l | l c | l c | l c |}
\hline
\multicolumn{7}{|c|}{Time-domain GSF results} \\
\hline
 & \multicolumn{2}{c|}{$\tilde{H}$} & \multicolumn{2}{c|}{$(M^2/\mu^2)F_t^{\text{self}}$} & \multicolumn{2}{c|}{$(M^2/\mu^2) F_r^{\text{self}}$} \\
\hline
\multirow{2}{*}{$r_0 = 6M$} & $-5.2355(6)$ & \multirow{2}{*}{$\e1$ } & $\msp 1.3299(2)$ &  \multirow{2}{*}{$\e3$ } & $\msp 3.6695(7)$ & \multirow{2}{*}{$\e2$} \\
	          &  $-5.23602$ & & $\msp 1.32984$ & & $\msp 3.66992$ &  \\
\hline
\multirow{2}{*}{$r_0 = 7M$} & $-4.0314(4)$ & \multirow{2}{*}{$\e1$ } & $\msp 5.293(2)$ &  \multirow{2}{*}{$\e4$ } & $\msp 3.0096(3)$ & \multirow{2}{*}{$\e2$} \\
	          &  $-4.03177$ & & $\msp 5.29358$ & & $\msp 3.00985$ &  \\
\hline
\multirow{2}{*}{$r_0 = 8M$} & $-3.3022(3)$ & \multirow{2}{*}{$\e1$ } & $\msp 2.482(2)$ &  \multirow{2}{*}{$\e4$ } & $\msp 2.4475(4)$ & \multirow{2}{*}{$\e2$} \\
	          &  $-3.30239$ & & $\msp 2.48055$ & & $\msp 2.44769$ &  \\
\hline
\multirow{2}{*}{$r_0 = 10M$} & $-2.4462(6)$ & \multirow{2}{*}{$\e1$ } & $\msp 7.348(7)^{*}$ &  \multirow{2}{*}{$\e5$ } & $\msp 1.6736(2)$ & \multirow{2}{*}{$\e2$} \\
	          &  $-2.44630$ & & $\msp 7.35254$ & & $\msp 1.67369$ &  \\
\hline
\multirow{2}{*}{$r_0 = 14M$} & $- 1.6270(4)$ & \multirow{2}{*}{$\e1$} & $\msp 1.2583(9)^{*}$ & \multirow{2}{*}{$\e5$} & $\msp 9.0685(3)$ & \multirow{2}{*}{$\e3$} \\
	          &  $-1.62705$ & & $\msp 1.25872$ &  & $\msp 9.06858$ &  \\
\hline
\multirow{2}{*}{$r_0 = 20M$} & $-1.0889(3)$ & \multirow{2}{*}{$\e1$} & $\msp 2.033(4)^{*}$ & \multirow{2}{*}{$\e6$} & $\msp 4.620(1)$ & \multirow{2}{*}{$\e3$} \\
	          &  $-1.08893$ &  & $\msp 2.02994$ &  & $\msp 4.61896$ &  \\
\hline
\end{tabular}
\caption{
Numerical results for the temporal and radial components of the GSF in Lorenz gauge, $F_t^{\text{self}}$ and $F_r^\text{self}$, and for the Detweiler MP invariant $\tilde{H}$ [defined in Eq.~(\ref{Htil-def})], for a range of orbital radii $r_0$.  The table compares results from our 2+1D $m$-mode implementation (upper entries), with results from the 1+1D $l$-mode implementation of Barack and Sago \cite{Barack:Sago:2007} and Barack, Sago and Detweiler \cite{Sago:Barack:Detweiler:2008} (lower entries). Parenthetical figures in the upper values indicate the estimated error bar on the last quoted decimals; in the lower entries all figures are significant. The values of $F_t^{\text{self}}$ marked with an asterisk were computed by replacing the $m=1$ contribution (which suffers from large numerical error) with the crude estimate $F_t^{(m=1)} \approx 25M F_t^{(m=2)} / (896 r_0)$ from Ref.\ \cite{Finn:Thorne:2000}. In the large-$r_0$ limit, $\tilde H$, $F^{\rm self}_r$ and $F^{\rm self}_t$ scale as $r_0^{-1}$, $r_0^{-2}$ and $r_0^{-5}$, respectively, which explains the relative loss of accuracy in the dissipative component $F_t$. All values were computed from runs with resolutions $\Delr = M/n$, $\Delq = \pi / (8n)$ where $n = 4, 6, 8, 12, 16$, and $\tmax = 300M$.}
\label{tbl:Force}
\end{table}

\section{Summary and Outlook\label{sec:outlook}}

This paper---the third in our series---represents a significant step towards the goal of computing the GSF on the Kerr spacetime. Let us take this opportunity to review the achievements herein, and the challenges that remain. 

In this work we have reported on the first implementation of the $m$-mode scheme for a practical GSF computation. The implementation represents the first realisation of the approach outlined in 2005 in Sec.~V of Ref.~\cite{Barack:Lousto:2005}. In addition to demonstrating that the GSF may be computed with a 2+1D time-domain implementation utilizing the puncture formulation, we also have shown that the GSF may be computed \emph{entirely in the time domain} (at least for circular orbits). That is, whereas previous studies resorted to frequency-domain methods to obtain the low multipoles, $l=1$ and $l=0$, we have persevered to tackle the problem of obtaining these modes within a time-domain scheme. 

Our approach is based on a constraint-damped Z4 formulation of the linearized Einstein equations, with the constraints being provided by the Lorenz-gauge conditions. In Ref.~\cite{Barack:Lousto:2005} it was shown that this approach works well within the 1+1D setting, for modes $l\ge2$. Here, we have shown for the first time that the approach also works in the 2+1D setting, for modes $m \ge 2$, provided that we make a judicious choice of constraint damping. Furthermore, in our formulation, the modes $m = 0$ and $m=1$ are found to be free from exponential-in-$t$ instabilities. Nevertheless, these modes are still plagued by linear-in-$t$ instabilities, which are associated with the low-multipole $l < 2$ sector. In Sec.~\ref{sec:low-multipoles} we argued that the linear-in-$t$ growth is attributed to certain (Lorenz-)gauge vacuum modes, which are globally regular and have the correct behavior on the future horizon.  Such gauge modes are clearly ``unphysical'' (as they are inconsistent with the helical symmetry of the perturbed spacetime) and must be eliminated from our simulations. Unfortunately, this cannot be achieved via application of boundary conditions alone, because (at any finite time) the undesirable gauge modes satisfy the same physical boundary conditions as the physical solution. Instead, we have been forced to reconsider the foundations of the Lorenz-gauge formulation itself.

In Sec.~\ref{sec:stabilization} we investigated the idea of employing a Generalized Lorenz Gauge to promote stability, and we identified a class of gauge drivers which could render the axisymmetric ($m=0$) mode stable, in the sense that arbitrary initial data evolve towards a stationary ($\partial_t \hh_{\alp \beta} = 0$) solution. We picked a specific GLG with the special property that a static MP ($\partial_t \hh_{\alp \beta} = 0$ and $\hh_{ti} = 0$) would automatically be in Lorenz gauge. Then we demonstrated that (in some sense) there was only one degree of freedom remaining in the class of stationary solutions, so that it became straightforward to find initial data which led us to the static, Lorenz gauge solution.

Thus far we have not found it possible to identify a GLG which promotes stability in the even-parity $l=1$ sector. Instead, we have been forced to employ an ad-hoc ``trick'', whereby a frequency-filter is applied to the Lorenz-gauge evolution to eliminate the undesirable time-growing gauge modes.  We demonstrated that this trick works sufficiently well, insomuch as it allows us to compute $F_r$ and $\tilde{H}$ to an accuracy of a few parts in $10^{4}$, for strong-field circular orbits. Nevertheless, our handling of the $m=1$ mode remains the weakest point of our analysis, and the largest source of numerical error. 

Let us now contemplate the obstacles that remain to impede our pursuit of the first GSF calculation in Kerr geometry. On Kerr spacetime, we are fundamentally limited by the lack of separability of the tensorial equations; indeed, this was the primary motivation for the development of the $m$-mode approach. Inevitably, non-separability means that we have a more limited understanding of the low-multipole sector. For example, since we cannot separate the MP into $l$ modes, we cannot obtain an analytic solution for the Lorenz-gauge ``monopole'' piece (or even define what that piece is). However, the analytical techniques developed in this work can take us a long way toward being able to construct physical mass and angular momentum perturbations in Kerr, in the Lorenz gauge. Building on a key theme of Sec.~\ref{sec:low-multipoles}, we may seek a basis of homogeneous Lorenz-gauge solutions, by considering gauge vectors satisfying $\Box \xi_\alp = 0$, as well as mass perturbations akin to the ``conformal'' solution of Eq.\ (\ref{SolA}). The energy and angular momentum content of such solutions may be assessed by the method of Sec.~\ref{subsec:conservation}. Armed with a sufficiently large basis of homogeneous solutions, we may apply post-hoc corrections to our numerical results, in order to seek a globally regular solution which has the correct physical content (i.e.~energy and angular momentum). Once more, an insistence upon Lorenz gauge will lead us to a MP which is not asymptotically-regular in the $tt$ component (and possibly in $t\phi$). 

Inevitably, we are left with some questions that can only be answered by attempting a Kerr implementation. Are linear-in-$t$ gauge modes present on Kerr? We presume yes, since the scalar-type gauge modes (which are straightforward to find on Kerr) are implicated. Does the GLG approach, generalized to Kerr, restore stability to the $m=0$ mode? If so, is the class of stationary solutions just as simple, in the sense described above? 

Let us now outline some areas for future research. The lack of stability in the low modes strikes to the core of the feasibility of the time-domain Lorenz-gauge approach. Addressing stability is a high priority, not least because (we believe) the issue will also affect the 3+1D time-domain approach of Vega \emph{et al.}.  In 3+1D one is not so easily able to isolate the $m=0$ and $m=1$ modes from the rest of the system, and the problem may well hinder any naive attempt to evolve the 3+1D system. We may seek lessons from other approaches to the Z4 formulation in Numerical Relativity. Using a range of (usually implicit) gauge drivers, various authors have shown that stable evolutions may be achieved, even for the non-linear problem. One key difference is that their approach is usually based around a Cartesian grid and the harmonic gauge, rather than spherical coordinates and the Lorenz gauge. A further complication is that moving to more general gauges will require some careful thought about how regularization of the GSF may be achieved and how physically-meaningful quantities may be extracted \cite{Gralla:2011zr}. Nevertheless, there is clearly much scope for a fertile exchange of ideas.  

There are a number of ways in which the efficiency of our numerical scheme could be improved. First, the method of hyperboloidal slicing, in combination with compactification of the domain \cite{Zenginoglu:2008}, shows much potential to dramatically reduce the computational burden for time-domain calculations. Second, the method of mesh refinement under development by Thornburg \cite{Thornburg:2009,Thornburg:2010} is ripe for application to this problem. Third, the fourth-order puncture formulation of Wardell {\it et al.}~\cite{Wardell:Vega:Thornburg:Diener:2011} may be easily adapted to our $m$-mode scheme. Fourth, a careful optimization of the initial data could help us minimize the seeding of junk radiation that decays away only slowly in the low multipoles.

The $m$-mode GSF program has four immediate priorities: (i) to find a GLG which renders the $m=1$ mode stable; (ii) to find a GLG which is appropriate for generic orbits (bearing in mind that we want the transformation from the GLG to Lorenz gauge to be regular); (iii) to implement a fourth-order puncture scheme, which will yield both a more rapid convergence with grid resolution and a more rapid convergence of the $m$-mode sum, leading to higher numerical accuracy; and (iv) to press ahead with the an implementation in Kerr.  
In the longer term we have two aims: (i) to develop the method up to the point where it can handle self-consistent orbital evolutions, in which the influence of the GSF is applied at every time step to correct the orbital trajectory, and (ii) to incorporate GSF effects which are second-order in $\mu/M$, based on the recent formulation by Pound \cite{Pound:2012nt,Pound:2012dk} and Gralla \cite{Gralla:2012db}.

\acknowledgments

SD acknowledges support from EPSRC through grant No.~EP/G049092/1. LB acknowledges support from STFC through grants No.~ST/J00135X/1, and from the European Research Council under grant No.~304978.  We are grateful for the use of the {\sc Iridis} 3 HPC facility at the University of Southampton. We thank David Hilditch for discussions on generalized gauges within Z4 schemes, and Niels Warburton and Sarp Akcay for providing numerical data for comparison and validation.

\appendix

\section{Key quantities in the 2+1D field equations\label{appendix:fieldeq}}

We give here explicit expressions for the matrices $\MM_{\alp \bet}^{(m)}$ and $\Mtil_{\alp \bet}^{(m)}$ appearing in Eq.~(\ref{no-gcd}) and (\ref{sam-gcd}), respectively. 
Recall $\dot{u}$ denotes differentiation with respect to $t$, and $u^\prime$ denotes differentiation  with respect to the tortoise coordinate $r_\ast$ defined in Eq.~(\ref{rstar-def}). For brevity we set $M=1$ throughout this appendix; the correct factors of $M$ can be easily recovered, if needed, by considering the dimensionality of the various terms. Also for brevity we omit the suffix $(m)$ in $u_{\alpha\beta}^{(m)}$. 

The coupling terms in the original 2+1D field equations (\ref{no-gcd}), prior to the imposition of gauge constraint damping, are given by
\beq
\MM_{tt}^{(m)} = \frac{2\left( 2 r^2 ( \dot{u}_{tr}  - u'_{tt} ) + u_{tt} - u_{rr} \right)}{r^4} + \frac{4f \left(u_{tt} - u_{rr} \right)}{r^3} + \frac{2 f^2 \left(u_{\theta\theta} + u_{\phi\phi} \right)}{r^3}, \label{MM00}
\eeq
\beq
\MM_{tr}^{(m)} = - \frac{2 f^2 \left( \cosq u_{t\theta} + i m u_{t\phi} \right)}{r^2 \sinq} + \frac{2 (\dot{u}_{tt} + \dot{u}_{rr} - 2 u'_{tr})}{r^2} - \frac{2 f^2 (u_{tr} + \dq u_{t\theta} )}{r^2},
\eeq
\beq
\MM_{t\theta}^{(m)} = -\frac{f(u_{t\theta} + 2im\cosq u_{t\phi})}{r^2 \sinqsq} + \frac{2(\dot{u}_{r\theta} - u'_{t\theta})}{r^2} + \frac{f[ (4+r) u_{t\theta} + 2 r \dq u_{tr} ]}{r^3} - \frac{f^2 u_{t\theta}}{r^2},
\eeq
\beq
\MM_{t\phi}^{(m)} = -\frac{f (u_{t\phi} - 2im\cosq u_{t\theta})}{r^2 \sinqsq} + \frac{2 f i m u_{tr}}{r^2 \sinq} + \frac{2 (\dot{u}_{r\phi} - u'_{t\phi})}{r^2} + \frac{f (4+r) u_{t\phi}}{r^3} - \frac{f^2 u_{t\phi}}{r^2},
\eeq
\begin{eqnarray}
\MM_{rr}^{(m)} &=&  -\frac{4f^2 ( \cosq u_{r\theta} + i m u_{r\phi} )}{r^2 \sinq} + \frac{2 [ 2r^2 (\dot{u}_{tr} - u'_{rr} ) + u_{rr} - u_{tt} ]}{r^4} - \frac{4f (u_{tt} - u_{rr})}{r^3} \nn \\
 && - \frac{2f^2 ( 2r u_{rr} + u_{\theta\theta} + u_{\phi\phi} + 2 r \dq u_{r\theta} )}{r^3} + \frac{2 f^3 (u_{\theta\theta} + u_{\phi\phi})}{r^2}, 
\end{eqnarray}
\begin{eqnarray}
\MM_{r\theta}^{(m)} &=& -\frac{f (u_{r\theta} + 2 i m \cosq u_{r\phi} ) }{r^2 \sinqsq} - \frac{2 f^2 [\cosq (u_{\theta\theta} - u_{\phi\phi}) + i m u_{\theta\phi}] }{r^2 \sinq} + \frac{2 (\dot{u}_{t\theta} - u'_{r\theta})}{r^2} \nn \\
 && + \frac{f [(4+r) u_{r\theta} + 2r \dq u_{rr}]}{r^3} - \frac{f^2(5 u_{r\theta} + 2 \dq u_{\theta\theta})}{r^2},
\end{eqnarray}
\begin{eqnarray}
\MM_{r\phi}^{(m)} &=& -\frac{f (u_{r\phi} - 2 i m \cosq u_{r\theta})}{r^2 \sinqsq} - \frac{2f [2 f \cosq u_{\theta\phi} +  i m (f u_{\phi\phi} - u_{rr})  ] }{r^2 \sinq} + \frac{2 (\dot{u}_{t\phi} - u'_{r\phi})}{r^2} \nn \\ 
&& + \frac{f(4+r)u_{r\phi}}{r^3} - \frac{f^2(5 u_{r\phi} + 2 \dq u_{\theta\phi})}{r^2},
\end{eqnarray}
\beq
\MM_{\theta\theta}^{(m)} = -\frac{2f [ u_{\theta\theta} - u_{\phi\phi} + 2 i m \cosq u_{\theta\phi} ] }{r^2 \sinqsq} + \frac{2 (u_{tt} - u_{rr})}{r^3} + \frac{2 f (u_{rr} + u_{\theta\theta} + 2 \dq u_{r\theta} )}{r^2} - \frac{2f^2 (u_{\theta\theta} + u_{\phi\phi})}{r^2},
\eeq
\beq
\MM_{\theta\phi}^{(m)} = -\frac{2f [ 2u_{\theta\phi} - i m \cosq (u_{\theta\theta} - u_{\phi\phi} )  ]}{r^2 \sinqsq} - \frac{2 f (\cosq u_{r\phi} - i m u_{r\theta})}{r^2 \sinq} + \frac{2 f (u_{\theta\phi} + \dq u_{r\phi})}{r^2},
\eeq
\begin{eqnarray}
\MM_{\phi\phi}^{(m)} &=&  \frac{2f (u_{\theta\theta} - u_{\phi\phi} + 2 i m \cosq u_{\theta\phi}) }{r^2 \sinqsq} + \frac{4f (\cosq u_{r\theta} + i m u_{r\phi})}{r^2 \sinq} + \frac{2(u_{tt} - u_{rr})}{r^3} + \frac{2f (u_{rr} + u_{\phi\phi})}{r^2} \nn \\ && - \frac{2f^2(u_{\theta\theta} + u_{\phi\phi})}{r^2} . \label{MM33}
\end{eqnarray}

The components of the $m$-decomposed divergence $Z_\mu = {\bar h}{_{\mu \nu} ^{\ \ ;\nu}}$ are given by
\beq
\ZZ_{t}^{(m)} \equiv \mu^{-1} f Z_t = \frac{f (\cosq u_{t\theta} + i m u_{t\phi})}{r^2 \sinq} - \frac{\dot{u}_{tt} - u'_{tr}}{r} + \frac{f (u_{tr} + \dq u_{t\theta}) }{r^2},   \label{gauge-t}
\eeq
\beq
\ZZ_{r}^{(m)} \equiv \mu^{-1} f^2 Z_r = \frac{f (\cosq u_{r\theta} + i m u_{r\phi})}{r^2 \sinq} - \frac{\dot{u}_{tr} - u'_{rr}}{r} + \frac{u_{tt} - u_{rr}}{r^3} + \frac{f(u_{rr} + \dq u_{r\theta})}{r^2} - \frac{f^2 (u_{\theta\theta} + u_{\phi\phi})}{r^2},   \label{gauge-r}
\eeq
\beq
\ZZ_{\theta}^{(m)} \equiv \mu^{-1} (f/r) Z_\theta = \frac{f [\cosq (u_{\theta\theta} - u_{\phi\phi}) + i m u_{\theta\phi}]}{r^2 \sinq} - \frac{\dot{u}_{t\theta} - u'_{r\theta}}{r} + \frac{f(2 u_{r\theta} + \dq u_{\theta\theta})}{r^2},
\eeq
\beq
\ZZ_{\phi}^{(m)} \equiv \mu^{-1} [f/(r\sinq)] Z_\phi = \frac{f (2 \cosq u_{\theta\phi} + i m u_{\phi\phi})}{r^2 \sinq} - \frac{\dot{u}_{t\phi} - u'_{r\phi}}{r} + \frac{f (2 u_{r\phi} + \dq u_{\theta\phi})}{r^2} . \label{gauge-p}
\eeq
The factors of $f$ are introduced so that the components $\ZZ_{\alpha}$ are regular at the event horizon---see the discussion in the main text.

The coupling terms in the eventual 2+1D field equations (\ref{sam-gcd}), after the imposition of gauge constraint damping, are given by
\begin{eqnarray}
\Mtil_{tt}^{(m)} &=& \frac{4f(\cosq u_{t\theta} + i m u_{t\phi})}{r^3 \sinq} + \frac{2 \left[ u_{tt} - u_{rr}  \right]}{r^4} + \frac{4 \left[ \dot{u}_{tr} + u'_{tr} - ( \dot{u}_{tt} + u'_{tt} ) \right]}{r^2} \nn \\ && + \frac{4f (u_{tt} + u_{tr} - u_{rr} + \dq u_{t\theta})}{r^3} + \frac{2 f^2 (u_{\theta\theta} + u_{\phi\phi})}{r^3}, \label{Mtil00}
\end{eqnarray}
\begin{eqnarray}
\Mtil_{tr}^{(m)} &=& -\frac{2f \left[ \cosq ( (r-3) u_{t\theta} - u_{r\theta} ) + i m ( (r-3) u_{t\phi} - u_{r\phi} )  \right]}{r^3 \sinq} + \frac{2(u_{tt} - u_{rr})}{r^4} + \frac{2 [\dot{u}_{rr} + u'_{rr} - (\dot{u}_{tr} + u'_{tr})]}{r^2} \nn \\
 && + \frac{2f (u_{rr} + u_{tr} + \dq u_{r\theta} + \dq u_{t\theta})}{r^3} - \frac{2 f^2 (r u_{tr} + r \dq u_{t\theta}^{(m)} +  u_{\theta\theta} + u_{\phi\phi}  )}{r^3},
\end{eqnarray}
\begin{eqnarray}
\Mtil_{t\theta}^{(m)} &=& -\frac{f (u_{t\theta} + 2 i m \cosq u_{t\phi})}{r^2 \sinqsq}  +  \frac{2 f [\cosq (u_{\theta\theta} - u_{\phi\phi}) + i m u_{\theta\phi}]}{r^3 \sinq} + \frac{2 [\dot{u}_{r\theta} + u'_{r\theta} - (\dot{u}_{t\theta} + u'_{t\theta})]}{r^2} \nn \\
&& + \frac{f [(4+r)u_{t\theta} + 4u_{r\theta} + 2 \dq u_{\theta\theta} + 2r \dq u_{tr}]}{r^3} - \frac{f^2 u_{t\theta}}{r^2},
\end{eqnarray}
\begin{eqnarray}
\Mtil_{t\phi}^{(m)} &=& -\frac{f (u_{t\phi} - 2 i m \cosq u_{t\theta})}{r^2 \sinqsq} + \frac{2f [2 \cosq u_{\theta\phi} + i m (r u_{tr} + u_{\phi\phi}) ]}{r^3 \sinq} + \frac{2 [\dot{u}_{r\phi} + u'_{r\phi} - (\dot{u}_{t\phi} + u'_{t\phi})]}{r^2}  \nn \\
&& + \frac{f ((4+r) u_{t\phi} + 4 u_{r\phi} + 2 \dq u_{\theta\phi})}{r^3} - \frac{f^2 u_{t\phi}}{r^2} ,
\end{eqnarray}
\begin{eqnarray}
\Mtil_{rr}^{(m)} &=& - \frac{4 f (r - 3) (\cosq u_{r\theta} + i m u_{r\phi})}{r^3 \sinq} + \frac{2(u_{tt} - u_{rr})}{r^4} - \frac{4f[u_{tt} - 2 u_{rr} - \dq u_{r\theta}]}{r^3} \nn \\ 
&& - \frac{2f^2 [3(u_{\theta\theta} + u_{\phi\phi}) + 2r(u_{rr} + \dq u_{r\theta})]}{r^3} + \frac{2f^3 (u_{\theta\theta} + u_{\phi\phi})}{r^2},
\end{eqnarray}
\begin{eqnarray}
\Mtil_{r\theta}^{(m)} &=& -\frac{f (u_{r\theta} + 2 i m \cosq u_{r\phi} ) }{r^2 \sinqsq} - \frac{2f(r-3) [ \cosq (u_{\theta\theta} - u_{\phi\phi}) + i m u_{\theta\phi} ]}{r^3 \sinq}  \nn \\  && + \frac{f ((8+r)u_{r\theta} + 2 \dq u_{\theta\theta} + 2r \dq u_{rr})}{r^3} 
 - \frac{f^2 (5 u_{r\theta} + 2 \dq u_{\theta\theta})}{r^2},
\end{eqnarray}
\begin{eqnarray}
\Mtil_{r\phi}^{(m)} &=& -\frac{f (u_{r\phi} - 2 i m \cosq u_{r\theta})}{r^2 \sinqsq} - \frac{2 f (2(r-3)\cosq u_{\theta\phi} + i m (r-3)u_{\phi\phi} - i m r u_{rr})}{r^3 \sinq} \nn \\  && + \frac{f( (8+r) u_{r\phi} + 2 \dq u_{\theta\phi})}{r^3} 
- \frac{f^2(5 u_{r\phi} + 2 \dq u_{\theta\phi})}{r^2},
\end{eqnarray}
\begin{eqnarray}
\Mtil_{\theta\theta}^{(m)} &=& \MM_{\theta\theta}^{(m)} , \quad
\Mtil_{\theta\phi}^{(m)} = \MM_{\theta\phi}^{(m)} , \quad
\Mtil_{\phi\phi}^{(m)} = \MM_{\phi\phi}^{(m)}. \label{Mtil33}
\end{eqnarray}

\section{Elliptic integrals in the puncture and effective source\label{appendix:elliptic}}

In this appendix we list relevant elliptic integrals that go into the construction of the $m$-mode puncture function [Eq.\ (\ref{hatchi})] and effective source [Eqs.\ (\ref{Zsc}) and  (\ref{DeltaZ})]. Below, $\ellK(\cdot)$ and $\ellE(\cdot)$ denote complete elliptic integrals of the first and second kinds, respectively, defined by
\begin{eqnarray}
\ellK(k) = \int_0^{\pi/2} \left(1 -  k^2 \sin^2 x \right)^{-1/2} dx, \quad  \label{ellipK} \\
\ellE(k) = \int_0^{\pi/2} \left(1 -  k^2 \sin^2 x \right)^{1/2} dx .  \label{ellipE}
\end{eqnarray}

The relevant integrals are
\begin{eqnarray}
I_{0}^m &\equiv&  \int_{-\pi}^{\pi} \epsilon_{\mathcal{P}}^{-1} \, e^{-i m \delta \phi} d(\dph)   
 =  \frac{\gamma}{B^{1/2}} \left[ p_{0K}^m(\p) \ellK(\gamma) + p_{0E}^m(\p) \ellE(\gamma) \right]   ,\label{integralI0}  \\
I_{1}^m &\equiv&  \int_{-\pi}^{\pi} \epsilon_{\mathcal{P}}^{-3} \, e^{-i m \dph} d(\dph)   
 =  \frac{\gamma}{B^{3/2}} \left[ p_{1K}^m(\p) \ellK(\gamma) + \p^{-2} p_{1E}^m(\p) \ellE(\gamma) \right]   ,\label{integralI1}  \\
I_{2}^m &\equiv&  \int_{-\pi}^{\pi} \epsilon_{\mathcal{P}}^{-3} \, \cos \dph \, e^{-i m \dph} d(\dph)  
 =  \frac{\gamma}{B^{3/2}} \left[ p_{2K}^m(\p) \ellK(\gamma) + \p^{-2} p_{2E}^m(\p) \ellE(\gamma) \right]  ,  \\
I_{3}^m &\equiv&  \int_{-\pi}^{\pi} \epsilon_{\mathcal{P}}^{-5} \cos^2 \left( \dph / 2 \right) \, e^{-i m \dph} d(\dph)  
 =  \frac{\gamma}{\p^2 B^{5/2}} \left[ p_{3K}^m(\p) \ellK(\gamma) + \p^{-2} p_{3E}^m(\p) \ellE(\gamma) \right]  , \label{integralI3}  \\
I_{4}^m &\equiv& \int_{-\pi}^{\pi} \epsilon_{\mathcal{P}}^{-5} \sin^2 \left( \dph \right) \, e^{-i m \dph} d(\dph) 
 =  \frac{\gamma}{B^{5/2}} \left[ p_{4K}^m(\p) \ellK(\gamma) + \p^{-2} p_{4E}^m(\p) \ellE(\gamma) \right] ,  \\
I_{5}^m &\equiv& \int_{-\pi}^{\pi} \epsilon_{\mathcal{P}}^{-5} \sin^2 \left(\dph / 2 \right) \, e^{-i m \dph} d(\dph) 
=  \frac{\gamma^3}{B^{5/2}} \left[ p_{5K}^m(\p) \ellK(\gamma) + \p^{-2} p_{5E}^m(\p) \ellE(\gamma) \right]  ,\label{integralI5} \\
I_{6}^m &\equiv& \int_{-\pi}^{\pi} \epsilon_{\mathcal{P}}^{-3} \sin^2 \left(\dph \right) \, e^{-i m \dph} d(\dph) 
=  \frac{\gamma}{B^{3/2}} \left[ p_{6K}^m(\p) \ellK(\gamma) + p_{6E}^m(\p) \ellE(\gamma) \right]  ,
\label{integralI6} \\
I_{7}^m &\equiv&  \int_{-\pi}^{\pi} \epsilon_{\mathcal{P}}^{-1} \, \cos \dph \, e^{-i m \delta \varphi} d( \dph )   
 =  \frac{\gamma}{B^{1/2}} \left[ p_{7K}^m(\p) \ellK(\gamma) + p_{7E}^m(\p) \ellE(\gamma) \right]   , \label{integralI-1}  
\end{eqnarray}
and
\begin{eqnarray}
J_0^m &\equiv& \int_{-\pi}^{\pi} \eps_{\mathcal{P}}^{-1} \sin \dph \, e^{-i m \dph} d(\dph) = \frac{-i \rho}{B^{1/2}}  \left[q_{0K}^m \ellK(i / \rho) +  q_{0E}^m \ellE(i / \rho) \right] , \label{integralJ0} \\
J_1^m &\equiv& \int_{-\pi}^{\pi} \eps_{\mathcal{P}}^{-3} \sin \dph \, e^{-i m \dph} d(\dph) = \frac{-i}{B^{3/2} \rho} \left[q_{1K}^m \ellK(i / \rho) + \rho^2 q_{1E}^m \ellE(i / \rho) \right] , \label{integralJ1} \\
J_2^m &\equiv& \int_{-\pi}^{\pi} \eps_{\mathcal{P}}^{-3} \sin \dph \, \cos \dph \, e^{-i m \dph} d(\dph) = \frac{-i \gam}{B^{3/2}} \left[q_{2K}^m \ellK(\gam) +  q_{2E}^m \ellE(\gam) \right] , \label{integralJ2} \\
J_3^m &\equiv& \int_{-\pi}^{\pi} \eps_{\mathcal{P}}^{-5} \sin \dph \, \cos^2( \dph / 2) \, e^{-i m \dph} d(\dph) = \frac{-i \gam}{B^{5/2}} \left[ q_{3K}^m \ellK(\gam) +  \rho^{-2} q_{3E}^m \ellE(\gam) \right] , \label{integralJ3} \\
J_4^m &\equiv& \int_{-\pi}^{\pi} \eps_{\mathcal{P}}^{-5} \sin \dph \, \sin^2( \dph ) \, e^{-i m \dph} d(\dph) = \frac{-i}{B^{5/2} \rho} \left[ q_{4K}^m \ellK(i/\rho) +  \rho^{2} q_{4E}^m \ellE(i / \rho)  \right] , \label{integralJ4} \\
J_5^m &\equiv& \int_{-\pi}^{\pi} \eps_{\mathcal{P}}^{-5} \sin \dph \, \sin^2( \dph / 2) \, e^{-i m \dph} d(\dph) = \frac{-i \gam^2}{B^{5/2} \rho} \left[ q_{5K}^m \ellK(i/\rho) +  \rho^{2} q_{5E}^m \ellE(i / \rho)  \right] . \label{integralJ5}
\end{eqnarray}
Here 
\beq
\rho^2 \equiv A / (4 B) ,    \quad \quad \gamma \equiv \left( 1 + \p^2 \right)^{-1/2},   \label{rhotilde-def}
\eeq
\beq
A \equiv P_{rr} \delta r^2 + P_{\theta\theta} \delta\theta^2 + Q_{rr} \delta r^3 + Q_{\theta\theta} \delta r \delta\theta^2 , \quad \quad B \equiv P_{\phi\phi} + Q_{\phi\phi} \delta r,  \label{AB-def}
\eeq
with
\beq
\Prr = f_0^{-1}, \quad\quad P_{\theta\theta} = r_0^2, \quad\quad P_{\phi\phi} = \frac{r_0^3 f_0}{r_0-3M}, \label{Pco-def}
\eeq
\beq
Q_{rr} = - \frac{M}{r_0^2 f_0^2}, \quad\quad Q_{\theta\theta} = r_0, \quad\quad Q_{\phi\phi} = r_0 \, \left( \frac{r_0 - M}{r_0 - 3M} \right).   \label{Qco-def}
\eeq
We remind $f_0=1-2M/r_0$. The quantities $p_{nK}^m$ and $p_{nE}^m$ in the above expressions are polynomials in $\rho^2$. Some of these polynomials (for $m=0, \ldots, 5$) were tabulated in Tables I and II of Ref.~\cite{Barack:Golbourn:2007} (where the notation $p_{K/E}^m$ was used in place of our $p_{0K/E}^m$), and a few others in Table VIII of Paper I. All remaining polynomials relevant for the current work are given in Tables \ref{table:p-polynomials}, \ref{table:q-polynomials1} and \ref{table:q-polynomials2} below, again for $m=0, \ldots, 5$. Higher-$m$ polynomials can be similarly obtained using a symbolic algebra package.

\begin{table}
\begin{tabular}{l l}
\hline
\hline
$m$ & $p_{6K}^m(\rho)$ \\
$0$ & $0$ \\
$1$ & $\tfrac{2}{3}\left(4\p^2+1\right)\left(4\p^2+3\right)$  \\
$2$ & $\tfrac{2}{5}\left(2\p^2 + 1\right)\left(64\p^4+64\p^2+5\right)$ \\
$3$ & $\tfrac{8192}{35}\p^8 + \tfrac{16384}{35}\p^6 + \tfrac{6176}{21}\p^4 + \tfrac{6304}{105}\p^2 + 2$ \\
$4$ & $\tfrac{2}{315} \left(2 \p^2 + 1\right)\left(81920\p^8 + 163840\p^6 + 97408\p^4 + 15488\p^2 + 315\right)$ \\
$5$ & $\tfrac{1048576}{231}\p^{12} + \tfrac{1048576}{77}\p^{10} + \tfrac{17686528}{1155}\p^8 + \tfrac{9158656}{1155}\p^6 + \tfrac{427744}{231}\p^4 + \tfrac{180832}{1155}\p^2 + 2$  \\
\hline
\vspace{0.2cm} \\
\hline
$m$ & $p_{6E}^m(\rho)$ \\
$0$ & $0$ \\
$1$ & $-\tfrac{16}{3}\left(2\p^2 + 1\right)(\p^2+1)$ \\
$2$ & $-\tfrac{4}{5}\left(\p^2 + 1\right)\left(64 \p^4 + 64\p^2 + 9\right)$ \\
$3$ & $-\tfrac{16}{105}\left(2\p^2+1\right)\left(\p^2+1\right)\left(768\p^4+768\p^2+53\right)$ \\
$4$ & $-\tfrac{4}{315}\left(\p^2+1\right)\left(81920\p^8+163840\p^6+102528\p^4+20608\p^2+683\right)$ \\
$5$ & $-\tfrac{16}{1155}\left(2\p^2+1\right)\left(\p^2+1\right)\left(163840\p^8+327680\p^6+194304\p^4+30464\p^2+659\right)$ \\
\hline
\vspace{0.2cm} \\
\hline
$m$ & $p_{7K}^m(\rho)$ \\
$0$ & $4\p^2 + 2$ \\
$1$ & $\tfrac{16}{3}\p^4+\tfrac{16}{3}\p^2+2$ \\
$2$ & $\tfrac{2}{15}\left(8\p^2+5\right)\left(2\p^2+1\right)\left(8\p^2+3\right)$ \\
$3$ & $\tfrac{2048}{35}\p^8 + \tfrac{4096}{35}\p^6 + \tfrac{1712}{21}\p^4 + \tfrac{2416}{105}\p^2 + 2$ \\
$4$ & $\tfrac{2}{315}\left(2\p^2+1\right)\left(16384\p^8+32768\p^6+21632\p^4+5248\p^2+315\right)$ \\
$5$ & $\tfrac{524288}{693}\p^{12}+\tfrac{524288}{231}\p^{10}+\tfrac{9181184}{3465}\p^8+\tfrac{5255168}{3465}\p^6 + \tfrac{100624}{231}\p^4 + \tfrac{192496}{3465}\p^2 + 2$ \\
\hline
\vspace{0.2cm} \\
\hline
$m$ & $p_{7E}^m(\rho)$ \\
$0$ & $-4 \left(\p^2+1\right)$ \\
$1$ & $-\tfrac{8}{3}\left(2\p^2+1\right)\left(\p^2+1\right)$ \\
$2$ & $-\tfrac{4}{15}\left(\p^2+1\right)\left(64\p^4+64\p^2+19\right)$ \\
$3$ & $-\tfrac{8}{105}\left(2\p^2+1\right)\left(\p^2+1\right)\left(384\p^4+384\p^2+79\right)$ \\
$4$ & $-\tfrac{4}{315}\left(\p^2+1\right)\left(16384\p^8+32768\p^6+22656\p^4+6272\p^2+523\right)$ \\
$5$ & $-\tfrac{8}{3465}\left(2\p^2+1\right)\left(\p^2+1\right)\left(163840\p^8+327680\p^6+215424\p^4+51584\p^2+3079\right)$ \\
\hline
\hline
\end{tabular}
\caption{The polynomials $p_{nK}^m$ and $p_{nE}^m$ appearing in Eqs.~(\ref{integralI6}) and (\ref{integralI-1}), for $n = 6, 7$ and $m=0, \ldots, 5$.}
\label{table:p-polynomials} 
\end{table}

\begin{table}
\begin{tabular}{l l}
\hline
\hline
$m$ & $q_{0K}^m(\rho)$ \\
$0$ & $0$ \\
$1$ & $-\tfrac{16}{3}\left(\p^2 + 1\right)$  \\
$2$ & $-\tfrac{128}{15}\left(2\p^2 + 1\right)\left(\p^2 + 1\right)$ \\
$3$ & $-\tfrac{16}{35}\left(\p^2 + 1\right)\left(128 \p^4 + 128\p^2 + 27\right)$ \\
$4$ & $-\tfrac{1024}{315}\left(2\p^2+1\right)\left(\p^2+1\right)\left(32\p^4+32\p^2+5\right)$ \\
$5$ & $-\tfrac{16}{693}\left(\p^2+1\right)\left(32768\p^8+65536\p^6+44160\p^4+11392\p^2+875\right)$  \\
\hline
\vspace{0.2cm} \\
\hline
$m$ & $q_{0E}^m(\rho)$ \\
$0$ & $0$ \\
$1$ & $\tfrac{8}{3}\left(2 \p^2 + 1 \right)$  \\
$2$ & $\tfrac{16}{15}\left(16\p^4 + 16\p^2 + 1\right)$ \\
$3$ & $\tfrac{8}{35}\left(2\p^2 + 1\right)\left(128\p^4 + 128\p^2 + 3\right)$ \\
$4$ & $\tfrac{65536}{315}\p^8 + \tfrac{131072}{315}\p^6 + \tfrac{26624}{105}\p^4 + \tfrac{2048}{45}\p^2 + \tfrac{32}{63}$ \\
$5$ & $\tfrac{8}{693}\left(2\p^2 + 1\right)\left(32768\p^8+65536\p^6+38016\p^4+5248\p^2+35 \right)$  \\
\hline
\vspace{0.2cm} \\
\hline
$m$ & $q_{1K}^m(\rho)$ \\
$0$ & $0$ \\
$1$ & $2\left(2\p^2 + 1\right)$  \\
$2$ & $\tfrac{4}{3}\left(4\p^2+1\right)\left(4\p^2+3\right)$ \\
$3$ & $\tfrac{2}{5}\left(2\p^2+1\right)\left(128\p^4 + 128\p^2 + 15\right)$ \\
$4$ & $\tfrac{16384}{35}\p^8 + \tfrac{32768}{35}\p^6 + \tfrac{12800}{21}\p^4 + \tfrac{14848}{105}\p^2 + 8$ \\
$5$ & $\tfrac{2}{63}\left(2\p^2+1\right)\left(32768\p^8 + 65536\p^6 + 40576\p^4 + 7808\p^2 + 315\right)$  \\
\hline
\vspace{0.2cm} \\
\hline
$m$ & $q_{1E}^m(\rho)$ \\
$0$ & $0$ \\
$1$ & $-4$  \\
$2$ & $-\tfrac{32}{3}\left(2\p^2+1\right)$ \\
$3$ & $-\tfrac{512}{5}\p^4 - \tfrac{512}{5}\p^2 - \tfrac{92}{5}$ \\
$4$ & $-\tfrac{256}{105}\left(2\p^2 + 1\right)\left(96\p^4 + 96\p^2 + 11\right)$ \\
$5$ & $-\tfrac{131072}{63}\p^8 - \tfrac{262144}{63}\p^6 - \tfrac{18944}{7}\p^4 - \tfrac{5632}{9}\p^2 - \tfrac{2252}{63}$  \\
\hline
\vspace{0.2cm} \\
\hline
$m$ & $q_{2K}^m(\rho)$ \\
$0$ & $0$ \\
$1$ & $\tfrac{2}{3}\left(4\p^2+1\right)\left(4\p^2+3\right)$  \\
$2$ & $\tfrac{4}{5}\left(2\p^2+1\right)\left(32\p^4+32\p^2+5\right)$ \\
$3$ & $\tfrac{8192}{35}\p^8 + \tfrac{16384}{35}\p^6 + \tfrac{2208}{7} \p^4 + \tfrac{2848}{35}\p^2 + 6$ \\
$4$ & $\tfrac{8}{315}\left(2\p^2+1\right)\left(20480\p^8+40960\p^6+26368\p^4+5888\p^2+315\right)$ \\
$5$ & $\tfrac{1048576}{231}\p^{12} + \tfrac{1048576}{77}\p^{10} + \tfrac{3645440}{231}\p^8 + \tfrac{2048000}{231}\p^6 \tfrac{568544}{231}\p^4 + \tfrac{22944}{77}\p^2 + 10$  \\
\hline
\vspace{0.2cm} \\
\hline
$m$ & $q_{2E}^m(\rho)$ \\
$0$ & $0$ \\
$1$ & $-\tfrac{16}{3}\left(2\p^2+1\right)\left(\p^2+1\right)$  \\
$2$ & $-\tfrac{8}{5}\left(\p^2+1\right)\left(32\p^4+32\p^2+7\right)$ \\
$3$ & $-\tfrac{16}{35}\left(2\p^2+1\right)\left(\p^2+1\right)\left(256\p^4+256\p^2+41\right)$ \\
$4$ & $-\tfrac{16}{315}\left(\p^2+1\right)\left(20480\p^8+40960\p^6+27648\p^4+7168\p^2+533\right)$ \\
$5$ & $-\tfrac{16}{231}\left(2\p^2+1\right)\left(\p^2+1\right)\left(32768\p^8+65536\p^6+42240\p^4+9472\p^2+519\right)$  \\
\hline
\hline
\end{tabular}
\caption{The polynomials $q_{nK}^m$ and $q_{nE}^m$ appearing in Eqs.~(\ref{integralJ0}) -- (\ref{integralJ2}), for  $n=0,1,2$ and $m=0, \ldots, 5$.}
\label{table:q-polynomials1}
\end{table}

\begin{table}
\begin{tabular}{l l}
\hline
$m$ & $q_{3K}^m(\rho)$ \\
$0$ & $0$  \\
$1$ & $-\tfrac{1}{6}\left(8\p^2+5\right)$ \\
$2$ & $-\tfrac{1}{3}\left(32\p^4+40\p^2+11\right)$  \\
$3$ & $-\tfrac{1}{30}\left(8\p^2+5\right)\left(256\p^4+304\p^2+63\right)$ \\
$4$ & $-\tfrac{8192}{21}\p^8 - \tfrac{96256}{105}\p^6 - \tfrac{77824}{105}\p^4 - \tfrac{8256}{35}\p^2 - \tfrac{70}{3}$ \\
$5$ & $-\tfrac{1}{126}\left(8\p^2+1\right)\left(32768\p^8+90112\p^6+89344\p^4+37520\p^2+5565\right)$ \\
\hline
\vspace{0.2cm} \\
\hline
$m$ & $q_{3E}^m(\rho)$ \\
$0$ & $0$ \\
$1$ & $\tfrac{1}{6}\left(8\p^2+1\right)\left(\p^2+1\right)$  \\
$2$ & $\tfrac{1}{3}\left(\p^2+1\right)\left(32\p^4+24\p^2+1\right)$ \\
$3$ & $\tfrac{1}{30}\left(\p^2+1\right)\left(2048\p^6 + 2688\p^4 + 808\p^2 + 15\right)$ \\
$4$ & $\tfrac{2}{105}\left(\p^2+1\right)\left(20480\p^8 + 37888\p^6 + 21248\p^4 + 3488\p^2 + 35\right)$ \\
$5$ & $\tfrac{1}{126}\left(\p^2+1\right)\left(262144\p^{10} + 622592\p^8 + 509952\p^6 + 165248\p^4 + 17080\p^2 + 105\right)$  \\
\hline
\vspace{0.2cm} \\
\hline
$m$ & $q_{4K}^m(\rho)$ \\
$0$ & $0$ \\
$1$ & $\tfrac{2}{3}\left(4\p^2+1\right)\left(4\p^2+3\right)$  \\
$2$ & $\tfrac{4}{15}\left(2\p^2+1\right)\left(128\p^4+128\p^2+15\right)$ \\
$3$ & $\tfrac{8192}{21}\p^8 + \tfrac{16384}{21}\p^6 + \tfrac{10592}{21}\p^4 + \tfrac{800}{7}\p^2 + 6$ \\
$4$ & $\tfrac{8}{315}\left(2\p^2+1\right)\left(40960\p^8+81920\p^6+50048\p^4+9088\p^2+315\right)$ \\
$5$ & $\tfrac{1048576}{99}\p^{12} + \tfrac{1048576}{33}\p^{10} + \tfrac{24977408}{693}\p^8 + \tfrac{13254656}{693}\p^6 + \tfrac{1091936}{231}\p^4 + \tfrac{318496}{693}\p^2 + 10$  \\
\hline
\vspace{0.2cm} \\
\hline
$m$ & $q_{4E}^m(\rho)$ \\
$0$ & $0$ \\
$1$ & $-\tfrac{16}{3}\left(2\p^2 + 1\right)$  \\
$2$ & $-\tfrac{1024}{15}\p^4 - \tfrac{1024}{15}\p^2 - \tfrac{184}{15}$ \\
$3$ & $-\tfrac{16}{21}\left(2\p^2+1\right)\left(256\p^4 + 256\p^2 + 27\right)$ \\
$4$ & $-\tfrac{131072}{63}\p^8 - \tfrac{262144}{63}\p^6 - \tfrac{280576}{105}\p^4 - \tfrac{26624}{45}\p^2 - \tfrac{9328}{315}$ \\
$5$ & $-\tfrac{16}{693}\left(2\p^2+1\right)\left(229376\p^8+458752\p^6+278784\p^4+49408\p^2+1697\right)$  \\
\hline
\vspace{0.2cm} \\
\hline
$m$ & $q_{5K}^m(\rho)$ \\
$0$ & $0$ \\
$1$ & $\tfrac{1}{6}\left(\p^2+1\right)\left(8\p^2+3\right)$  \\
$2$ & $\tfrac{1}{3}\left(\p^2+1\right)\left(32\p^4+24\p^2+3\right)$ \\
$3$ & $\tfrac{1}{30}\left(\p^2+1\right)\left(8\p^2+3\right)\left(256\p^4+208\p^2+15\right)$ \\
$4$ & $\tfrac{2}{105}\left(\p^2+1\right)\left(20480\p^8+33792\p^6+17408\p^4+2976\p^2+105\right)$ \\
$5$ & $\tfrac{1}{126}\left(\p^2+1\right)\left(8\p^2+7\right)\left(32768\p^8+40960\p^6+15616\p^4+1904\p^2+45\right)$  \\
\hline
\vspace{0.2cm} \\
\hline
$m$ & $q_{5E}^m(\rho)$ \\
$0$ & $0$ \\
$1$ & $-\tfrac{1}{6}\left(8\p^2 + 7\right)$  \\
$2$ & $-\tfrac{32}{3}\p^4 - \tfrac{40}{3}\p^2 - 3$ \\
$3$ & $-\tfrac{1024}{15}\p^6 - \tfrac{576}{5}\p^4 - \tfrac{788}{15}\p^2 - \tfrac{51}{10}$ \\
$4$ & $-\tfrac{8192}{21}\p^8 - \tfrac{88064}{105}\p^6 - \tfrac{8704}{15}\p^4 - \tfrac{14528}{105}\p^2 - \tfrac{258}{35}$ \\
$5$ & $-\tfrac{131072}{63}\p^{10} - \tfrac{16384}{3}\p^8 - \tfrac{320512}{63}\p^6 - \tfrac{125248}{63}\p^4 - 292\p^2 - \tfrac{1231}{126}$  \\
\hline
\end{tabular}
\caption{The polynomials $q_{nK}^m$ and $q_{nE}^m$ appearing in Eqs.~(\ref{integralJ3}) -- (\ref{integralJ5}), for $n=3,4,5$ and $m=0, \ldots, 5$.}
\label{table:q-polynomials2}
\end{table}

\section{$m$-mode decomposition of the effective source\label{appendix:source}}

In Sec.~\ref{subsec:mmode-source} we introduced the effective source, and in Eq.~(\ref{SeffZeff}) we gave an expression for its $m$-mode decomposition. Here we complete the description by giving explicit expressions for the coefficients $c^k_{\alpha\beta}$ and $d^k_{\alpha\beta}$ featuring in Eq.~(\ref{DeltaZ}) for $\Delta Z_{\alp \bet}^{(m)}$.  Here $\Omega$ is the orbital frequency, $A_r$, $A_\theta$, $B_{r}$, $B_{\theta}$, etc., denote the partial derivatives of $A$ and $B$ defined in Eq.~(\ref{AB-def}), and we define $E_{\alp \bet} = u_{\alp}u_{\bet} + D_{\alp \bet} \dr$ where $D_{\alp \bet}$ were given in (\ref{Dtt})--(\ref{Dpp}). After setting $M=1$ the ($m$-independent) coefficients $c^k_{\alpha\beta}$ and $d^k_{\alpha\beta}$ read
\begin{eqnarray}
c_{tt}^0 &=& \frac{2 (r-1)}{r^4 f} E_{tt} + \frac{2(r-3)}{r^2} D_{tt} + \frac{2 f}{r^5 \sin^2\theta} E_{\phi\phi} , \\
c_{tt}^1 &=& \left(2 E_{tt} / r^2 -  f D_{tt} \right) \left(A_r + 2 B_r \right) ,\\
c_{tt}^2 &=& -2 \left( 2 E_{tt} / r^2 - f D_{tt} \right) B_r ,\\
c_{tt}^6 &=& 4 D_{tr} \Omega B / r^2 ,\\
c_{tt}^7 &=& -4 D_{tr} \Omega / r^2 ,\\
d_{tt}^0 &=& -8 (r - 1) D_{tr} / r^4 ,\\
d_{tt}^1 &=& 4 \Omega B E_{tt} / (f r^2)  + 4 B / (r^4 \sin^2 \theta) E_{t\phi} + 2 f (A_r + 2 B_r) D_{tr} / r^2  ,\\
d_{tt}^2 &=& -4 f B_r D_{tr} / r^2 ,\\
c_{tr}^0 &=& 2 E_{tt} / (r^4 f^2) - 2 E_{\phi\phi} / (r^5 \sin^2 \theta) ,\\
c_{tr}^6 &=& -2  B \left[ \Omega D_{tr} / (r^2 f) + D_{r\phi} / (r^4 \sin^2 \theta) \right] ,\\
c_{tr}^7 &=& 2 \Omega D_{tr} / (r^2 f) + 2 D_{r\phi} / (r^4 \sin^2 \theta) ,\\
d_{tr}^0 &=& \left[ (-r^4 \Omega^2 + 2 r^2 - 6 r + 4)/(r^4 f) +  (r^2 \sin^2 \theta)^{-2}  \right] D_{tr} ,\\
d_{tr}^1 &=& (A_r + 2 B_r) D_{tr} / r^2 - 2 (r - 3) B E_{t\phi} / (r^4 f \sin^2 \theta) ,\\
d_{tr}^2 &=& \left[ - 2 (B r^3 \Omega^2 + r f B_r) / (r^3 f) + 2 B / (r^2 \sin^2 \theta) \right] D_{tr} ,\\
c_{t\theta}^0 &=& -2 \cos \theta E_{\phi\phi} / (r^4 \sin^3 \theta) ,\\
d_{t\theta}^1 &=& -2 \cos \theta B E_{t\phi} / (r^2 \sin^3 \theta)  + f A_\theta D_{tr} / r ,\\
c_{t\phi}^0 &=& 5 E_{t\phi} / r^3 ,\\
c_{t\phi}^1 &=& A_\theta E_{t\phi} \cos \theta / (r^2 \sin \theta)  + \left( (r - 1) E_{t\phi} / r^2 - f D_{t\phi} \right)\left(A_r + 2 B_r \right) ,\\
c_{t\phi}^2 &=& - 2 B_r \left( (r - 1)E_{t\phi} / r^2  -  f D_{t\phi} \right) ,\\
c_{t\phi}^6 &=& -2 B \left( f D_{tr} / r - \Omega D_{r\phi} / r^2 \right) ,\\
c_{t\phi}^7 &=&  2 \left( f D_{tr} / r - \Omega D_{r\phi} / r^2 \right) ,\\
d_{t\phi}^0 &=& -4 (r - 1) D_{r\phi} / r^4 ,\\
d_{t\phi}^1 &=& 2 B E_{\phi\phi} / (r^4 \sin^2 \theta) + f (A_r + 2 B_r) D_{r\phi} / r^2 + 2 \Omega B E_{t\phi} / (r^2 f) ,\\
d_{t\phi}^2 &=& -2 f B_r D_{r\phi} / r^2  ,\\
c_{rr}^0 &=& -2 (3r - 7) E_{tt} / (r^4 f^3) + 2 (r - 5) E_{\phi\phi} / (r^5 f \sin^2 \theta) ,\\
c_{rr}^6 &=& 4(r-3) B D_{r\phi} / (r^4 f \sin^2 \theta) ,\\
c_{rr}^7 &=& -4(r - 3) D_{r\phi} / (r^4 f \sin^2 \theta) ,\\
c_{r\theta}^0 &=& 2 \cos \theta E_{\phi\phi} (r-3) / (r^4 f \sin^3 \theta) ,\\
c_{r\theta}^6 &=& 2 B D_{r\phi} \cos \theta / (r^2 \sin^3 \theta) ,\\
c_{r\theta}^7 &=& -2 D_{r\phi} \cos \theta / (r^2 \sin^3 \theta)  ,\\
d_{r\phi}^0 &=& D_{r\phi}  \left[ -(r^4 \Omega^2 - 4r^2 + 19r - 18)/(r^2 f) + 1 / \sin^2 \theta \right] / r^2 ,\\
d_{r\phi}^1 &=& -D_{r\phi} \left[ (r-4)(A_r + 2B_r) + A_\theta \cot \theta \right] / r^2 - 2 B (r-3) E_{\phi\phi} / (r^4 f \sin^2 \theta) ,\\
d_{r\phi}^2 &=& 2 D_{r\phi} \left[ (r- 4) B_r - r^3 \Omega^2 B / (fr) +  B / \sin^2 \theta \right] / r^2 ,\\
c_{\theta\theta}^0 &=& 2 E_{tt} / (r f) + 2 E_{\phi\phi} / (r^2 \sin^2 \theta) \left[ - f + 1 / \sin^2 \theta \right] ,\\
d_{\theta\phi}^0 &=& 4 f \cos\theta D_{r\phi} / (r \sin \theta) ,\\
d_{\theta\phi}^1 &=& A_\theta f D_{r\phi} / r - 2 B E_{\phi\phi} \cos \theta / (r^2 \sin^3 \theta)  ,\\
c_{\phi\phi}^0 &=& 2 E_{tt} \sin^2\theta / (r f) - 4 E_{\phi\phi} / r^3 - 2(r-3)D_{\phi\phi} / r^2 + 2 E_{\phi\phi} / (r^2 \sin^2 \theta) ,\\
c_{\phi\phi}^1 &=& f (2 E_{\phi\phi} - r D_{\phi\phi} ) (A_r + 2 B_r) / r  + 2 A_\theta E_{\phi\phi} \cos \theta / (r^2 \sin \theta) ,\\
c_{\phi\phi}^2 &=& -2 f (2 E_{\phi\phi} - r D_{\phi\phi}) B_r / r ,\\
c_{\phi\phi}^6 &=& - 4 f B D_{r\phi} / r ,\\
c_{\phi\phi}^7 &=& 4 f D_{r\phi} / r .
\end{eqnarray}
It should be noted that the appropriate effective source depends upon the choice of gauge constraint damping (Sec.~\ref{subsec:gauge-choice}); the expressions above correspond to the choice made in Eq.~(\ref{AC-eq}).

\section{Projection onto a tensor-harmonic basis\label{appendix:lmmodes}}

We give below formulas for constructing the multipole $lm$-mode functions $h^{(i)}_{lm}(r,t)$ (as defined by Barack and Lousto in \cite{Barack:Lousto:2005}) from our $m$-mode variables $u^{(m)}_{\alp\beta}(t,r,\theta)$. We omit the suffix $(m)$ for brevity.
\begin{eqnarray}
\bar h^{(1)}_{lm}(r,t) &=& 2 \pi \int_0^\pi \sin \theta \left( u_{tt} + u_{rr} \right) \Y d\theta  \label{hilm-eq1} ,\\
\bar h^{(2)}_{lm}(r,t) &=& 4 \pi \int_0^\pi \sin \theta \, u_{tr} \Y d\theta  ,\\
\bar h^{(3)}_{lm}(r,t) &=& 2 \pi \int_0^\pi \sin \theta \left( u_{tt} - u_{rr} \right) \Y d\theta ,\\
\bar h^{(4)}_{lm}(r,t) &=& 4 \pi \int_0^\pi \left[ \sin \theta \, u_{t\theta} \, \partial_\theta  - i m u_{t\phi} \right] \Y d\theta ,\\
\bar h^{(5)}_{lm}(r,t) &=& 4 \pi \int_0^\pi \left[ \sin \theta \, u_{r\theta} \, \partial_\theta  - i m u_{r\phi} \right] \Y d\theta ,\\
\bar h^{(6)}_{lm}(r,t) &=& 2 \pi \int_0^\pi \sin \theta \left( u_{\theta\theta} + u_{\phi\phi} \right) \Y d\theta ,\\
\bar h^{(7)}_{lm}(r,t) &=& 2 \pi \int_0^\pi \left[ \sin \theta (u_{\theta\theta} - u_{\phi\phi}) \hat{D}_2 + 2 u_{\theta\phi} \hat{D}_1 \right] \Y d\theta  ,\\
\bar h^{(8)}_{lm}(r,t) &=& 4 \pi \int_0^\pi \left[ -i m u_{t\theta} - \sin \theta u_{t\phi} \partial_ \theta \right] \Y d\theta ,\\
\bar h^{(9)}_{lm}(r,t) &=& 4 \pi \int_0^\pi \left[ -i m u_{r\theta} - \sin \theta u_{r\phi} \partial_ \theta \right] \Y d\theta  ,\\
\bar h^{(10)}_{lm}(r,t) &=& 2 \pi \int_0^\pi \left[ (u_{\theta\theta} - u_{\phi\phi}) \hat{D}_1 - 2 \sin \theta u_{\theta\phi} \hat{D}_2 \right] \Y d\theta .\label{hilm-eq10} 
\end{eqnarray}
Here $Y_{lm}(\theta,\phi)$ are the usual spherical harmonics, an asterisk denotes complex conjugation, $\hat{D}_1 \equiv 2 \left(\partial_\theta - \cot \theta\right) \partial_\phi$ and $\hat{D}_2 \equiv \partial_{\theta\theta} - \cot \theta \partial_\theta - (\sin \theta)^{-2} \partial_{\phi \phi}$.

\end{document}